\documentclass[12pt]{book}

\usepackage{cranfieldthesis}

\usepackage[british]{babel}

\usepackage{array}

\usepackage{CJKutf8}

\usepackage{mathtools}

\usepackage[backend=biber,style=authoryear,giveninits=true,maxbibnames=99,alldates=long,dateabbrev=false]{biblatex}

\renewbibmacro*{volume+number+eid}{%
  \printfield{volume}%
  \printfield{number}%
  \setunit{\addcomma\space}%
  \printfield{eid}}
\DeclareFieldFormat[article]{number}{\mkbibparens{#1}}

\renewbibmacro{in:}{%
  \ifboolexpr{%
     test {\ifentrytype{article}}%
     or
     test {\ifentrytype{inproceedings}}%
  }{}{\printtext{\bibstring{in}\intitlepunct}}%
}

\usepackage{pdflscape}

\usepackage{booktabs} 

\usepackage{longtable}

\newcolumntype{F}{>{\raggedright\arraybackslash}p{2cm}}
\newcolumntype{L}{>{\raggedright\arraybackslash}p{6cm}}

\usepackage{etoolbox}
\AtBeginEnvironment{longtable}{\singlespacing}
\usepackage{graphicx}

\graphicspath{ {img/} }

\title{Potential applications of Distributed Ledger Technology to the Defence Support Network}
\author{Lt Cdr J D Hackman RN}
\date{July 2018}
\school{Defence and Security}
\course{Information Capability Management}
\degree{MSc}
\academicyear{2017--2018}
\supervisor{Dr Duncan Hodges}
\copyrightyear{2018}

\usepackage[bookmarks]{hyperref}

\hypersetup{
    colorlinks=true,
    linkcolor=blue,
    filecolor=magenta,      
    urlcolor=cyan,
    pdftitle={Potential applications of Distributed Ledger Technology to the Defence Support Network},
    pdfauthor={Lt Cdr J D Hackman RN},
    pdfkeywords={Blockchain; Distributed Ledger Technology; Smart Contracts; Defence Support Network; Supply Chain},
    pdfsubject={MSc Information Capability Management},
}

\usepackage[all]{hypcap}

\begin{document}

\frontmatter

\maketitle

\begin{abstract}
Proponents of Distributed Ledger Technology (DLT) claim it could have an impact greater than the internet; a breakthrough defying organisational boundaries by securely storing data across trustless entities.  This would allow decisions to be made on verifiable data in an automated manner without the costs imposed by middlemen, with a corresponding economy-wide impact.  Despite this potential, real-world application is embryonic with public and private sectors rapidly seeking exploitation opportunities.  This research seeks to understand how DLT might apply to the Defence Support Network (DSN), the mechanism used to sustain UK Armed Forces with materiel and equipment. Drawing on academic and commercial models, a framework was produced for evaluating DLT use cases which measures utility, ease of implementation and impact.  Using a functionalist research paradigm, interviews were conducted with DLT and DSN experts on potential use cases, the data from which was then analysed against a lightweight version of the evaluation framework.  Results show that use cases involving codification, certification and supply chain provenance merit further investigation.  The research concluded with recommendations that the DSN should pilot DLT use cases, but these should be carefully selected utilising an evaluation framework due to DLT's emergent nature.

    \section*{Keywords}
    Blockchain; Distributed Ledger Technology; Smart Contracts; Defence Support Network; Supply Chain
\end{abstract}

\chapter{Acknowledgements}
My first thanks must go to my dissertation supervisor: Dr Duncan Hodges.  His contribution to this research can only be adequately described as `Nelsonian' - in that he provided me with battle winning guidance whilst allowing me the research freedom to close and engage the enemy.  I appreciate his wisdom and advice, and have enjoyed working with him.  His advice on \LaTeX ~also saved a number of frustrating hours.  Next I thank my interviewees who were kind of enough to give me their time and knowledge, there was not a single interview where I did not learn something new.  I also thank colleagues who engaged with my research, especially those within Navy Command Information Warfare Division, the Defence Logistics Directorate and DSTL; of particular note is Gary Glennon-Alty.  The patience and support of my line management during my entire MSc was also appreciated: Graham Cheshire and Captains P Waterhouse, M Rance, K Nicholson and A Parry Royal Navy.\\

\tableofcontents
\sslistoffigures

\sslistoftables

\begin{listofabbreviations}
    \abbrev{ACTO}{Attractive to Criminal or Terrorist Organisations}
    \abbrev{ALIS}{Autonomic Logistics Information System}
    \abbrev{AM}{Additive Manufacturing}
    \abbrev{BOSE}{Blockchain Oriented Software Engineering}
    \abbrev{CfA}{Contracting for Availability}
    \abbrev{CSIS}{Codification Support Information System} 
    \abbrev{DAO}{Decentralised Autonomous Organisation}
    \abbrev{DApp}{Decentralised Application} 
    \abbrev{DAR}{Defence Application Register}
    \abbrev{DE\&S}{Defence Equipment and Support}
    \abbrev{DL}{Distributed Ledger}
    \abbrev{DLT}{Distributed Ledger Technology}
    \abbrev{DSTL}{Defence Science and Technology Laboratory}
    \abbrev{DSN}{Defence Support Network}
    \abbrev{DT}{DE\&S Delivery Team}
    \abbrev{E\&AM}{Engineering and Asset Management}
    \abbrev{EEL}{Electronic Equipment Logbook}
    \abbrev{FOCJ}{Functional, Overlapping and Competing Jurisdictions}   
    \abbrev{GDPR}{General Data Protection Regulation}
    \abbrev{HMRC}{Her Majesty's Revenue and Customs}
    \abbrev{IS}{Information System}
    \abbrev{ITAR}{International Traffic in Arms Regulations}
    \abbrev{JAMES}{Joint Asset Management and Engineering Solutions}
    \abbrev{JSF}{Joint Strike Fighter}
    \abbrev{KSI}{Keyless Signature Infrastructure}
    \abbrev{Log IS}{Logistics Information System}
    \abbrev{MDL}{Mutual Distributed Ledger}
    \abbrev{MoD}{Ministry of Defence}
    \abbrev{NAO}{National Audit Office}
    \abbrev{NATO}{North Atlantic Treaty Organisation}
    \abbrev{NMCRL}{NATO Master Catalogue of References for Logistics}
    \abbrev{NSN}{NATO Stock Number}
    \abbrev{OEM}{Original Equipment Manufacturer}
    \abbrev{PAC}{Public Accounts Committee}
    \abbrev{PoA}{Proof-of-Authority}
    \abbrev{PoS}{Proof-of-Stake}
    \abbrev{PoW}{Proof-of-Work}
    \abbrev{SC}{Smart Contract}
    \abbrev{SCIS}{Support Chain Information Services}
    \abbrev{SCISRA}{Support Chain Information Services Architectural Repository}
    \abbrev{UKNCB}{United Kingdom National Codification Bureau}
    \abbrev{UoA}{Unit of Analysis}

\end{listofabbreviations}

\mainmatter

\chapter{Introduction}

\section{Overview}
\label{sec:Overview}

Bitcoin, the best known example of Distributed Ledger Technology (DLT), has been heralded as bigger than the internet by Silicon Valley entrepreneurs \parencite{Carlson2015}.  The UK Government's Chief Scientific Advisor believes the technology behind Bitcoin could underpin ``potential explosions of creative potential that catalyse exceptional levels of innovation'' \parencite{Walport2016}.  Not surprisingly a host of enterprises are now attempting to exploit this much trumpeted innovation hoping that, as SAP declare, DLT is ''ideally suited for... complex industry processes involving many untrusted parties taking part in reading and writing a multitude of transactions, decisions, and documents'' \parencite{Galer2017}.

At its most basic, a Distributed Ledger (DL) is  a database distributed across more than one organisational entity that makes use of cryptography \parencite{Kello2017}; typically it is used to store transactions of anything that might be considered an asset (e.g. cash, physical items, personal data, etc).  DL's unique value lies in sharing or validating the database across many different people or organisations, who do not necessarily trust each other.  This is achieved through a cryptographic consensus between all organisations, so that the data held is verified and represents one immutable version of the truth; once recorded, no party can amend the data. 

If this represents DLT's uniqueness, its promise is in reducing the friction between organisations. When each organisation holding its own siloed version of events share or agree data across organisations, a transactional cost is entailed.  Such cost is reflected by the existence of roles and processes whose function is to verify transactions, e.g solicitors, notaries, audits and performance measurement.   Given that once data is in a DL it is accepted as an accurate reflection of events, DLT eliminates the agreeing stage and the transactional cost, thereby increasing the efficiency of inter-party transactions \parencite{DAVIDSON2018}. This dissertation analyses how DLT might be applied to Defence, specifically to examine how it might benefit the Defence Support Network (DSN).

\section{Definitive articles}

\begin{figure}
\includegraphics[width=\textwidth]{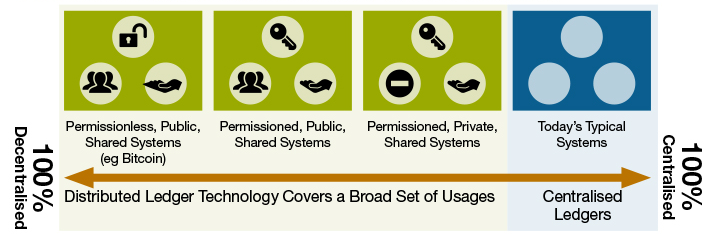}
\caption[Variants of DLT]{Variants of DLT \parencite{Walport2016}}
\label{fig:WalportTypesDLT}
\centering
\end{figure}

Given that DLT is an emergent technology, there is currently little agreement to standard terms used, although international standards are being developed \parencite{isoblockchain2017}. In view of this DLT, should not be referred to monolithically, as one might to the concept of relational databases.   For the purpose of this dissertation the definition used by the UK's Chief Scientific Officer \parencite{Walport2016} will be adopted as at Figure \ref{fig:WalportTypesDLT}.

This shows DLT is a wide term covering many applications, used to contrast this technology from predecessors - namely organisationally centred ledgers, whether paper ledgers, relational databases or NoSQL.  Walport \parencite*{Walport2016} further defines a DL as ``a type of database that is spread across multiple sites, countries or institutions, and is typically public'' - although as Figure \ref{fig:WalportTypesDLT} shows private versions are equally valid.  This definition is not universally accepted - Swanson \parencite*{Swanson2015} asserts a DL has to involve a legal entity; while Mainelli \parencite*{Mainelli2017b} prefers the term Mutual Distributed Ledger, usefully emphasising that the data is ``held in common or owned by no one.''

In the wider media the term `blockchain' is better known than DLT  \parencite{Deshpande2017}.  Blockchain, the technology underlying the cryptocurrency Bitcoin, is a list of transactions recorded in a block that is linked or `chained' through cryptography to previous blocks of transactions.  As Section \ref{sec:GenesisBlock} examines, Bitcoin is the first example of a blockchain and DL.  Bitcoin is not though the only blockchain, many others have utilised this format, for instance IBM markets a ``blockchain for business'' \parencite{Hartman2017}. 

However not all DLs are blockchains.  For instance, IOTA designed for the internet of things, uses ``the tangle'' where instead of transactions being recorded in blocks they are recorded in a directed acyclic graph \parencite{Popov2016}.  Due to the prevalence of the term, even DLT that do not rely on blockchains use the term liberally.  An example is Guardtime's Keyless Signature Infrastructure which is referred to as blockchain technology even though it does not make use of blocks of transactions, but rather a Merkle tree culminating in an internet published hash calendar \parencite{Buldas2013}.  Such is the market hype that even vendors selling an implementation of Git version-control (a common tool) claim to have a DLT product \parencite[p.~112]{Gerard2017}.  This semantic flexibility is partly because Bitcoin's blockchain, although unique in itself, is based on the synergy of already existing components \parencite{Narayanan2017}, further discussed in Chapter \ref{ch:LitReview}.

The UK's Chief Scientific Officer at Figure \ref{fig:WalportTypesDLT} makes a distinction between public and private DLTs, the former being subdivided further into permissionless versus permissioned.  At the far left of the diagram is the blockchain (i.e. Bitcoin's ledger) as an example of both public and permissionless DLT - anyone can view the transactions contained therein (public) and anyone can act as nodes on the network (permissionless) \parencite{Huberman2017}.  Ripple is an example of a public, permissioned DL \parencite{Schwartz2014} - that is transactions are available for all to see but nodes (or `gateways' in Ripple terminology) are run by trusted parties (i.e. financial institutions).  Corda, a DLT intended for use by banks, sits on the right of Figure \ref{fig:WalportTypesDLT}'s DLT spectrum being private and permissioned - this network is used only by authorised organisations, with transaction details  only accessible to the parties involved and regulators \parencite{Brown2016a}.  Indeed Corda is attempting to implement hardware-level privacy via chip-manufacturer Intel \parencite{Hearn2017}.  Lundb\ae k and Huth \parencite*{Lundbaek2017} however argue these terms, although used extensively in strategy papers, have little technical implementation consensus - and furthermore cannot be verified due to proprietary source code and documentation.

\section{Code is law}
\label{sec:codeislaw}

Smart contracts (SC), a concept that predates Bitcoin \parencite{Szabo1997}, is frequently associated with DLT.  Contracts constitute a written agreement between two parties where services or items are transacted based on conditions being met.   SCs are the evolution of this agreement into a logic-based form implemented by software.  Szabo \parencite*{Szabo1997} provides the example of a vending machine - a user provides cash and the machine returns an item  with no further human interaction. A SC could  be layered over a DL and conducts activity when certain conditions are met.  If a DL was used for land registry (as Sweden and India are evaluating \parencite{Bal2017}), then a SC could be instigated so that when a change of ownership was recorded, the contract executes a monetary transaction from the buyer instantly.  If cryptocurrency was used as the medium of exchange, a phenomenon already observed \parencite{Prynn2017}, this becomes even more plausible. SC are also relevant for more complex transactions, e.g. a farmer seeks insurance cover for temperature fluctuations outside an agreed range over an agreed time period \parencite{Mainelli2017}.  By automating transactions, SC promotes efficiency and could lead to structural changes in sectors that manage transactions.
 
Building on SCs are Decentralised Autonomous Organisations (DAO) \parencite{Johnston2013}.  DAOs are established by entrepreneur-programmers founding businesses comprised entirely of software; these entities  perform transactions dependent on algorithmic business plans.  They would receive payments from clients, execute trades based on a DL and then reimburse owners with profits; all without human agency.  Ethereum is a DL created to act as a platform for DAOs \parencite{Morris2015}. 

A prototype DAO is Bitbarista \parencite{Pschetz2017}, a coffee machine paid in cryptocurrency (although not unique in this aspect \parencite{Beck2016}).  The Bitbarista  uses the freedoms offered by cryptocurrencies to automate business processes such as ordering refills online and rewarding people for maintenance tasks (i.e. refilling coffee beans).  Using DLT removes barriers associated with traditional financial structures, such as accounts accessible to legal entities only, as opposed to machines.  Although unlikely to threaten Starbucks, it is an illustration of how removing frictions might shift how enterprises operate.

\section{Defence Support Network} \label{sec:DefenceSupportNetwork}

This dissertation examines how DLT might be applied in the environment of the DSN.  UK military doctrine \parencite[p.~9]{MoDJDP2015} defines the DSN as:

\begin{quotation}``a flexible set of supply chains connecting points of production and use, ensuring the most appropriate and efficient use of resources across the Whole Force, maximising information and technology to assure logistic support to operational commanders. The DSN consists of a series of linked nodes through which support is delivered in an agile manner, giving end-to-end visibility and control.''\end{quotation}

Although this statement opens by referring to tangibles - i.e. supply chains, production and use; the latter sentence talks more generally of `support'.  Support, in a Defence context, is defined by the same doctrine as encompassing, as well as physical items such as logistics and equipment, more abstract areas such as legal, medical and infrastructure.  This dissertation focuses on the concrete embodiment of this support, such as equipment and items of supply.

\begin{figure}
\includegraphics[width=\textwidth]{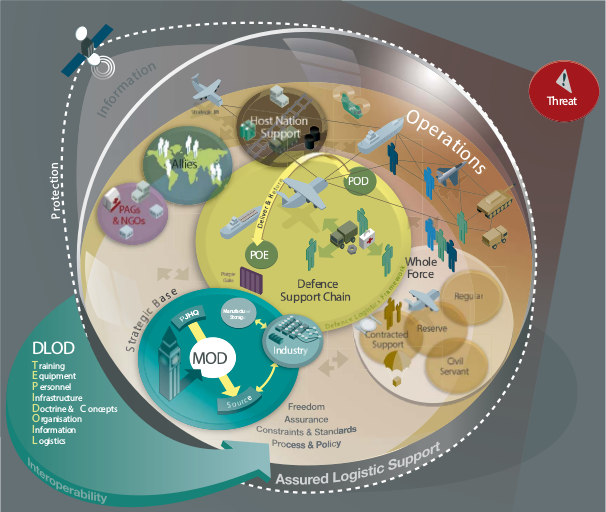}
\caption[The Defence Support Network]{The Defence Support Network \parencite[p.~10]{MoDJDP2015}}
\label{fig:DSNRichPicture}
\centering
\end{figure}

The definition also refers to the `Whole Force,' an acknowledgement of Defence's dependence on a wider pool than uniformed personnel - including civil servants, other government departments, contractors and external parties.  This is illustrated by the rich picture at Figure \ref{fig:DSNRichPicture} visualising the many organisations comprising the DSN.  The recent Afghanistan and Iraq conflicts brought into contrast MoD's increased reliance on contractor support \parencite[p.~34]{MinistryofDefence2015}.  This ties strongly into the DLT concept - if DLs are concerned with sharing information across organisational boundaries, then the more boundaries, the more impact DLT stands to make.

Information and technology is given a central role in the DSN according to the above definition.  The MoD has been extensively criticised \parencite{NationalAuditOffice2011} for weaknesses in exploiting logistics information, with multiple failures in understanding the complex interplay of assets and supply chains.  This makes a strong case for investigating how DLT, a new paradigm for managing information, might be beneficially applied to the DSN.

\section{Research question, aims and objectives}
\label{sec:researchQAO}

\subsection{Research question}
Understanding the background that has led to this research the question that this dissertation sets out to answer is:

\begin{quote}
How could DLT and SC be beneficially applied to the DSN?
\end{quote}

\subsection{Research aim}
A logical conclusion of this Research Question is the following Research Aim:

\begin{quote}
The aim of this research is to understand how DLT and SC might be applied to the DSN and the potential benefits.
\end{quote}

\subsection{Research objectives}
The Research Aim breaks  down into the following objectives: :

\begin{enumerate}
\item To define DLT (Section \ref{sec:Overview}) and smart contracts (Section \ref{sec:codeislaw}).

\item To define the DSN (Section \ref{sec:DefenceSupportNetwork}) and how DLT might address the challenges it faces (Section \ref{sec:challengesaddressed}).

\item To create a framework for evaluating the utility of DLT against use cases (Section \ref{sec:DSNevaluationframework}), drawing from academic or business models (Section \ref{sec:extantframeworks}).

\item To assess generic DSN use cases against a lightweight version of the evaluation framework identified in Objective 3 (Section \ref{sec:Resultsofquestionnaire}), by gathering quantitative and qualitative evidence from subject matter experts (Section \ref{sec:genericusecaseexploration}).

\item To explore further use cases of how DLT might apply to the DSN beyond the generic use cases of Objective 4 (Section \ref{sec:Widerusecaseexploration}).

\end{enumerate}

\chapter{Literature review}
\label{ch:LitReview}

A literature review of the concept of DLT is presented first, followed by a thematically structured analysis of its potential wider impact.

\section{Genesis block} 
\label{sec:GenesisBlock}

No discussion of DLT can be complete without referring to its genesis: the Bitcoin white paper by the pseudonymous Satoshi Nakamoto \parencite*{Nakamoto2008}.  This paper was distributed outside of either academic or commercial circles, having its roots in the cryptoanarchist community \parencite[p.~36]{Frisby2014}. 

Bitcoin introduced four key concepts as proposed by Antonopoulos \parencite*[p.~40]{Antonopoulos2014}:

\begin{enumerate}
\item The Bitcoin protocol itself - a decentralised peer-to-peer network.

\item The blockchain - a public transaction ledger. 

\item Rules for establishing consensus for validating independent transactions and issuing currency. 

\item A proof-of-work algorithm - the mechanism by which global consensus is reached on which ledger is valid.
\end{enumerate}

This combination for the first time allowed the creation of electronic cash without relying on financial institutions to serve as trusted third parties.  Satoshi argues that removing these institutions is beneficial as it will reduce transaction costs, which are created forming and enforcing agreements \parencite[p.~605]{Cooper2011}.   When trusted third parties are involved, they have the power to reverse payments or alter balances which leads to an increase in the requirement for trust.  As greater trust is needed, greater amounts of information must be accumulated by participants in the network, thus increasing costs.  Szabo states the point more forcefully: ``trusted third parties are security holes'' \parencite{Szabo2001}; arguing that the most expensive and vulnerable part of any security system that relies on trusted third parties, will be that third party itself.  North \parencite*{North1987} shows transaction costs prevent economic development, i.e. the cost used to fulfil this `trust' function (e.g. lawyers, auditors) could be utilised more productively in creating goods or services.     

Bitcoin was not the first attempt to introduce electronic cash (excluding fiat money accounted for electronically), failed previous examples were e-gold and beenz \parencite{Eiland2017}. These endeavours were unsuccessful because they had not solved three fundamental questions \parencite[p.~40]{Antonopoulos2014}:

\begin{enumerate}
\item Is the money authentic, i.e. not counterfeit?
\item Is the money unspent - also known as double-spend where somebody spends the same money twice?
\item Is this money claimed by me (as opposed to someone else)?
\end{enumerate}

\begin{figure}
\includegraphics[width=\textwidth]{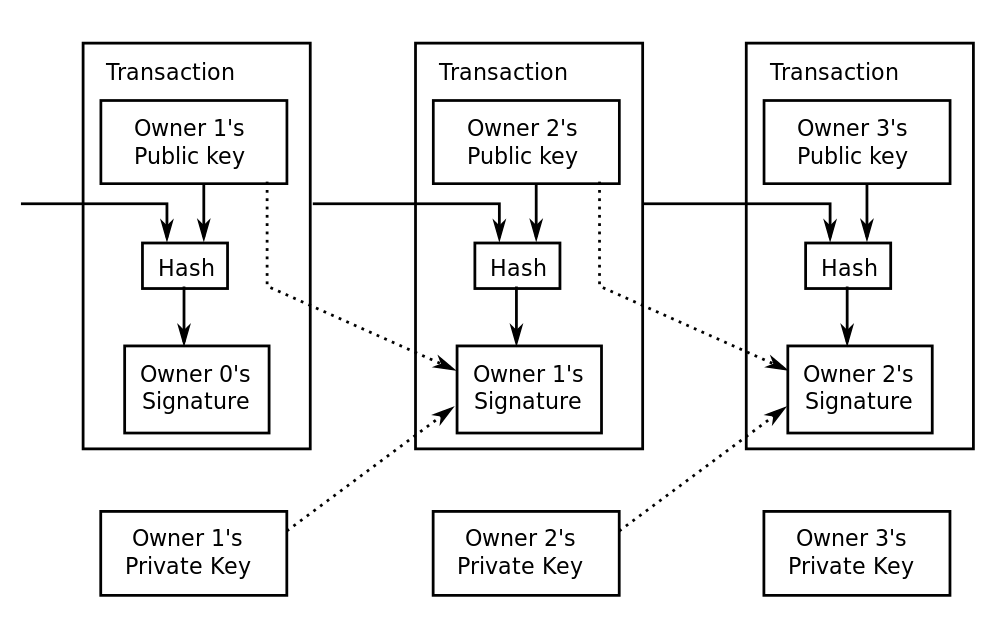}
\caption[Bitcoin transactions]{Bitcoin transactions \parencite{Nakamoto2008}}
\label{fig:BitcoinTransaction}
\centering
\end{figure}

An explanation of the technology is necessary to understand how Bitcoin solves these problems.  Imagine a situation where Alice and Bob wish to exchange bitcoin.  Each will have a wallet - software that is able to conduct Bitcoin transactions.  Each wallet will contain a number of addresses; addresses simply being containers to hold any amount of bitcoin.  Each address will have an associated private and public key.  Assuming Alice wished to pay Bob one bitcoin she would transfer this by using her private key to digitally sign what is known as a hash of the previous transaction and Bob's public key, as at figure \ref{fig:BitcoinTransaction}.

A hash is an algorithm that takes an arbitrary amount of digital data; which could be anything from a Microsoft word document, jpeg photo or text file; and returns a fixed length value (known as a digest).  As an example if the text from the abstract of this dissertation is run through a SHA1 function the resulting value is:
\begin{quote}
f4f3c70d905f007f6c069def9b66082aaab22bee
\end{quote}
Every time the abstract of this dissertation is run through the SHA1 algorithm it will produce the same value.  However it is computationally difficult to retrieve the original text (the abstract)  if one is simply provided with the hash value.  Should any change be made in the original text, the resulting hash will be different.  For instance if the abstract is run again through the algorithm, but with the final full-stop omitted the result will be:
\begin{quote}
d21a01538b8c87394d44567269f6ac5aa22335e0
\end{quote}
Despite the input text differing by only one character, the resulting hash is completely different.  Hashing therefore allows a piece of data to be reduced to a shorter fixed length value.  This can be used to quickly determine whether the original data has been altered - by hashing the suspect data and checking it matches the hash value recorded earlier.

To return to Alice and Bob's bitcoin transfer - this transaction is now broadcast to the network.  At the same time other transactions are also being broadcast to the network; which are bundled up into blocks together.  Other users, known as miners, will now create cryptographic hashes of the data contained within those blocks.  The first miner to create a hash that conforms to a certain format (specifically that it begins with a prescribed number of zeros) is rewarded with newly minted bitcoins.  The creation of this winning hash however is computationally difficult - it is accomplished by adding a random segment (known as a nonce) to the block of transactions, so eventually resulting in a hash matching the required format.  In this way a new block is created on the ledger.  This new block will contain the winning hash of the last block, and the next block the winning hash of this one and so forth.  Because the hash of the last block is in the current block, changing any transactions in the last block will cause a hash mismatch and alert all to the attack.  

\label{sec:Genesis}

In this way the transaction of one bitcoin between Alice and Bob cannot be altered in any way.  Alice cannot change her mind and take it back, so defeating the double-spend problem where Alice tries to give the same one bitcoin to both Bob and Charlie - as the first transaction of this bitcoin (to Bob) is recorded immutably in the ledger and considered the `valid' transaction.  It also means a malicious actor, such as Mallory, cannot change the transaction to divert Alice's bitcoin to her.  As more blocks are created, each containing the hash of the last, the previous transactions are less vulnerable to attack, due to the increasing number of blocks that would need to be changed.  

To attack the network successfully Mallory would have to take over 51\% of the nodes so allowing her version of the chain to be accepted.  However to do this the amount of computing power (and associated costs such as electricity, hardware, etc) would be so great, it would make more financial sense to use those resources to support the network and reap the benefits of mining.  Indeed any successful attack would likely cause the value of Bitcoin to drop, again removing the motivation of financial gain.

Although a considerable technical achievement, it does not necessarily explain the excitement around Bitcoin.  Reasons behind this are varied; many early libertarian advocates welcomed it as a transfer of government financial power to individuals with the prospect of states no longer controlling the money supply \parencite[p.~152]{Frisby2014}.  Others foresee its role in transforming the economy and questioning the assumptions of the industrial age - Antonopoulos for instance asks whether in a truly digital economy salaries should be streamed by the minute rather than arriving in monthly instalments \parencite{Dale2017}.  Alternatively cynics suggest that interest is primarily fuelled by speculation \parencite{Baur2017}, comparing it to the seventeenth century tulip craze which bankrupted many investors \parencite{Jones2017}.

\section{The root of all evil}

The applications of DLT go beyond cryptocurrency.  Ledgers have been a fundamental feature of trade since ancient times \parencite{Gray1996} so it is unsurprising that this technological shift could have wider impact.

Although it is clear why Bitcoin is a novel approach to digital cash, Wenger \parencite*{Wenger2014} provides a good explanation of why blockchain represents a discontinuum with previous technology as a means of organising information, as shown at Table \ref{Wenger2014Table}.

\begin{table}
\centering
\begin{tabular}{ m{6em} m{7em} m{7em}  } 
\toprule  
& \begin{flushleft}Organisationally centralised\end{flushleft} & \begin{flushleft}Organisationally decentralised\end{flushleft} \\ 
 \midrule 
  \begin{flushleft}Logically centralised\end{flushleft} & e.g. Paypal & \textbf{***new***} blockchain \\
  \begin{flushleft}Logically decentralised\end{flushleft} & e.g. Excel & e.g. email \\ 
\bottomrule
\end{tabular}
\caption[Foundational innovation of the blockchain]{Foundational innovation of the blockchain \parencite{Wenger2014}}
\label{Wenger2014Table}
\end{table}
Wenger posits that DLT represents a new category of information management - logically centralised, but organisationally decentralised.  Using the Wenger classifications there are many examples of organisationally centralised information technologies in the world - Paypal and Excel are two examples, both owned by organisations who chose what version to release and their retail price.  However they differ in their `logical' centralisation - it is possible for Alice to send Bob an Excel spreadsheet and for Bob to edit that spreadsheet independently of Alice.  Excel is therefore logically decentralised.  

Paypal however is logically centralised - if Alice sends Bob \textsterling1 via Paypal the accounts of Alice, Bob and the system as a whole have to correlate.  Whereas email is both organisationally and logically decentralised; no single organisation owns email and Alice sending Bob an email is unrelated to other people's emails.  

Blockchain is a new category in that there is no central organisation, no permission is needed to write a new software wallet or run a network mining node, however there is a logical centralisation - when Alice sends Bob bitcoin the entire system is aware of that. 
Ludwin \parencite*{Ludwin2016} takes this point further by arguing that it is unhelpful to think of Bitcoin as a currency but rather as a ``new asset class that enables decentralised applications.''  The utility of a decentralised application (DApp) being that it is organisationally decentralised - no one person owns it.  Ludwin however proposes that DApps, although useful, have disadvantages due to inefficiency.

Not all DLT has to use coins or tokens.  Bitcoin uses a Proof-of-Work (PoW) consensus mechanism meaning miners are rewarded with the ability to add blocks to the chain (so earning bitcoin) dependent on the amount of compute they have undertaken \parencite{Bonneau2015}.  However there are other consensus mechanisms, such as Proof-of-Stake (PoS), where the more tokens you have (e.g. stake) the greater your probability of creating the next block \parencite{Bentov2016}.  PoS is less energy intensive and more environmentally sustainable than PoW \parencite{Dwyer2014}; although PoS has been criticised for unfairly rewarding those who already have amassed the most \parencite{Mamoria2017}.  Meanwhile Proof-of-Authority (PoA) is a consensus mechanism where those involved in establishing the network have decided which nodes are deemed reliable \parencite{Cachin2016} and is synonymous with permissioned blockchains.  PoA has been criticised as removing one of the central aspects of blockchain - achieving trust without a central authority - and as such has been criticised as ``probably not [a] real blockchain'' \parencite[p.~28]{Bashir2017}.  There are yet other mechanisms for consensus; e.g. Proof-of-Burn, Proof-of-Capacity \parencite{Tasca2017}; which are outside this dissertation's scope.  

\section{Two sides to the coin} 
\label{sec:twosidestothecoin}

DL is a new category of technology that could lead to a swathe of different business models \parencite{Tapscott2017a}.  Mainelli and Gupta \parencite*{Mainelli2017b} point to an earlier technological shift that occurred with the rise of digital mapping, combined with GPS, which allowed the real-world to be visualised on computers leading to challenger upstarts such as Uber.  DLT could allow a similar shift with the digitalisation of business fundamentals; if the transactions and contracts that are the lifeblood of business, previously locked within company silos but now shared between enterprises, are able to be manipulated digitally new worlds of possibility could emerge \parencite{Iansiti2017}.   

\begin{figure}
\includegraphics[width=\textwidth]{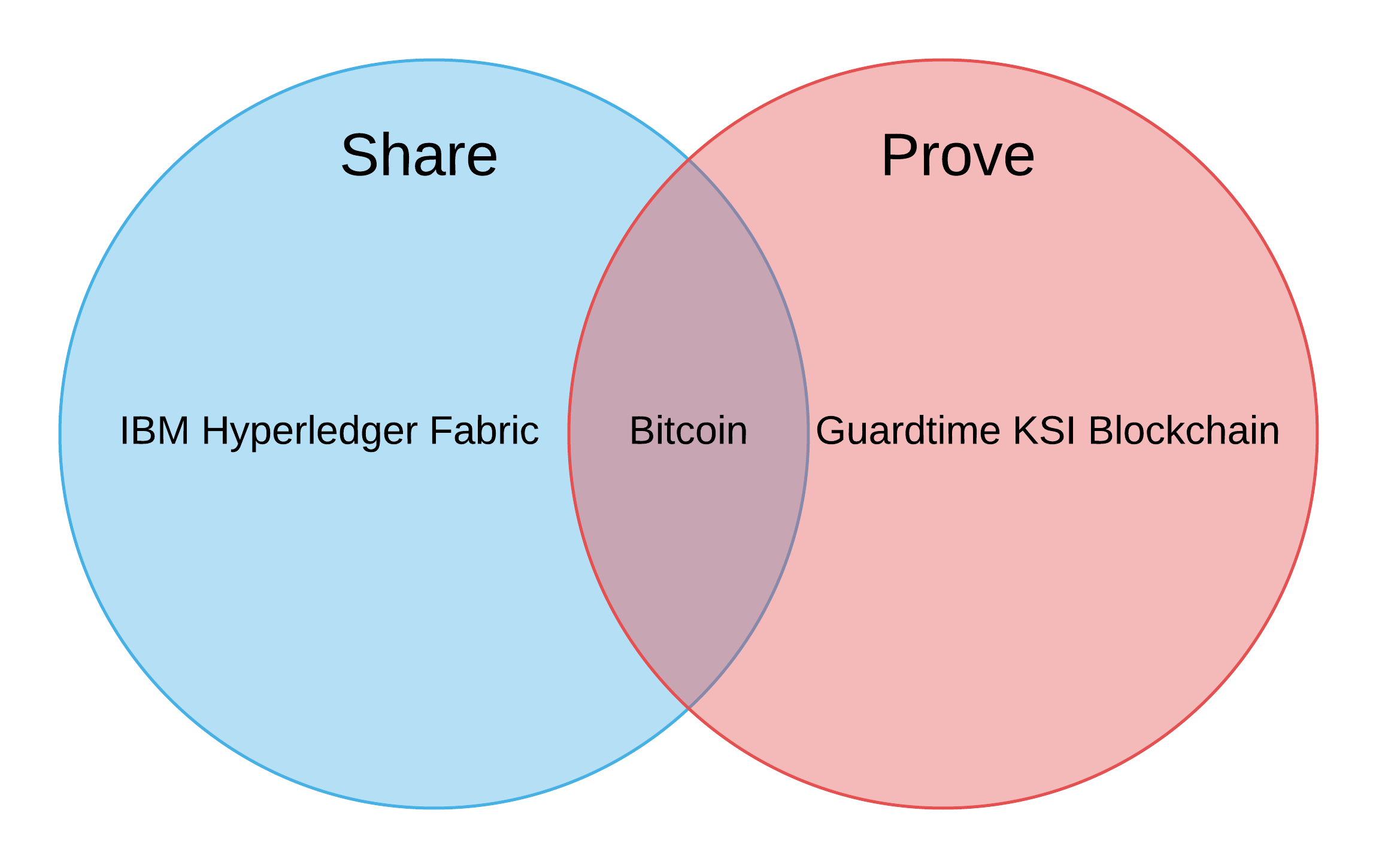}
\caption[Sharing vs Proving DLT]{Sharing vs Proving DLT (Author's own work)}
\label{fig:ShareProveVenn}
\centering
\end{figure}

However as has already been covered there is considerable variety in what can be thought of as DLT and correspondingly great variety in what DLT might be used for.  Ultimately however all the technologies in this area are considered with one or other business problem: Sharing or Proving, and in some circumstances both.  This is illustrated at Figure \ref{fig:ShareProveVenn}.

Bitcoin sits in the middle of this Venn diagram being a DL that both proves and shares data.  When Alice sends Bob one bitcoin the technology is both used to \textbf{prove} that Alice has one bitcoin to send and \textbf{share} the data that Alice has transferred one bitcoin to Bob to all other participants on the network.  However it is not necessary to have both aspects present within DLT.   

The \textbf{share} use case is exemplified by an example provided by Hyperledger Fabric \parencite{IBM2016a} where a blockchain is established for a consortium of companies involved in car leasing.  The participants; such as the vehicle manufacturer, dealerships and scrap merchant; all use the ledger to access information such as the Vehicle Identification Number and maintenance logs.  All participants are now sharing one view of the vehicle history - IBM gives the example that if a recall had to be organised this would be far more efficient and effective if all participants were using the same blockchain.  However although we can see data \textbf{share} occurring, there is no \textbf{prove}.  For instance if the manufacturer were to erroneously ascribe the wrong Vehicle Identification Number to the car record there would be no way of verifying that fact by looking at the information contained on-chain.  Rather physical verification would need to take place that off-chain reality (i.e. the number written on the car) matched the data on-chain.  The situation further complicates if there are malignant actors within the consortium.  For instance if the scrap merchant is involved in an illegal `cut and shut' scheme where two halves of old cars are welded together to form a `new' vehicle \parencite{BBC2000}; this blockchain would not guarantee that cars marked as destroyed had been so.  This can be considered the digital-physical gap: the difficulty of achieving a link between the immutable digital object on-chain and its twin mutable physical object in the real world.  

\begin{figure}
\includegraphics[width=\textwidth]{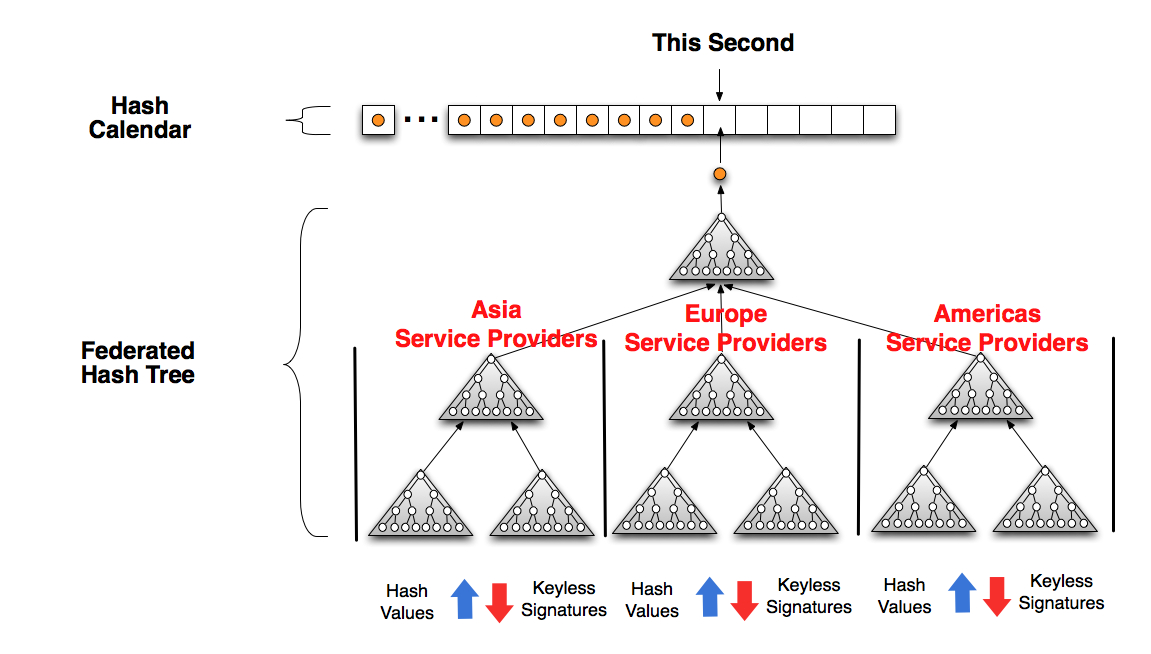}
\caption[KSI Federated Hash Tree]{KSI Federated Hash Tree \parencite{Zatyko2015}}
\label{fig:KSIFederatedHashTree}
\centering
\end{figure}

The right side of the Venn diagram is entirely \textbf{prove} and no \textbf{share}.  Guardtime's Keyless Signature Infrastructure (KSI) Blockchain creates chain-of-custody information for digital assets - any time a protected file (e.g. MS Word document, JPEG image) is modified, created, deleted or transmitted there is forensic evidence of that activity, admissible in a court of law \parencite{Johnston2014}.  This technology works by taking a hash of any protected asset, combining those hashes with other hashes using a Merkle-tree and then publishing that data in a hash calendar (whereby the current hash is combined with the previous hash) as demonstrated at Figure \ref{fig:KSIFederatedHashTree}.  Although this is far removed from Bitcoin, it is described as blockchain technology by the creators \parencite{Guardtime2016} and Walport \parencite*{Walport2016} counts it as DLT.  The Estonian government, who employ this proprietary technology, elaborate by stating that KSI Blockchain was being tested in Estonia in 2008, prior to the Bitcoin white-paper, at which point the term blockchain was known as ''hash-linked time stamping'' \parencite{eestoniacom2017}.  In this use case however the utility comes entirely from the \textbf{prove} aspect, there is no \textbf{share} as with the car lease demonstration.  Rather here a distributed ledger acts purely to provide provenance, not to distribute data.  A similar initiative is Archangel which looks to verify the contents of the National Archive to prove items have not changed over time \parencite{Thereaux2017}.  This share versus prove dichotomy illustrates how radically different approaches to solving different business problems are still covered under the umbrella of DLT.  In some ways DLT can be compared to a clawhammer: both Guardtime KSI Blockchain and IBM Hyperledger Fabric share a common stem (the clawhammer shaft) but are used for purposes as radically different as hammering versus removing nails.

Praise for use cases which move away from Bitcoin's original conception of both share and prove is not universal.  Antonopoulos \parencite*{Antonopoulos2017b} for instance mocks how the conversation has turned increasingly anodyne as development has moved from bitcoin to blockchain to DLT, which he believes is led by vested corporate interests attempting to head off disruption.  Here he makes a parallel between blockchain's maturity and the internet circa 1997; although there were attempts to use the internet for ambitious plans (e.g. grocery deliveries) these failed until a sufficiently dense level of adoption had been established through the relatively simple application of email.  Similarly he argues DLT will not be used for ambitious plans, such as real estate title, until cryptocurrencies are used for everyday transactions; at which point a tipping point will have been reached which challenges the establishment (e.g. banks are out-competed in their core business of banking).  Similarly Song \parencite*{Song2018} asserts that blockchain without bitcoin is equivalent to selling ``snake oil.''   These viewpoints contrast strongly with that of Walport \parencite*{Walport2016} which is about working within current systems; this conflict between revolutionary and evolutionary paths is a recurring theme and will be examined more closely in the literature search.    

\section{Search strategy}
\label{sec:searchstrategy}

DLT can therefore be seen as a unique concept that has emerged from counter-cultural roots and has generated considerable debate on how it can be best utilised.  In understanding how it might serve the needs of the DSN it would be helpful to survey the academic literature, as well as this being a required Individual Learning Outcome of an ICM Research Project.  A structured search using metadata only was conducted on IEEE Xplore's Digital Library for the search terms shown in Table \ref{table:SearchTerms}.  Because the term `blockchain' (or derivatives) is used in other academic disciplines, the search is limited to a computer-science relevant database.  No time period was specified, and no restriction on  materiel (e.g., journals, books etc) was set. The search was limited to English material only due to resources available.

The Alan Turing Institute notes the prodigious rate of output on the subject of blockchains (averaging 250 papers per year since 2014), but that most of these have appeared in the form of white papers outside traditional academic peer-reviewed literature \parencite{Bano2017}.  Thus grey literature has been included separately in the results of the literature search. 
 
Interest in DLT is not limited to the English speaking world.  For instance China is involved in cryptocurrency mining and speculation - in 2015 88\% of total Bitcoin trades took place there \parencite{Pel2015}; while political support for DLT research comes from President Xi Jinping himself \parencite{Cheng2018}.  Japan and South Korea share similar enthusiasm for the technology \parencite{Price2017}. Searching the China Academic Journals Database using the sinographs for bitcoin, \begin{CJK*}{UTF8}{gbsn}比特币\end{CJK*}, yielded 669 articles; a higher count than that of IEEE Xplore.  This finding is borne out by the China Intellectual Property Office filing nearly 140 DLT related patents in the last three years \parencite{Zhao2018}. Thus future literature review should endeavour to cover non-English language publications. 

\begin{table}
\centering
\begin{tabular}{ l l } 
\toprule
Search Term & Frequency \\
\midrule
blockchain & 236 \\
bitcoin & 213 \\
distributed ledger & 22 \\
\bottomrule
\end{tabular}
\caption{Search term frequency using IEEE Xplore dated 14 Nov 17}
\label{table:SearchTerms}
\end{table}

Table \ref{table:SearchTerms} shows 55\% of the search results included the term `blockchain' but not `bitcoin' (with the remaining 45\% having the terms `blockchain' and `bitcoin').  This is in contrast to an earlier literature review \parencite{Yli-huumo2016} which found that 80.5\% of academic output focused on Bitcoin rather than wider applications.  This difference may be due to the stricter search criteria for this study (which selected 41 papers).  Alternatively the greater number of blockchain minus bitcoin results, may represent the increasingly popularity view that DLT not Bitcoin is where potential lies \parencite{Knight2017}.  

\begin{figure}
\includegraphics[width=\textwidth]{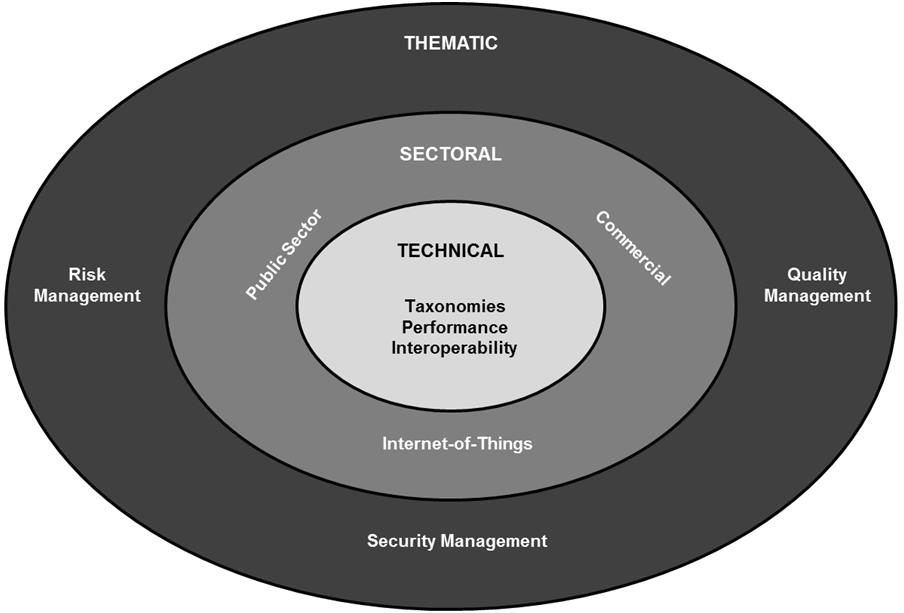}
\caption[Representation of the standards environment for DLT]{Representation of the standards environment for DLT \parencite{Mainelli2016b}}
\label{fig:MainelliStandards}
\centering
\end{figure}

 It is worth noting the UK government's preferred term, `distributed ledger technology' \parencite{Walport2016} is less frequently used in the literature.  This term having such little traction shows stakeholders have not coalesced around agreed definitions, although there have been attempts in the literature to define a DLT ontology \parencite{Tasca2017}.  The British Standards Institution \parencite{Deshpande2017} has highlighted that the lack of industry standards for this emergent technology makes comparison or categorisation challenging.  They also believe that standards would drive other benefits such as addressing the concerns of security, privacy and data governance.  Insurance industry research \parencite{Mainelli2016b} tallies with this, proposing that the introduction of voluntary standards markets would allow organisations to manage risk and reduce regulatory uncertainty via establishing compliance and verification regimes.  Figure \ref{fig:MainelliStandards} illustrates what standards might apply.

Although how these standards might apply to DL goes beyond this dissertation's scope, one area from Figure \ref{fig:MainelliStandards} that is worth exploring are standards for `interoperability.'  There are real-world examples of attempts to integrate DLs - e.g. IBM's cross-border trade payments project \parencite{DelCastillo2017} uses both private (Hyperledger) and public (Stellar) blockchains in conjunction; the former for transaction clearing, the latter for settlement payment.  Not that this represents a standard, simply that agreement on interoperability can be reached.  

Common interoperability standards  underpins today's networked world: Hypertext Transfer Protocol and Simple Mail Transfer Protocol standards allow users to communicate across networks and applications, but no similar standard exists for DLT \parencite{Strajnar2015}.  Of course not all technologies have common standards.  For instance DLT's fundamentals are compared to a database; yet the recent big-data NoSQL databases are a competing set of technologies without a common standard \parencite[p.~9]{Sadalage2012}.  This though could be a false comparison - DLT's power is sharing data across organisational boundaries.  Scenarios can be envisaged where an organisational grouping using a DL, e.g. a fishing co-operative verifying their environmentally sustainable processes \parencite{Provenance2016}, have a requirement to pass information to another organisational grouping using a different DL, e.g. to grocery retailers selling the seafood and using a DL for food-safety \parencite{Kharif2016}.  Hardjono, Lipton and Pentland \parencite*{Hardjono2018} illustrates how in the same way the internet is neutral as to which network your packet travels on, applications might be agnostic as to which DL your transaction is recorded on.  Common standards would allow interoperability between these types of DL to become routine, rather than being an expensive ad-hoc process.

Indeed, Mougayar \parencite*{Mougayar2017} argues that before DLT can become as ubiquitous as the web, common standards are required to enable interoperability. The opposing case is that standards might damage this emerging technology, freeze-framing it prematurely, which Mougayar \parencite*{Mougayar2016} himself previously posits.  Either way momentum is building for standards, with the International Organization for Standardization drafting proposals \parencite*{isoblockchain2017}.

This lack of standards represents a challenge for literature reviews, as Webster and Watson \parencite*{Webster2002} argue the first step of academic enquiry is classification.  As yet there have been few comprehensive reviews, and much of this focuses on the technical \parencite{Tasca2017} rather than on enterprise applications.  Additionally defence, by its very nature, is guarded on revealing how they might employ new technologies.  Although there are exceptions to this; for instance US Air Force cyber-security orientated studies \parencite{Barnas2016}, Washington think-tank papers \parencite{Hsieh2017} and Canadian defence research \parencite{Willink2018}; the sub-set of the literature specifically considering how defence might use DLT is small.

\section{Thematic analysis}

This lack of precedence as well being a challenge, can be seen as an opportunity to create a small contribution to the field.  A thematic analysis was chosen because Saunders, Lewis and Thornhill \parencite*[p.~ 80]{Saunders2016} conclude that those papers that contribute most to an area of study follow this approach.

Three themes in the literature were identified: that of supporting the `revolutionary', `evolutionary' or `reactionary' paradigm.

\subsection{Revolutionary} 

As previously discussed DLT's roots lay within Bitcoin, emerging from the cryptoanarchist and cypherpunk movements.  The former movement is a political philosophy recognising no laws apart from those enforced by code, the latter a technological vision of socio-political change limiting the power of authorities \parencite{Narayanan2013}.  These aligned movements took early inspiration from Chaum \parencite*{Chaum1985}, who proposed a decentralised digital cash that empowers individuals, rather than governments or corporations.  Later seminal texts `A declaration of the independence of cyberspace' and `A Cypherpunk's Manifesto'  \parencite[pp.~27-30, 81-83]{Ludlow2001} saw an opportunity in technological progress to break with the industrial nations' established order.  Early victories came in defeating US Government bans on the export of encryption software PGP (Pretty Good Privacy) software \parencite[p.~152]{Gellman2011}.  

Despite these successes, by the 2010s Narayanan \parencite*{Narayanan2013a} declared that ``cypherpunk crypto [has] failed to materialise,'' while Zittrain \parencite*{Zittrain2012} foresaw ``the end of crypto.''  They argued that the difficulties of implementing cryptography, plus the resistance of corporates and democratic governments, meant that the cypherpunk movement would not have the impact imagined.

However Bitcoin's arrival marks a resurgence of the cypherpunk movement; both from the original proponents who always had digital cash central to their vision \parencite{Torpey2016}, and from a new wave of adherents \parencite{Bartlett2016}.  Links between these communities are often explicit - for instance the Bitcoin subredit directs new joiners to `A Cypherpunk's Manifesto' \parencite{Reddit.com2018}.

\begin{figure}
\includegraphics[width=\textwidth]{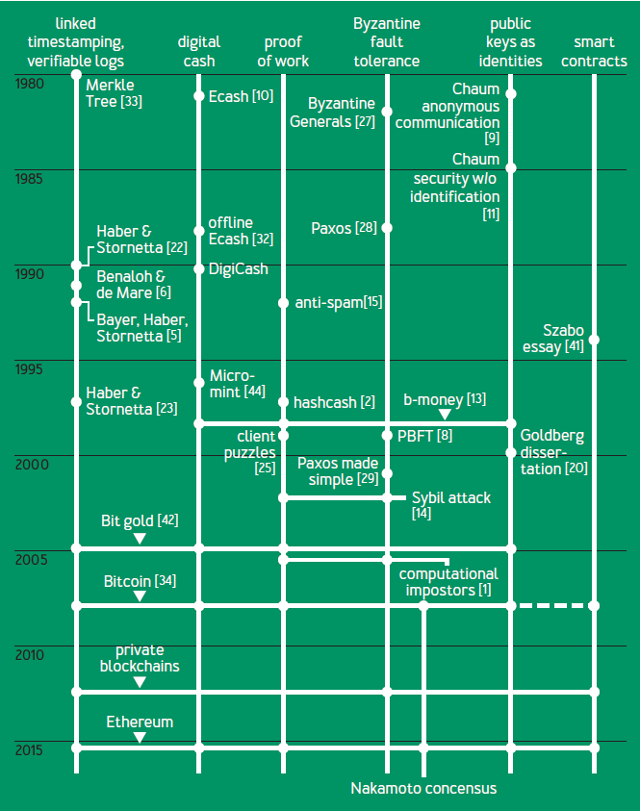}
\caption[Chronology of Bitcoin's technological precursors]{Chronology of Bitcoin's technological precursors \parencite{Narayanan2017}}
\label{fig:BitcoinChronology}
\centering
\end{figure}

Following cypherpunk's near-death; why then did Bitcoin, followed by the DLT paradigm, emerge when it did?  The creator's secrecy, the pseudonymous Satoshi Nakamoto, thwarts a definitive answer; although there appear to be two drivers: technical and political.  From a technical standpoint, as acknowledged by Nakamoto \parencite*{Nakamoto2008}, Bitcoin built upon much previous work \parencite{Bonneau2015}, as illustrated at Figure \ref{fig:BitcoinChronology}, such as the Hashcash project which used Proof-of-Work to counter email spam \parencite{Back2002}. 

Therefore Bitcoin partly emerged when it did because the technical foundations had caught up with cypherpunk's aim of digital gold.  The second driver, political, is something Nakamoto explains less: ``it's very attractive to the libertarian viewpoint if we can explain [Bitcoin] properly. I'm better with code than with words though'' \parencite{Jansen2012}.  Nakamoto was familiar with Austrian economics (libertarianism's fountainhead) \parencite{Davis2011}, which proposes government interference rarely benefits economic systems.  Bitcoin explicitly references 2008's Global Financial Crisis, inserting this phrase in the genesis block: ``The Times 03/Jan/2009 Chancellor on brink of second bailout for bank'' \parencite{Maurer2013}.  Libertarians would object to governments creating money, as in The Times headline, to save financial institutions.  Bitcoin with its limit of 21 million coins and resistance to any central authority, is a strong reaction to these events by those who question the state's role in economic affairs.  The MoD's Development, Concepts and Doctrine Centre even forecasts longer term scenarios where cryptocurrencies might challenge the state's dominance \parencite[p.~78]{MoDDCDC2014}.  The second reason for Bitcoin's emergence was therefore the political environment which Nakamoto, and allies, were reacting to. 

Given this it is unsurprising much debate around DLT (especially in cryptocurrencies) has been couched in the language of opposition to established order, bluntly put by Antonopoulos \parencite{Bundrick2015}:

\begin{quotation}``You put an open, decentralized ecosystem: open source, open standards, open networking and the intelligence and innovation pushed all the way to the edge — put that against a closed system, controlled by a central provider, whose permission you need in order to innovate and who will only innovate at the exclusion and competition of all of the other companies — and we will crush them.''\end{quotation}

Even the less colourful Bank of Finland uses the term ``revolutionary'' to describe this ``marvellous structure'' \parencite{Huberman2017} thereby sharing common ground with firebrands such as Antonopoulos.  

The adjective `disruptive' is often applied to emergent technology; however given the regularity of disruptors being bought out by the very firms they are trying to disrupt \parencite{Faktor2016}, this hardly captures DLT's transformative vision.  Rather `revolutionary' - seeking to upend the established order - is used as one thematic grouping in this dissertation instead.

\subsection{Evolutionary} 

\begin{figure}
\includegraphics[width=\textwidth]{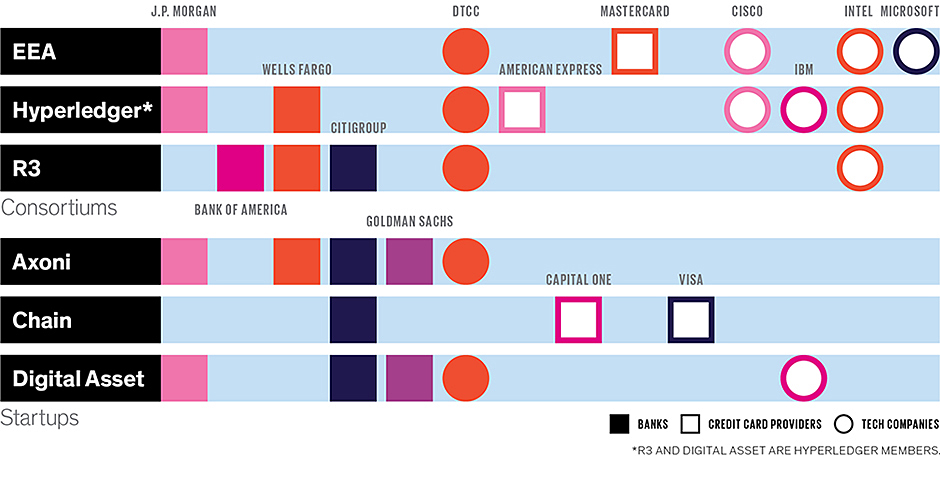}
\caption[Financial industry involvement with DLT]{Financial industry involvement with DLT \parencite{Nordrum2017}}
\label{fig:ieeeSpectrum2017}
\centering
\end{figure}

Despite its roots in radical thought, DLT is  being increasingly co-opted by established firms.  Figure \ref{fig:ieeeSpectrum2017}  shows the heavy involvement of multi-national finance and technology firms within DLT, often with one firm backing several initiatives. Companies such as J.P. Morgan and Goldman Sachs are hardly edgy upstarts wanting to turn the system upside down.  

In a similar vein `evolutionary' also covers the phenomenon of IT departments promoting DLT simply to make ``boring back-office coordination work sexy'' so prompting funding \parencite{Levine2016}.  Indeed the Centre for Evidence Based Management claim the most successful use of DLT is where it is a ``catalyst'' to implement benefits such as common data standards and automation, even though much of this work could have been done without DLT; or where DLT is the clarion-call for change but the eventual final product shares more similarities with conventional systems \parencite{Parliament.HouseofCommons.2018}.  One might argue an example of this is R3 Corda's product which began as a ``blockchain for finance'' but engineering choices moved it away from Bitcoin's architecture \parencite{GendalBrown2016}.  Although critics such as Gerard \parencite*[p.~123]{Gerard2017} might reach the conclusion that this is a ``blockchain product'' that does not contain a blockchain; it is also possible to see this as the evolving nature of DLT which, as Section \ref{sec:searchstrategy} discussed, lacks a common definition.  For these reasons `evolutionary' is the second thematic grouping.

\subsection{Reactionary}

Both the revolutionary and evolutionary paradigm agree that DLT has potential benefits.  Others are less optimistic: one argument is that mass take-up of DLT is unlikely given the existing good alternatives, with comparisons being made between DLT and open-source Linux as a Windows replacement for consumers \parencite{Evans2017}.  Conte de Leon et al \parencite*{ContedeLeon2017} goes further suggesting DLT is inferior to existing systems: the interactions of independent agents are by definition complex so DLT proves difficult to verify it will work correctly.  Even proponents acknowledge DLT is typically slow \parencite{Greenspan2016}.  The Centre for Evidence Based Management concludes that much of the hype is  due to the disconnect between senior management and IT departments; with DLT simply being the latest fad that offers ``magic wand pixie dust'' to solve enterprise problems \parencite{Parliament.HouseofCommons.2018}.  Gerard \parencite*[p.~17]{Gerard2017} expresses the maximally negative view that the technology as a whole is as unstable as its originating anarchist sub-culture, with the primary driver greed and naivety.  `Reactionary' therefore defines the final thematic grouping. 

\subsection{Units of Analysis}

These three broad themes cover wide ground.  Therefore to make the critical literature review more meaningful a further dimension was added - units of analysis (UoA) \parencite{Webster2002}.  This was needed as DLs that are revolutionary in one sense, for instance a radical method of allocating domain names so challenging internet governance \parencite[p.~30-33]{Swan2015}, might have minimal impact elsewhere (e.g. socio-cultural).  An analysis framework could therefore usefully sub-categorise the impact DLT might have.  A number of frameworks were considered, including commercial, for instance Porter's Five Forces model which analyses market competition \parencite{Porter2008}.  Academic DLT literature reviews which used conceptual groupings were also studied, specifically: keyword analysis \parencite{Notheisen2017a, Holub2018}, social media context \parencite{Risius2017}, engineering layers \parencite{Hawlitschek2018} and technical challenges \parencite{Yli-huumo2016}.  The UoA eventually selected is recommended by the British Computer Society for business analysis \parencite{Cadle2010}, and is established in information systems research \parencite{Peng2007, Bakri2012} - PESTLE (Political, Economic, Social, Technological, Legal, Ecological) analysis.  One criticism of a PESTLE approach is it lacks a Defence research focus, however understanding impacts wider than Defence enables discovery of cross-cutting or unanticipated applications.  It is therefore using PESTLE, coupled with the revolutionary, evolutionary and reactionary themes that the literature will be reviewed.

\section{Matrix}

Items relevant to the research question of how DLT might apply to the DSN were selected from the IEEE Xplore database (Section \ref{sec:searchstrategy}).  These papers then had their citations reviewed and, using the Web of Science, forward-citations were also reviewed.  Additionally where it was felt the IEEE Xplore database had not provided enough coverage in any UoA, databases from Cranfield University, Scopus and Google Scholar were also utilised, using the same search terms.  Papers selected on this basis were then mapped against the thematic analysis and UoA so creating the matrix at Table \ref{table:LitReviewThematicAnalysis}.

The papers are listed in chronological order of publication; where two papers are published in the same year then date of submission is used to further sort.  Where non-academic papers (e.g. government reports) contributed to the topic, they were included at the end of the matrix under the heading `grey literature.'

\begin{landscape}
\centering
\begin{longtable}{ l  l l l l l l  l l l l l l  c }
\toprule
Theme
&
\multicolumn{6}{l}{Revolutionary}
&
\multicolumn{6}{l}{Evolutionary}
&
Reactionary\\
Unit of Analysis	&P &E &S &T &L &E &P &E &S &T &L &E &- \\
\midrule
\endhead
\cite{Maurer2013} 	&  &  &X &  &  &  &  &  &  &  &  &  &  \\

\cite{Dwyer2014} 	&  &  &  &  &  &  &  &  &  &  &  &X &X \\
\cite{Jenssen2014} 	&X &X &  &  &  &  &  &  &  &  &  &  &  \\
\cite{Benet2014} 	&  &  &  &X &  &  &  &  &  &  &  &  &  \\

\cite{Swan2015} 	&X &X &X &X &X &X &  &  &  &  &  &  &  \\
\cite{Wright2015} 	&  &  &  &  &X  & &  &  &  &  &  &  &  \\
\cite{Golumbia2015} 	&X &  &  &  &  &  &X &  &  &  &  &  &X \\
\cite{Zyskind2015} 	&  &  &X &  &  &  &X &  &  &  &  &  &  \\
\cite{Dupont2015}       &  &  &  &  &X &  &  &  &  &  &  &  &  \\
\cite{Dennis2015} 	&  &  &X &  &  &  &  &  &  &  &  &  &  \\

\cite{Haubo2016}        &  &X &  &  & &  &  &  &  &  &  &  &  \\
\cite{Abramowicz2016}   &  &  &  &  &X &  &  &  &  &  &  &  &  \\ 
\cite{Yli-huumo2016}	&  &  &  &X &  &  &  &  &  &X &  &  &  \\
\cite{Yasin2016}	&  &  &X &  &  &  &  &  &  &X &  &  &  \\
\cite{Ammous2016}       &  &X &  &  &  &  &  &  &  &  &  &  &X \\
\cite{Fu2016}	        &  &  &X &  &  &  &  &  &  &X &  &  &  \\
\cite{Natoli2016}	&  &  &  &X &  &  &  &  &  &  &  &  &X \\
\cite{Lajoie-mazenc2016} &  &  &  &X &  &  &  &  &  &  &  &  &  \\

\cite{Atzori2017} 	&  &  &  &  &  &  &X &  &  &  &  &  &  \\
\cite{Allen2017}        &X &X &  &  &X &  &  &  &  &  &  &  &X \\
\cite{Pazaitis2017}     &  &X &X &  &  &  &  &  &  &  &  &  &  \\
\cite{Yermack2017}	&  &  &  &  &  &  &  &X &  &  &X &  &  \\
\cite{Beck2017} 	&  &X &  &  &  &  &  &X &  &  &  &  &  \\
\cite{Risius2017} 	&  &X &X &X &  &  &  &  &  &X &  &  &  \\
\cite{Vranken2017}  	&  &  &  &  &  &  &  &  &  &  &  &X &  \\
\cite{Nowinski2017}     &  &X &  &  &  &  &  &X &  &  &  &  &  \\
\cite{Chapron2017} 	&  &  &  &  &  &  &  &  &  &  &  &X &  \\
\cite{Porru2017}        &  &  &  &X &  &  &  &  &  &  &  &  &  \\
\cite{Sullivan2017} 	&  &  &  &  &  &  &X &  &  &  &X &  &  \\
\cite{Natoli2017}        &  &  &  &X &  &  &  &  &  &  &  &  &X \\
\cite{Hsueh2017}        &  &  &  &X &  &  &  &  &  &  &  &  &X \\
\cite{Hwang2017} 	&  &  &  &  &  &X &  &  &  &  &  &  &  \\
\cite{Wu2017}           &  &  &  &  &  &  &  &X &  &  &  &  &  \\
\cite{Velasco2017}      &X &X &  & &  &  &  &  &  & &  &  &  \\
\cite{Pazaitis2017}     &  &X &X &  &  &  &  &  &  &  &  &  &  \\
\cite{Mathews2017}     &  &X &  &  &  &  &  &  &  &  &  &  &  \\
\cite{Fink2017}         &  &  &  &  &X &  &  &  &  &  &X &  &  \\
\cite{Werbach2017a}     &  &  &  &  &  &  &  &  &  &  &X &  &X \\

\cite{Mengelkamp2018}  	&  &  &  &  &X &  &  &  &  &  &  &  &  \\
\cite{Kshetri2018}      &  &  &  &  &  &  &  &X &  &  &  &  &  \\
\cite{Aniello2017}      &  &  &  &  &  &  &  &  &  &X &  &  &  \\
\cite{Hawlitschek2018}  &  &  &  &  &  &  &  &X &  &  &  &  &  \\
\cite{Holub2018}        &  &X &  &X &  &  &  &X &  &X &  &  &  \\
\cite{Saberi2018} 	&  &  &  &  &  &  &  &  &  &  &  &X &  \\
\cite{Reyna2018}        &  &X &  &  &  &  & &  &  &X &  &  &  \\
\cite{Matzutt2018}      &  &  &  &  &  &  &  &  &  &  &X &  &X \\
\cite{Peterson2018}     &  &X &X &  &  &  &  &  &  &  &  &  &  \\
\cite{Destefanis2018}   &  &  &  &X &  &  &  &  &  &  &  &  &X \\
\cite{Banerjee2018}     &  &  &  &  &  &  &  &X &  &  &  &  &  \\
\cite{Garcia2018}       &  &  &  &  &  &  &  &  &X &  &  &  &  \\
\cite{Angrish2018}      &  &  &  &  &  &  &  &  &  &  &  &X &  \\
\cite{Wang2018}         &  &  &  &  &  &  &X &  &  &  &  &  &  \\
\cite{Chakraborty2018}  &  &  &  &  &  &  &  &  &  &X &  &  &  \\
\cite{Juskalian2018}    &  &X &  &  &  &  &X &X &  &  &  &  &  \\

 \multicolumn{1}{ c }{\textbf{Grey literature}} \\

\cite{Rosenfeld2012}	&  &X &  &X &  &  &  &  &  &  &  &  &  \\

\cite{McCook2014} 	&  &  &  &  &  &X &  &  &  &  &  &  &  \\
\cite{Schwartz2014} 	&  &  &  &  &  &  &  &X &  &  &  &  &  \\

\cite{Ross2015}	        &  &  &  &X &  &  &  &  &  &  &  &  &  \\
\cite{Swanson2015}	&  &  &  &  &  &  &  &X &  &  &  &  &  \\
\cite{Swanson2015a}	&  &  &  &  &  &  &  &X &  &  &  &  &X \\

\cite{Walport2016} 	&  &  &  &  &  &X &X &X &  &  &X &X &  \\
\cite{Weber2016a}      	&  &X &  &  &  &  &  &  &  &  &  &  &X \\
\cite{Brown2016a}	&  &  &  &  &  &  &  &X &  &  &  &  &  \\

\cite{Strobele2017} 	&  &  &  &  &  &  &X &  &  &  &X &  &  \\
\cite{Gupta2017} 	&  &  &  &  &  &  &X &X &  &  &X &  &  \\
\cite{MercyCorps2017} 	&  &  &X &  &  &  &  &  &X &  &  &  &  \\
\cite{Grigg2017} 	&  &  &  &X &  &  &  &  &  &  &  &  &  \\
\cite{Maupin2017}       &  &  &  &  &  &  &  &  &  &  &X &  &X \\
\cite{Holmes2017} 	&  &  &  &  &  &  &X &X &  &  &X &  &  \\
\cite{Mazet2017} 	&  &  &  &  &  &  &  &  &X &  &  &  &  \\
\cite{Ernst2017} 	&  &  &  &  &  &X &  &  &  &  &  &  &  \\

\cite{Catalini2018}     &  &X &  &  &  &  &  &  &  &  & &  &  \\
\cite{Greene2018} 	&  &  &  &  &  &  &  &X &  &  & &  &  \\
\cite{Baird2018} 	&  &  &  &X &  &  &  &  &  &  & &  &  \\
\cite{Gammar2018} 	&  &  &  &  &  &  &X &  &  &  &  &  &  \\
\cite{Murphy2018}       &  &  &  &X &  &  &  &  &  &X &  &  &  \\
\bottomrule

\caption{Literature review thematic analysis matrix}
\label{table:LitReviewThematicAnalysis}

\end{longtable}
\end{landscape}

\section{Matrix analysis}

Examination of the matrix at Table \ref{table:LitReviewThematicAnalysis}, illustrates two phenomena.  Firstly the academic literature is of a more revolutionary bent than the grey.  This is unsurprising; grey literature is produced by industry, government or related institutions (e.g. think-tanks) - industry profit-maximises and is unlikely to innovate in such a way that their company, or industry sector, disappears.  Similar logic could apply to government - public choice theory suggests that governments are motivated as much by self-preservation as the common good \parencite[p.~80]{Buchanan1984}.  Academics meanwhile, although subject to funding and institutional pressures, have leeway to radically re-imagine all the PESTLE factors.
The second phenomena is chronological: within academia less revolutionary papers appear to be published over time.  This could be confirmation of a previously expressed idea (Figure \ref{fig:ieeeSpectrum2017}): revolutionary concepts are co-opted by established players.  However the time-span of this review is short (due to DLT's recent emergence), so this conclusion is rudimentary.

If these observations are valid then this could herald DLT breaking with its founder's vision.  Bitcoin was fundamentally based on disintermediation; the trends above suggest that the ideas of the central parties are gaining ground.  This should not be over-stated however; partly because this review is not exhaustive \parencite[p.~21]{Booth2016}, and partly because of the history of DLT - its most well-known proponent, Bitcoin, emerged almost entirely outside of government and academic scrutiny; its development may well continue in this vein.

Even given the limitations of this review, the matrix clearly demonstrates that DLT could impact across all areas of PESTLE; each area will now be examined in turn.

\subsection{Political}

Political analysis within the literature has largely focussed on the politics of the blockchain and cryptocurrencies themselves, likely due to interest in the counter-cultural cypherpunk movement which gave birth to DLT.  Golumbia \parencite*{Golumbia2015} posits that Bitcoin is merely a form of political expression because it is neither a money (as it lacks support of the state) nor a currency (as it is not useful as a means of exchange).  Specifically Golumbia characterises that expression as ``right wing extremism'' with elements of anti-semitism.  This analysis, and similar \parencite{Maurer2013}, is based on online articles or web forums, a source that is considered compromised for rational debate \parencite{Mitchelstein2011}.  The problem though is deeper: in a distributed system where anyone can partake and the enigmatic creator disappeared; finding a concrete political philosophy that is more than sweeping cultural analysis, will likely frustrate any academic.

Politics is not just internal to DLT however, DLT could affect how politics is conducted.  State research, such as the UK Chief Scientific Officer \parencite{Walport2016} and the House of Lords \parencite{Holmes2017}, proposes `evolutionary' uses for DLT: suggesting how the government improve its offering to citizens.  Although such reports veer towards hype (Artificial Intelligence and Internet of Things being ever present) they also emphasise the unglamorous ``minimising... costs and redundant work... in administration and back office operations'' \parencite{Holmes2017} where DLT is best suited.  Lord Holmes \parencite*{Holmes2017} refers to `algorithmic government' which rests on authentication (proof you are \textit{x}), authorisation (proof you have permission to do \textit{x}) and accountability (proof that \textit{x} has completed an action).  Estonia, Switzerland \parencite{Strobele2017} and the UAE \parencite{Gupta2017} are investigating DLT in e-government initiatives like identity authentication and online voting.  The impact of DLT for government efficiency might be even greater in developing countries. A Swiss firm, Agora, has trialled voting on the blockchain in Sierra Leone: to record votes immutably at voting offices to combat corruption \parencite{Gammar2018}; while others examine DLT for voter privacy \parencite{Zyskind2015, Wang2018}.  DLT has also been used by supranational organisations, such as United Nations Office for Project Services \parencite{UN2018} and the World Food Programme using DLT to provide identification and payment to refugees \parencite{Juskalian2018}.

Outside of government the views on DLT can be more `revolutionary' - as Atzori \parencite*{Atzori2017} asks: with the advent of blockchain is the state still required?  Embryonic virtual states are now emerging; here Estonia leads the way with its e-residency scheme and partnership with `Bitnation' \parencite{Sullivan2017} which styles itself as: 

\begin{quote}
``a decentralized, open-source movement, powered by the Bitcoin blockchain 2.0 technology, in an attempt to foster a peer-to-peer voluntary governance system, rather than the current `top-down’, ‘one-size-fits-all’ model, restrained by the current nation-state-engineered geographical apartheid, where your quality of life is defined by where you were arbitrarily born.'' \parencite{Sullivan2017}
\end{quote}

This truly revolutionary idea is synonymous with the cyberanarchists rejection of ``governments of the industrial world... weary giants of flesh and steel'' \parencite[p.~28]{Ludlow2001} with free-agents instead voluntarily deciding what governance regime they wish to adopt.  This concept pre-dates DLT with Eichenberger and Frey's \parencite*{Eichenberger2006} Functional, Overlapping and Competing Jurisdictions (FOCJ), but one might argue technology is only now catching up.  Sullivan and Burger \parencite*{Sullivan2017} provide a practical example with a Spanish couple using the Estonian e-Residency to register a marriage on the Bitnation blockchain, thereby stepping outside traditional regulatory routes.  They however highlight the ultimate issue - if no territory recognises your Bitnation marriage then what good is it?

Velasco \parencite*{Velasco2017} continues this `revolutionary' stance arguing the fractional-reserve banking system is a technical device used politically, but that the passing of trust to DLT brings new forms of politics which current traditional actors of state and market cannot define.  Central to his thesis however is DLT's economic role, which will be examined next.

\subsection{Economic}

Economic research on the earliest DLT, cryptocurrencies, is considerable, covering positions from how these markets work \parencite{Haubo2016}, to how they do not \parencite{Greene2018}, to how they challenge current financial systems \parencite{Jenssen2014}.  Research naturally followed to DLT's role in non-cryptocurrency financial practices, such as clearing \parencite{Schwartz2014, Swanson2015, Brown2016a}.  The view that DLT can assist in financial markets is not homogeneous: Ammous \parencite*{Ammous2016} argues Bitcoin is the sole use case for DLT - and that foisting this slow, over-engineered mechanism elsewhere will prove futile.  Much analysis is macroeconomic, the ultimate expression of this being Bitcoin as a new gold standard which all other currencies are backed against, although Weber \parencite*{Weber2016a} appraises if this was achieved it would fail for the same reasons the original did.

At the microeconomic level Beck and Muller-Bloch \parencite*{Beck2017} present a case study of a large financial organisation adopting DLT, focusing on the pressure faced by incumbents whose business model is to act as trusted third parties.  Lessons drawn was that motivation was sparked by curiosity, that DLT specific knowledge had to be bought in (with start-ups being more competitive) and that adoption lowered intra- and inter-organisational boundaries.  Nowinski and Kozma \parencite*{Nowinski2017} add to the microeconomic understanding by categorising the various business models that DLT could foster from micropayments to eliminating forgery.

DLT could help all firms, not just those who adopt it, through improved corporate governance.  Yermack \parencite*{Yermack2017} examines how practices such as recording share ownership using DLT would improve transparency, reduce cost and minimise bad practices such as ``empty voting'' where participants acquire shares purely to vote malignantly against corporations.  Ironically though in the short term DLT is allowing governance to be disregarded as firms accumulate capital through Initial Coin Offerings \parencite{Catalini2018}.

DLT research focuses more on `Wall Street' than `Main Street'; but in reviewing supply chain DLT projects, Kshetri \parencite*{Kshetri2018} assesses these are a better fit, as DLT's strength is in solving problems of messaging rather than as a database.  Wu et al \parencite*{Wu2017} and Banerjee \parencite*{Banerjee2018} also propose DLT's use in supply chain; although both papers may have some commercial bias: the former features an author from Dow Chemicals who have experimented with DLT and may therefore wish to portray it positively, and the latter from Infosys - a software vendor seeking to market the technology.  A more concrete `Main Street' example is DLT's use in manufacturing and construction: Mathews, Robles and Bowe \parencite*{Mathews2017} argue that the 3D design files are well suited to the immutable nature of DLT, while Wang et al \parencite*{WANG2017} propose that DLT's ability to cross boundaries would help within construction's complex leasing arrangements.  Angrish et al \parencite*{Angrish2018} best demonstrate the use of DLT within this field.  Here automated machine tools broadcast their availability, capabilities and performance (as certified by a neutral third party) to a blockchain; whilst clients requiring manufacturing broadcast digital work packages they need completing and prices they are prepared to pay.  Smart contracts match supply and demand; so meaning that where previously a company would send work to a few trusted partners, now this trust cost is reduced they can source from a greater pool.  This would be a radical decentralisation with firms of all sizes being able to fairly compete.

Pazaitis, De Filippi and Kostakis \parencite*{Pazaitis2017}  see DLT as changing the basic fundamentals of  the economy.  Their study into `Backfeed' examines an organisation which uses DLT to form a governance structure allowing individual members to reward each other based on the size of their contribution to the group. Although one of the authors being an instigator of Backfeed is likely to have introduced bias, it highlights the potential impact of DLT on hierarchical structures and associated social interaction.  Hawlitschek, Notheisen and Teubner \parencite*{Hawlitschek2018} alternatively contrast that although DLT can replace trust on a narrow technical level, it cannot in complex social environments such as the sharing economy where it is a multilayered concept.  Judging then DLT has the potential to impact the economy multitudinously, what evidence exists it can do so with these social elements - the next element of this review will examine this.

\subsection{Social}

If DLT can be used to track diamonds \parencite{Everledger2017}, how about for what Socrates describes as one's `richest jewel' - reputation?  Dennis and Owens \parencite*{Dennis2015} suggest DLT could be used for recording a universal reputation score, a key element of much e-commerce e.g. Ebay, Uber, AirBnB.  Decentralising this would mean an end to a company's algorithm change eradicating overnight an individual's reputation, with financially negative implications in today's sharing economy; and potentially lower barriers to start-ups.

Zyskind, Nathan and A. S. Pentland \parencite*{Zyskind2015} go further in empowering users and inverting the power relationship, by proposing that keys to data shared with companies is held in a DL, meaning access can be revoked at any time.  The authors also propose social feedback could be added to a DL consensus model with nodes up and down-voted, Fu and Fang \parencite*{Fu2016} suggest improvements to this with a `Proof-of-Credit' hybrid-consensus model; although neither paper provides clear criteria on what determines a good node from a bad.

Another suggested DLT reputational system involves a comprehensive assessment of a user (including professional, social, academic, etc) \parencite{Yasin2016}, this stops short of Orwellian due to the user granting access rights.  It is noteworthy this research is from a Chinese university, where attitudes to social-media surveillance are markedly different, as evidenced with the proposed `social credit system' \parencite{Chin2016}. It is not clear how these systems would not be gamed however - if followers lead to upvotes, what is to stop reputation inflation by bot?  Nevertheless this illustrates that the benefit of DLT does not exist in a vacuum where only efficiency matters, rather cultural values such as privacy concerns also determine its use.

Underlying any reputational system is identification - proving to society you are who you say.  Garcia \parencite*{Garcia2018} addresses this suggesting biometrics coupled with distributed identifiers could allow government to write identification information to a blockchain, but then be highly selective as to what data is shared.  Caution must be taken when interpreting this academically-published research however, given Garcia's employ by a DLT vendor.

The literature therefore coalesces around DLT's impact on the individual's relationship with society; there is less on how DLT might change society itself, this mostly relates to increasing transparency in charitable giving to social enterprises \parencite{Mazet2017, MercyCorps2017}.  This limitation may be due to the difficulty of forecasting how technology changes society, given the complex two-way relationship between technology and society \parencite{Pinch1993}.  Although the literature might be reticent about predictions; DLT is not, with platforms hosting prediction markets where ``reputation tokens'' are staked on outcomes \parencite{Peterson2018}.  If the literature does not tell us a great deal on technology's impact on society, the next section will look at what it tells us about the technology itself.

\subsection{Technological}

As this is a technology under discussion, it is unsurprising the majority of the literature has focused on technical aspects of DLT \parencite{Yli-huumo2016, Risius2017, Holub2018}.

A common theme is DLT design weaknesses.  This is to be expected: security and cryptography move forward by the research community revealing flaws which subsequently get rectified. An example of this is Natoli and Gramoli \parencite*{Natoli2016}'s examination of how delays in messaging between nodes in Proof-of-Work blockchains could lead to circumstances where double-spend occurs.  Research such as this leads to both proposed improvements in the systems \parencite{Lajoie-mazenc2016, Hsueh2017} and how this might impact on real-world enterprise uses, such as the R3 Consortium \parencite{Natoli2017}.

Although highlighting individual flaws is useful, Porru et al \parencite*{Porru2017} argue that DLT is not going to progress without an academic examination of the art and science of developing these systems which they coin BOSE (Blockchain Oriented Software Engineering).  Similarly, Destefanis et al \parencite*{Destefanis2018} evaluates that innovations such as smart contracts and DAOs can only reach full potential with a specific discipline considering best practices, testing and design patterns.

The literature also examines how DLT might impact other areas.  Within the field of computer science, DLT applications are proposed in cloud storage \parencite{Benet2014}, operating systems \parencite{Grigg2017}, artificial intelligence \parencite{Murphy2018} and internet-of-things \parencite{Chakraborty2018, Reyna2018}; so offering users the ability to own their identity whilst interacting with the promised ``new economy'' DLT heralds \parencite{Swan2015}.  DLT's impact on technology and science may go beyond computer science - e.g. within medicine FoldingCoin examines how cryptocurrency can incentivise distributed networks to contribute compute to projects, such as folding proteins \parencite{Ross2015}.  

The technological aspect of the literature reminds us DLT is still emergent. The body of work is primarily concerned with narrow design issues, although this is changing with the rise of BOSE as a discipline and consideration how this might cross-fertilise wider technology and science.

\subsection{Legal}

How a technology relates to the world is much dependent on the legal context. It is thus important to review the literature that examines where the law and DLT intersect.

DLT's `immutability' represents a challenge for  legal systems - a court order instructing deletion is unenforceable against a censorship-resistant, decentralised system offering no method to erase.  Fink \parencite*{Fink2017} argues the European Union's General Data Protection Regulation (GDPR),  based on an old world order of centralised data silos, fundamentally conflicts with DLT; whilst also believing DLT might contribute to data protection regulations by providing citizens cryptographic control.  GDPR aside, another concern is illegal content, such as images of child sexual abuse, as Matzutt et al \parencite*{Matzutt2018} identifies on Bitcoin's blockchain.  Depending on legal interpretation this could make all full nodes illegal, so allowing unscrupulous authorities to clandestinely add illegal data and then ban Bitcoin use \parencite{Colyer2018}.  Moore \parencite*{Moore2018} however states this is over-emphasised, highlighting that illegal content on the blockchain requires considerable processing to make human-readable, and compares how many dollar bills contain some trace of illegal drugs.  In certain jurisdictions the issue is not how DLT interacts with existing law; but that specific DL have been declared illegal.  Venezuela for instance has banned Bitcoin, which can be interpreted as a defence to threats of ``crypto-secession'' \parencite{Allen2017}.  Maupin \parencite*{Maupin2017} criticises these blanket bans; instead jurisdictions should consider the merits of individual use cases and co-ordinate across borders.  Much of this law though applies to permissionless DLT, particularly cryptocurrency; permissioned DLT, often corporate established, is less constrained by these rulings.

The relationship between DLT and the law however is not one-way, Wright and De Filippi \parencite*{Wright2015} suggest DLT will lead to a reduction in authorities' control through legal mechanisms.  A subset of law, ``\textit{lex cryptographia}'' \parencite{Wright2015}, will emerge where smart contracts and decentralised autonomous organisations administer rules agreed by willing participants. Consequently the roles of centralised institutions as arbiter and notary will diminish \parencite{Dupont2015}.  Werbach and Cornell \parencite*{Werbach2017a} refute that law is replaceable as it arbitrates when parties disagree - a situation where inherently inflexible smart contracts fail.  This criticism  is answered by Abramowicz \parencite*{Abramowicz2016}, who demonstrates ``tacit coordination games'' can decentralise legal judgements.  Abramowicz proposes a thought-experiment where a person is offered \$10 to judge whether the next person who the same question is asked of, under the same conditions, will answer whether it is `hot' or `cold'; this could be expanded to more nuanced judgements.  The previous barrier to using this decision making mechanism: the requirement of a central party to remunerate winners, could be overcome by smart contracts leading to ``peer-to-peer governance'' structures \parencite{Abramowicz2016}.

In summary the literature shows that could DLT face legal challenges as the old world of centralised authority collides with the new decentralised one, but beyond that the impact might be the disruption of the entire legal system. 

\subsection{Ecological}

Given the permissionless nature of the network, estimating Bitcoin energy demand is not a simple task.  One estimate suggests it consumes the same amount of energy as Ireland: 3 - 6 gigawatts \parencite{Dwyer2014}. More recent studies \parencite{Vranken2017} estimate less: 0.1 - 0.5 gigawatts.  Bitcoin, which uses a Proof-of-Work consensus, relies on there being a mining cost to prevent a `51\% attack,' whereby a malicious party hijacks the network (Section \ref{sec:Genesis}); as all currencies entail a production and distribution cost when this is accounted for some counter that physical cash or gold is actually less sustainable \parencite{McCook2014}.  Although permissionless blockchains typically use PoW, most permissioned DL use other consensus mechanisms such as Proof-of-Authenticity, with similar energy cost as typical software.

DLTs impact on energy is far from solely negative - Saberi, Kouhizadeh and Sarkis \parencite*{Saberi2018} propose its use in product hazard and disposal management through recording information pertaining to the European Community Directive on Waste Electrical and Electronic Equipment.  Once could envisage a DL, with the Department for the Environment acting as regulator, which companies are mandated to complete for new products, and to which recyclers have access at the end of products' life; so retaining product information even when the item is no longer manufactured.  This scheme would allow stakeholders to understand both where hazards lie and how scarce resources could best be reclaimed.  

This life-cycle provenance could also apply to environmentally sensitive commodities. For example, using DLT to track diamonds, fish and pork allow consumers and retailers access to the items' ecological footprint \parencite{Chapron2017}.  Coupled with Internet-of-Things connected devices, information such as storage temperatures could be added for food safety management.

If this section started with DLT's electrical use, it is fitting to end with how it might directly reduce it.  Hwang et al. \parencite*{Hwang2017} identifies the use of blockchain in `prosumer' electrical generation, where individuals produce renewable energy.  DLT can be used to sell prosumers' surplus energy, recording who used what without relying on a central authority - smart contracts and cryptocurrency can be layered on top to enable autonomous payments.  Reporting on a microgrid utilising DLT in New York, Mengelkamp et al \parencite*{Mengelkamp2018} views this positively but suggests that real-world application will be constrained by this being a highly regulated market.   RightMesh \parencite{Ernst2017} intend to use DLT and tokens to incentivise mesh networks allowing subscribers to share their internet access and be recompensed, especially useful in deprived areas.  The use of DLT enabling consumers to be producers is another example of the revolutionary mindset, upending the traditional role of central authority, this time in the guise of big energy.

\section{Conclusions and implications}
\label{sec:conclusionsAndImplications}

The literature shows that DLT has the potential to impact across the entirety of PESTLE; in many ways radically.  Additionally it suggests that with the increasing adoption of DLT in established institutions, focus is switching to evolutionary from revolutionary.

There is a parallel here too within the technology - whereas the revolutionary can be seen to equate to Figure \ref{fig:WalportTypesDLT}'s early permissionless and public systems, the evolutionary counterpart is the later permissioned and private.  There are implications in this for the DSN - it is likely that permissioned systems will be the preferred route.

Lord Holmes \parencite*{Holmes2017} has already reached this judgement by advocating permissioned DLT in government.   Permissionless systems require a mechanism to ensure the honesty of nodes - Bitcoin's genius is in ensuring this through energy-intensive Proof-of-Work and on-chain financial incentives.  It is difficult to imagine what the equivalence of this would be with a system designed around the business of the DSN.  If the DSN were to use a DLT which required no permission to join or append information too, how would nodes be motivated to act in a trustworthy fashion?  Likewise a public DSN would put information, which you might not wish to share with military adversaries, in a public forum.  

That is not to rule out using permissionless DLT.  Recording hashes of a permissioned DLT on a permissionless DLT leverages up immutability, referred to as a two-layer blockchain architecture \parencite{Aniello2017}.  Alternatively Rosenfeld \parencite*{Rosenfeld2012} propose using meta-data in a permissionless blockchain to store information standing in for asset ownership (so called coloured coins), although this method has critics \parencite{Swanson2015a}.  Additionally newer forms of DLT, such as Hashgraph, a directed acyclic graph using ``permissionless consensus'' with ``permissioned governance'' \parencite{Baird2018}, might necessitate a reassessment of the basis for judgement.

This dissertation will also primarily examine DLT which enables the \textbf{share} functionality of Figure \ref{fig:ShareProveVenn}.  Although DLT that enables \textbf{prove} alone is useful in cyber-security, the efficiencies in the DSN will come through sharing data across organisational boundaries. However there will be no attempt to select a particular DL, this is a fast moving space and different use cases will draw on different technological models.

In the same way however that a permissioned DLT that follows the evolutionary path is more likely to be applicable for the DSN, the methodology of this dissertation will follow the sociology of regulation approach.  This paradigm considers iterative change rather than root and branch reform, and will be covered in the next chapter.

\chapter{Methodology}

\section{Research paradigm}

This dissertation takes a functionalist research paradigm \parencite[p.~80]{Johnson2000}.  That is an objectivist approach (the author is external to the study and is able to observe the facts in a way that is generalisable) combined with a `sociology of regulation’ perspective (that is the study will look to improve the organisation studied, rather than suggest radical change). The studies philosophical position is that of pragmatism: the research question of applying a technology to Defence takes centre stage, and concepts are  only relevant when they support this practical aim. The deductive approach is used – the literature will be studied, followed by suggestions of how DLT might be used within a defence context.  The time horizon will be cross-sectional, that is the study of DSN in the DLT at this moment, rather than change and development over time  \parencite[pp.~122-152]{Saunders2016}.  

This approach was selected as it matched the Research Question and Aim (Section \ref{sec:researchQAO}) which was not to establish a radical re-understanding of the DSN or the perception with which it is held, but rather a practical analysis as to how a technology might be applied.  It is acknowledged that the functionalist approach has limitations, namely its avoidance of both the role of conflict and the agency of individuals \parencite[p.~432-442]{King2011}, however this was considered less germane to the research in hand.

\section{Research strategy}

A simple form of Action Research is used – the context will be understood, problems will be identified and solutions will be proposed using the nominated technology. It will differ from typical Action Research in that proposals will not be implemented and iterated on; this is due both to time and financial constraints.  Given that this study focuses on understanding problems and providing solutions, potentially having practical relevance to further MoD work-streams, it falls closer to applied than basic on the research spectrum \parencite[p.~9]{Saunders2016}.  

\section{Data collection}

Data was collected using sequential mixed method research \parencite[p.6]{Ivankova2016}. Initial exploratory discussions with individuals was followed by semi-structured interviews combined with a questionnaire. These were held with two groups of interviewees: those with experience of DLT - primarily from the technology-sector; and those with knowledge of the DSN, primarily MoD employees. Exploratory discussions were open interviews enabling contextual data to be gathered and to facilitate further lines of enquiry. Follow-on semi-structured interviews were then held with experts on DLT and DSN processes.  Semi-structured interviews allowed exploration of interviewees' insight, a key requirement for an emerging (and therefore not fully understood) technology.  The questionnaire also acted as a prompt of discussion for qualitative data.  Data collection also occurred through analysis of documentary evidence \parencite[p.~107]{judithbell2014}, for instance policy instruction that governs the DSN (e.g. Defence Logistics Framework) and minutes of various committees (e.g. Defence Logistics Directorate IS Working Group).

Reliability and validity were key concerns throughout data collection; the former being whether a process produces the same results when duplicated and the latter being whether the process collects the correct evidence to support the research aim \parencite[p.~103-104]{judithbell2014}.  Interviews in particular can suffer from problems with reliability, validity and additionally bias - where either the personal views of the interviewer are introduced into the process or the interviewee does not provide a true representation of their views \parencite[p.~50]{White2014}.  Interviewee bias was minimised by explaining the research was not commercial in nature, offering anonymity and conducting interviews where participants could not be overheard, so minimising career harm and sales pitches - see Section \ref{EthicalConsiderations}.  Interviewer bias was minimised by using videos to explain the technology, although as discussed in Section \ref{sec:defenceSectorInterviewees}, it was difficult to find a video short enough to serve as an introduction, yet long enough to provide a balanced view of pros and cons.  The use of videos also increased reliability by ensuring all defence-sector interviewees were provided with the same base-line DLT information.  

Despite this bias was considered a particular problem for technology-sector interviewees.  Firstly, company employees are unlikely to disclose criticism of their products due to potential commercial implications, regardless of whether they are actively trying to sell.  Secondly, being actively involved with a technology and agreeing to an interview focused on such technology suggests a personal belief in DLT which might lead to over-optimism. Similarly, although effort was made to adopt a neutral stance, the present researcher's interest in this dissertation topic could lead to bias in interpreting data collected \parencite{Nickerson1998}.  Although this risk was difficult to mitigate, an attempt was made by encouraging interviewees to discuss negative aspects, as well as seeking out DLT-critical research.

Sampling-frame bias, where the selected interviewees constitute a non-representative sample so leading to an invalid understanding of either DLT or DSN \parencite[p.~82]{White2014}, was also considered. Defence interviewees were selected for their expertise in the generic use cases presented in the questionnaire, so enabling  assumptions to be thoroughly questioned.  Whereas technology-sector interviewees were selected to gain a broader understanding of DLT, therefore a variety of organisational types were sought out.  More technology-sector interviewees were interviewed than defence, as the research was being conducted from a position of experience with the DSN.  Table \ref{table:interviewees} illustrates these points.  The research did suffer from a lack of interviewee gender and ethnicity diversity \parencite[p.~126]{Mergaert2015}, efforts were made to counter this but were frustrated by dates available for interview; this may have been a consequence of neither the defence \parencite{Diversity2018} nor technology-sector \parencite{BritishComputerSociety2017} being exemplars of inclusivity.

Combining qualitative and quantitative data allowed triangulation, so increasing validity and reliability \parencite[p.~71-72]{Collis2013}.  Reliability was increased by using a Likert-scale questionnaire, meaning that the another interviewer would have received the same answers regardless of the  particular relationship between interviewer and interviewee.  Validity was increased by using semi-structured interviews which allowed a deeper appreciation of interviewees thought-processes, so allowing an understanding of why DLT would be applicable to questionnaire use cases; this also provided generalisability \parencite[p.~365-366]{Symon2012} as this understanding could then be considered against processes not present in the questionnaire.  Feedback on the questionnaire was also provided by a number of advisors with experimental experience.  Ultimately an awareness of reliability and validity concerns ensured some mitigation, however with a research team of one this was an inevitable limitation.

Surveys were considered but not used because DLT is an emergent technology and therefore by definition many will be unfamiliar with it.  Surveys are typically used with a ``sizeable population'' \parencite[p.~728]{Saunders2016}, yet the number of people who have understanding of how the technology might be applied to DSN amounts to a small sample.   As this research involves exploring a novel topic, the open ended nature of interviews afforded this latitude.

\section{Ethical considerations}
\label{EthicalConsiderations}

The primary ethical concern was causing interviewees  `career harm'  by publicising opinions which might damage their professional standing.  For instance a consultant revealing that they thought DLT was too immature for business application, might affect their future employment.  Therefore interviewees were offered the options of being identified personally (e.g. Joe Bloggs of IBM), by their employer (e.g. IBM) or anonymity.  In this latter case interviewees would be referred to by the sector they worked for (e.g. `a technology company employee...').  Care was also taken to ensure that interviewees were interviewed where they were not overheard by co-workers.  All interviewees were given an sequential number for the purpose of data collection and analyses.

Another ethical concern was conflict of interest.  Many interviewees from the technological sector work for DLT vendors.  Data collected would be biased if interviewees misinterpret the interview as a sales opportunity.  Therefore the purpose and nature of the research was clarified when recruiting  candidates. Not only was it clearly stated in the consent form that all discussions were of non-commercial nature without commitment or prejudice of the MoD, these were also preambled at the interview.

This research was awarded a Level 2b risk assessment level and authority to proceed on 20 Sep 17 by the Cranfield University Research Ethics System.  MoD Research Ethics Committee approval was not required.

\chapter{Method} 
\label{ch:method}

\begin{figure}
\includegraphics[width=\textwidth]{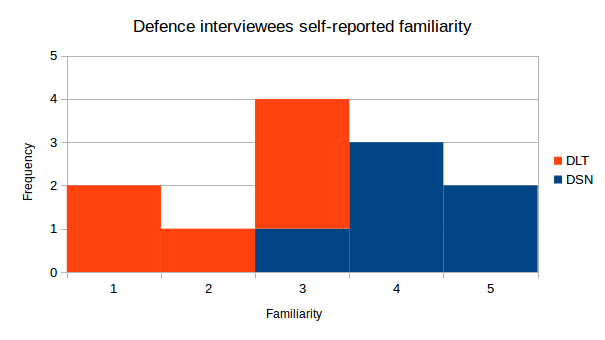}
\includegraphics[width=\textwidth]{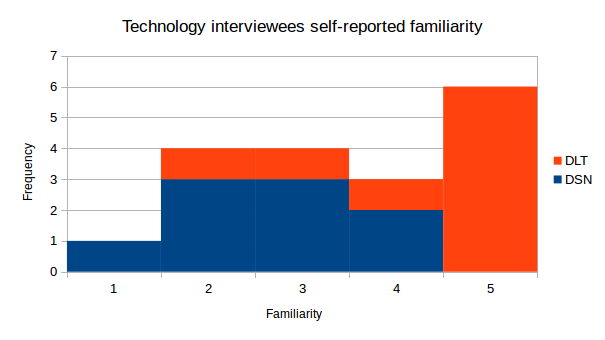}
\caption[Participants' familiarity with DLT and DSN]{Defence vs technology-sector interviewees self-reported familiarity with DLT and DSN (Author's own work)}
\label{fig:techFamiliarity}
\centering
\end{figure}

Participants were selected from two groups: the technology and defence sectors.  The former were chosen for their knowledge of DLT while the latter for their knowledge of the DSN.   An overlap was foreseen: some technology interviewees might be familiar with DSN (e.g reservists) and some defence interviewees with DLT (e.g. bitcoin media coverage).  To capture this, all interviewees reported their familiarity with both DSN and DLT on a five point Likert scale as shown in Figure \ref{fig:techFamiliarity}.  

Ten semi-structured face-to-face interviews were  conducted with technology interviewees and six with defence (with an in-depth exploratory interview being conducted with one interviewee from each sector prior to this to establish a research direction).  Interviews took place at the interviewees' workplace or at a mutually convenient location (e.g. cafe), locations where chosen to ensure interviewees could hear the videos clearly and were not overhead by co-workers .  Interviews typically lasted 45 minutes to one hour.  Technology and defence interviews followed different formats. 

\section{Technology-sector interviewees}

Participants were selected in two ways. Social networking within the MoD enterprise, trade events (e.g. Team Defence Information) and the MoD intranet were used to identify companies with an ongoing relationship with the MoD and an interest in DLT.  Companies were then approached and asked if any of their employees  would be willing to be interviewed.  Secondly interviewees were also recruited by asking at the end of an interview for additional contacts who would be willing to participate.  Typically, but not always, the former method led to interviewees from established technology companies, and the latter to start-ups. 

The format used for all interviewees from the technology-sector was:

\begin{enumerate}
\item Introduction to research and consent approval
\item Overview of DSN 
\item Discussion of company technology or organisational interest
\item Interviewee completes questionnaire - discussing with interviewee throughout 
\item Exploration of further defence use cases
\item Discussion of further avenues of research (either people, organisations or literature)
\end{enumerate}

Item 2, the overview of the DSN, was brief and featured an oral introduction similar to that provided in Section \ref{sec:DefenceSupportNetwork}.  Typically all interviewees had some knowledge (first-hand or otherwise) of supply chains, so this concept was easily relatable.  Figure \ref{fig:techFamiliarity} shows that technology-sector interviewees reported greater knowledge of the DSN, than defence-sector of DLT. 

\section{Defence-sector interviewees}
\label{sec:defenceSectorInterviewees}

Interviewees from the defence-sector (i.e. chosen for their knowledge of defence processes) were selected from networking within the enterprise and the MoD intranet.  While technology interviewees were selected from a wide pool, with any organisation that had a DLT undertaking considered admissible; Defence employees were selected on the basis of expertise in one of the generic use cases at Appendix \ref{ch:questionnaire}.

The defence interview format differs from the technology-sector in replacing the introduction to DSN with DLT:

\begin{enumerate}
\item Introduction to research consent approval
\item Discussion of role of interviewee
\item Introduction to DLT
\item Interviewee completes questionnaire - discussing with interviewer throughout 
\item Exploration of further defence use cases
\item Discussion of further avenues of research (either people, organisations or literature)
\end{enumerate}

Interviewees were introduced to DLT through two short videos.  Showing videos rather than reading a script helped prevent the researcher showing positive bias unconsciously (e.g. via tone of voice), whilst at the time ensuring consistency. Illustrating the relevant underlying concepts visually (for example animation of the chaining of data blocks) helped explain complicated concepts to those without an Information Systems (IS) background.  Videos might also make what was potentially a dry subject more interesting compared to reading or listening to a monotone script.

The videos were selected from YouTube.  The search term ``blockchain introduction'' was used and the resulted videos were sorted by YouTube ascribed `relevance.'  Videos longer than 10 minutes were excluded.  The resulting list was reviewed in order of relevance.  Videos with poor audio, graphics and non-professional delivery were discounted.  Remaining videos were checked for user-comments related to usefulness in explaining the technology to those previously unfamiliar with DLT before being given a researcher ascribed rating.  This process was repeated using the search term ``distributed ledger technology introduction.''  In total nineteen videos were given ratings.  

Following review it was decided that no single video sufficiently explained both the technology itself and applications of the technology, necessitating the selection of two.  The video selected to explain the technology was produced by the UK Government Office for Science \parencite{GOScience2016} and accompanied the Walport \parencite*{Walport2016} report; which although providing a good grounding in the technology concentrated more on finance - potentially leading defence-sector interviewees to think DLT had little relevance to the DSN.  The second video therefore selected was an IBM \parencite*{IBM2016a} demonstration of how the blockchain might be used in the car industry.  Choosing this video did risk interviewees being influenced by their attitude towards IBM; however it was felt  that the IBM video provided a far superior explanation on how the technology might be applied in a DSN-like environment.  Thus effort was made to reiterate that many vendors offer this technology besides IBM, and the Government video was shown first.  The videos last 5 min 15 seconds and 3 minutes respectively.  Interviewees had an opportunity to ask questions after watching - however typically interviewees reported that the videos had provided a satisfactory overview of DLT.

One risk that it was difficult to mitigate against was giving an overly positive impression of DLT.  Introductory videos by definition do not cover critical detail; this was especially the case in the IBM clip (a DLT vendor).  An attempt at balance was made by explaining that this was an emergent technology with few real world use cases.  Ultimately this trade off was accepted to ensure defence-sector interviewees had a basic understanding of DLT.  

\section{Questionnaire}

Both industry and Defence interviewees completed the same questionnaire, featuring the use cases at Appendix \ref{ch:questionnaire}.  These four generic DSN use cases had been selected following an initial review of the literature, and were: Codification, Revenue \& Customs, Engineering \& Asset Management and Contracting for Availability.  Questionnaires were structured as follows: a description of the four potential use cases, questions on familiarity with DSN and DLT, questions on utility and ease of implementation of each use case and questions on confidence in answers provided.  All operative words were defined at the start of the questionnaire.  All questions were answered using a Likert-scale rating of one to five.  Where one represented not at all useful, very difficult or not at all confident; five: very useful, very easy or very confident.  Although interviewees were asked to rate how confident they felt in rating each use case (e.g. I am very confident that Use Case A is very easy to implement), in the final analysis this confidence data was not used as it added little value, partly due to its self reported nature.

Interviewees were encouraged to read all use cases prior to answering the questions - this was to ensure they were able to compare use cases as they responded.  A discussion on the strengths and weaknesses of the use cases would normally take place as the questionnaire was completed, notes were taken by the interviewer and recorded via dictaphone with interviewees' consent. 

An open discussion followed the questionnaire, bringing out further issues with the use cases and exploring other use cases not covered in the questionnaire.

Table \ref{table:interviewees} lists semi-structured interviewees.  Sector was defined by expertise not organisation.  For instance on 1 Feb a MoD employee was interviewed who contracted DLT pilot projects; they were defined as tech-sector as the knowledge they were bringing to the research was of DLT.  One interviewee listed in Table \ref{table:interviewees}, although providing qualitative data, declined to answer the questionnaire - this was due to the ambiguity inherent in the term DLT and the variety of solutions that might be applied to each use case.  Although further clarification was provided by email after the interview, the interviewee did not follow up.

\begin{landscape}
\begin{table}
\centering
\begin{tabular}{ l l l l l} 
\toprule
Date & Sector & Location & Organisational Type & Reason for interview \\
\midrule
26 Sep 17 & Tech & London & Consultancy & DLT engineering use cases\\
26 Sep 17 & Tech & London & Consultancy & DLT engineering use cases\\
5 Oct 17 & Tech & London & Think tank & DLT finance use cases\\
28 Nov 17 & Tech & London & Technology start-up & DLT supply chain use cases\\
11 Dec 17 & Tech & London & Technology start-up & DLT supply chain use cases\\
11 Dec 17 & Tech & London & Consultancy & DLT generic use cases\\
1 Feb 18 & Tech & London & MoD Information Systems \& Services & DLT defence use cases\\
2 Feb 18 & Tech & Hampshire & Technology corporate (large cap) & DLT public sector use cases\\
9 Feb 18 & Defence & Portsmouth & Navy Command Headquarters & CfA \& E\&AM expertise\\
19 Mar 18 & Defence & Bristol & DE\&S Delivery Team & CfA \& E\&AM expertise\\
11 Apr 18 & Tech & Bristol & Engineering corporate (large cap) & DLT defence use cases\\
24 Apr 18 & Defence & Bristol & DE\&S Delivery Team & Customs expertise\\
24 Apr 18 & Defence & Bristol & DE\&S Delivery Team & Customs expertise\\
10 May 18 & Defence & Bristol & UK National Codification Bureau & Codification expertise\\
16 May 18 & Tech & Hampshire & Technology private company & DLT private sector use cases\\
5 Jun 18 & Defence & Portsmouth & Defence corporate (large cap) & CfA \& E\&AM expertise\\
\bottomrule
\end{tabular}
\caption{Semi-structured interviews}
\label{table:interviewees}
\end{table}
\end{landscape}

\chapter{Discussion}

\section{Extant evaluation frameworks}
\label{sec:extantframeworks}

Lord Holmes \parencite*{Holmes2017}, identified that although it is easy to find inefficient processes within government to serve as use cases, constructing viable DLT business cases would prove much harder.  Indeed an initial exploratory discussion with a technologist in a global IT company identified that a standard process to determine the validity and utility of use cases should be a pre-condition of any research into DLT.  This led to Research Objective 3 (Section \ref{sec:researchQAO}):

\begin{quote}
To create a framework for evaluating the utility of DLT against use cases, drawing from relevant academic or business models.
\end{quote}

A framework groups conceptual elements making general assertions and identifying key features on a relevant subject, in this case the utility of the application of DLT to DSN business processes; it differs from a theory in that is not sufficient to perform hypothesis-testing research \parencite[p.~2]{Anderson2014}.  By designing such a framework, this research adds value by enabling others to assess DLT DSN use cases against a standard process, or at least provide the basis for such a process.

Extant frameworks were reviewed to determine  relevant elements for the proposed framework.  Frameworks considered were generic technology frameworks that looked at the application of technology to an enterprise, and DLT-specific frameworks (divided into those originating from academia versus commerce).

\subsection{Generic technology evaluation frameworks}

A number of academic models explore a technology's utility within a business process.  Unfortunately from the perspective of this dissertation many of these models do not examine a technology directly, but rather users' perception of the technology.  For instance the Technology Acceptance Model \parencite{Davis1989} claims to be a ``robust, powerful, and parsimonious model for predicting user acceptance'' but acts through subjective questions, such as ``people who are important to me think that I should use the system'' \parencite{VenkateshDavis2000} without assessing the technology itself.  Likewise the Expectation-Confirmation Model uses consumer research to elicit similar findings, namely that users will accept a system if satisfied, and will be satisfied if it conforms to their expectations of utility \parencite{Bhattacherjee2001}.  Resultantly this `soft' social-science approach is difficult to use practically when analysing an emerging technology prior to adoption.  At the heart of this difficulty is that IS is an amalgam of technology, people and processes \parencite[p.~12]{stair2012information} so attempting to analyse one of these factors in isolation is difficult, particularly when software itself is abstract \parencite{dijkstra1989cruelty}.

The literature also looks at adoption within organisations, covering the concepts of technology diffusion (how widely technology is used in an organisation) and infusion (how deeply technology permeates an organisation).  Studies have been conduced using the Technology-Organization-Environment (TOE) framework to measure this \parencite{Zhu2005}.  Again however this is of limited benefit in the scenario where existing business processes are being mapped against an emergent technology.

\begin{figure}
\includegraphics[width=\textwidth]{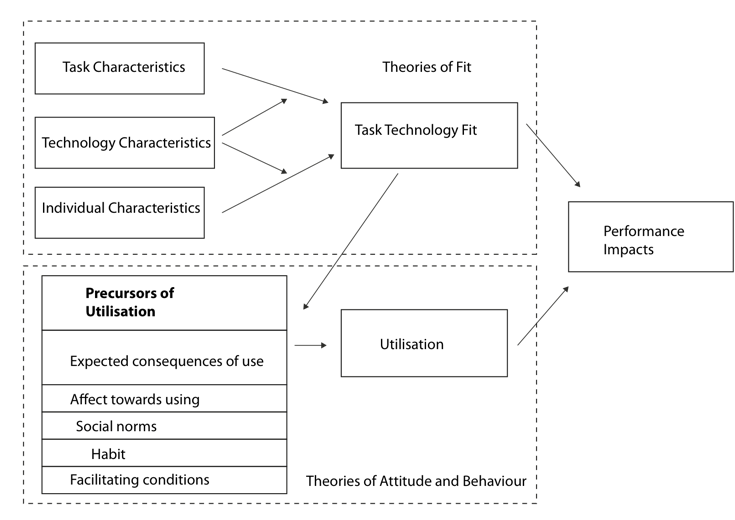}
\caption[Task Technology Fit]{Task Technology Fit \parencite{GoodhueThompson1995}}
\label{fig:ttfGoodhue}
\centering
\end{figure}

One academic model that does have relevance though is Task Technology Fit (TTF) theory.  TTF predicts that the impact on performance of a team will only be positive when the technology introduced meets the requirements of the team's task \parencite{Goodhue1995}.  Studies of TTF \parencite{Fuller2009} confirm the better the fit of the technology, the higher the initial performance of the team; expressed both in effectiveness (e.g. quality) and efficiency (e.g. time/cost).  There are three primary antecedents influencing this fit, as illustrated at Figure \ref{fig:ttfGoodhue}, and expanded below:

\begin{enumerate}
\item \textbf{Task}.  The actions conducted by individuals to transform inputs into outputs for the purposes of achieving a goal.
\item \textbf{Technology}.  The tools used by individuals to accomplish the task.  In the TTF context this refers to IS; which cover policies, training, etc, as well as the IT itself.
\item \textbf{Individual}.  Those using the tools to complete the actions required; influenced by their own specific characteristics (e.g. motivation). 
\end{enumerate}

Therefore TTF measures how a technology aids individuals in completing their tasks.  As Figure \ref{fig:ttfGoodhue} shows TTF is not the only determinant of performance, `utilisation' also matters which is influenced by attitudes, which is beyond the scope of this research.  TTF is a static model, when teams develop familiarity with IS they are used differently to the designer's intent - the Fit Appropriation Model \parencite{Fuller2009} considers this, but again is outside research scope.

Although not all the TTF factors are relevant to an assessment of DLT (for instance how reliable are the systems), Goodhue and Thompson \parencite*{GoodhueThompson1995} list some that are:

\begin{enumerate}
\item Task Equivocality
\begin{enumerate}
\item ADHC1 — I frequently deal with ill-defined business problems.
\item ADHC2 — I frequently deal with ad-hoc, non-routine business problems.
\item ADHC3 — Frequently the business problems I work on involve answering questions that have never been asked in quite that form before.
\end{enumerate}
\item Task Interdependence
\begin{enumerate}
\item INTR1 — The business problems I deal with frequently involve more than one business function.
\item INTR2 — The problems I deal with frequently involve more than one business function.
\end{enumerate}
\end{enumerate}

Task equivocality could be seen to negatively correlate with the utility of DLT.  Ultimately a distributed ledger is a form of database, and architectural decisions have to be made as to what information a database captures.  The more ill-defined the problem, the less certain we are that the right data is being captured or processed correctly, and the less useful the ledger will be in providing answers.  On the other hand task interdependence may positively correlate with the utility of DLT - after all a database able to cover more than one business unit is exactly what is meant by ``distributed.'' 

The literature also refers to more specific use case analysis tools such as Critical Success Factors (CSF) \parencite{Sebora2009, Chow2009} and Feasibility Analysis \parencite[p.~518]{stair2012information}.  CSF are those key elements that an enterprise must do well when implementing a technology to ensure success, which although useful, does not assist in actually selecting use cases, and typically requires a number of implemented use cases to study before CSF can be elicited.  Feasibility Analysis meanwhile tends to be a process where factors like Return on Investment are considered by an enterprise - but only once a use case has been selected.  Technology Readiness Levels \parencite{Mankins2009}, although initially promising due to their emphasis on emerging technology, were discounted due to the difficulty of using it with such a broadly defined technology.  Correspondingly these tools were not used in constructing a framework.

\subsection{DLT-specific evaluation frameworks} 

Turning from the more overarching literature of generic technology adoption, next DLT-specific papers were reviewed for assistance in designing an evaluation framework.  Examples of these can be found both within the academic literature, and as might be expected from commercial vendors seeking to sell DLT-based products.  

\subsubsection{Commercial DLT evaluation frameworks}

Clearly the profit seeking imperative should be considered when weighing commercial DLT evaluation Frameworks; vendors are ultimately looking to sell products.  But they should not be discounted due to this: during one interview an industry representative ventured that they were taking great pains to ensure that appropriate projects were chosen for DLT - as the technology is so new failures at this stage might poison adoption for a considerable period.

\begin{figure}
\includegraphics[width=\textwidth]{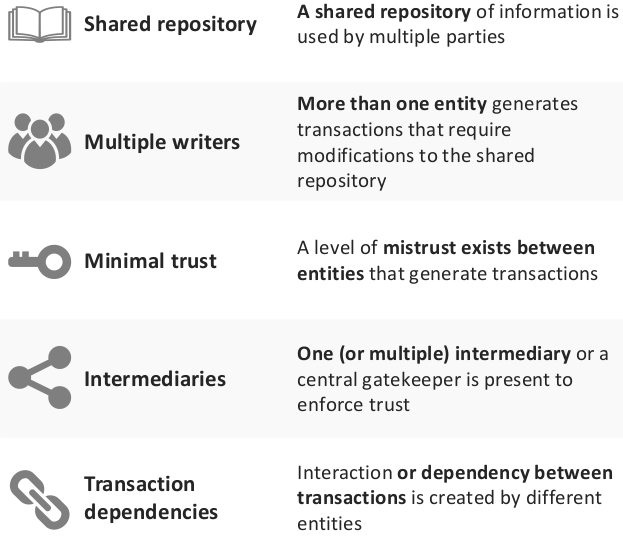}
\caption[Characteristics of high-potential use cases]{Characteristics of high-potential use cases \parencite{Mulligan2018}}
\label{fig:WEFUseCases}
\centering
\end{figure} 

Multichain, a DLT vendor, is specific about this point with their blog post entitled ``Avoiding the pointless blockchain project'' \parencite{Greenspan2015}.  Here they readily admit that DLT is, from a readiness standpoint, still in ``diapers'' and lays out a number of needed conditions where DLT should be used in preference to a standard database:

\begin{itemize}
\item \textbf{The database}.  Ultimately a DLT is a structured repository for information, therefore an enterprise needs to know why they are using a database - in the same way they would for any other store of information.
\item \textbf{Multiple writers}.  The above database needs to be modified by more than one entity.  In most cases, but not all, the writers will also keep a copy of the database themselves (i.e. run their own node).
\item \textbf{Absence of trust}.  An absence of trust must exist between the multiple entities specified above.  Greenspan usefully expands this by explaining that mistrust can exist within an organisation (i.e. between business units) and can be expressed as reluctance to let one entity modify database entities which another one ``owns.'' 
\item \textbf{Disintermediation}.  Amongst the multiple entities, who mistrust each other, there should be no ``trusted intermediary.''  That is a central gatekeeper who all parties trust and can verify and authenticate transactions.
\item \textbf{Transaction interaction}.  Transactions need to cross the boundaries of several organisational boundaries.  In other words there need to be situations of the sort where Alice's bank passes £1 to Bob's bank, who then loans that £1 to Charlie's bank.  If the ledger is simply being used to record Alice's balance, and no one else's, then DLT makes little sense.  
\end{itemize}

Although written with the aim of selling a proprietary DLT-product, this is a useful summary of where enterprises should seek to adopt DLT.  Greenspan's approach is validated by the adoption of these principles by Rodrigues, Bockek and Stiller  \parencite*{Rodrigues2018} and the World Economic Forum \parencite{Mulligan2018} (although the latter without attribution, so it is possible they have been formulated independently) as at Figure \ref{fig:WEFUseCases}.

Multichain is not alone in providing advice as to where firms should look to adopt DLT; established firms are providing similar advice in a consultancy role.  In contrast to the above which is technically orientated, much of this consultancy advice is more focussed on business use cases; Table \ref{table:DLTAdoptionComparison} is an example of this form of advice, comparing the criteria that IBM, SAP and Oracle suggest customers consider when adopting DLT.

\begin{landscape}
\begin{longtable}{ F L L L }
\toprule
\textbf{Factor}		&	\textbf{IBM}	&	\textbf{SAP}	&	\textbf{Oracle}\\ 
\midrule 
\endhead
Multiparty	&	Do we need to track transactions that involve more than two
Parties? & Multiparty collaboration: Are many different parties, and not just one, involved in the process or scenario, but one party dominates everything? For example, a company with many parties in the ecosystem that are all connected to it but not in a network or more decentralized structure. & Is my business process pre-dominantly cross-departmental / cross-organisational?
\\[3.5cm]
Auditability	&	Can the network benefit from increased trust, transparency,
and accountability in recordkeeping? & Transparency and auditability: Is it important to offer each party transparency (e.g., on the origin, delivery, geolocation, and hand-overs) and auditable steps? (e.g., How can I be sure that the wine in my bottle really is from Bordeaux?) & Is there a need to improve traceability or audit trail?
\\[4cm]
Manual process	&	Is the current system prone to errors due to manual
processes or duplication of effort? & Process optimization: Will blockchain massively improve a process that today is performed manually, involves multiple parties, needs to be digitized, and is very cumbersome to manage or be part of? & Can I improve business process by automating certain steps in it?
\\
Intermediary	&	Is the current system overly complex or costly, possibly due to the need for intermediaries or a central point of control? & - & Does it involve intermediaries, possibly corruptible? 
\\[2cm]
Trust		&	Does my business network need to manage contractual relationships? & - & Is there a trust issue among transacting parties?
\\[1cm]
Fraud		&	Is the current transaction system vulnerable to fraud,
cyber-attack, and human error? & Risk and fraud minimization: Does it help (or is there a need) to minimize risk and fraud for each party, or at least for most of them in the chain? (e.g., A company might want to know if its goods have suffered any shocks in transit or whether the predefined route was not followed.) & - 
\\[4.5cm]
Periodicity & - & - & Does it require periodic reconciliations?
\\[0.5cm]
Visibility & - & - & Do we need real time visibility of the current state of the transaction?\\
\bottomrule
\caption[Comparison of IBM, SAP and Oracle DLT use case adoption criteria]{Comparison of IBM \parencite[p.~37]{ManavGupta2017}, SAP \parencite{Roehricht2017} and Oracle \parencite{Goel2017} DLT use case adoption criteria}
\label{table:DLTAdoptionComparison}
\end{longtable}
\end{landscape}

Table \ref{table:DLTAdoptionComparison} shows that the vendors' advice has common themes.  It is possible that this is simple plagiarism; no company claiming to be at the forefront of technology would want to miss out on the next wave of innovation.  However this seems unlikely as a sole explanation considering the investment  these firms made, for instance IBM has donated 44,000 lines of code to an open-source DLT \parencite{Borek2016}.  

Assuming that this commonality exists because these are genuine areas where DLT might engage, how does this compare against the DSN?  More specifically, considering that Research Objective 2 (Section \ref{sec:researchQAO}) covers what challenges the DSN faces, this next section will consider these adoption criteria in light of these challenges.

\subsubsection{DSN challenges addressed by commercial DLT adoption criteria}
\label{sec:challengesaddressed}

\paragraph{Multiparty}  The DSN can be viewed as multiparty both externally and internally. As  illustrated in Figure \ref{fig:DSNRichPicture}, the DSN relies upon a network of external parties - equipment and items are purchased from, maintained and calibrated and subsequently disposed of by industry.  These are not merely external touch points - the use of contractors embedded on operations, for repair and resupply, has increased steadily since the 1990s as part of the `Whole Force' concept \parencite[p.~9]{uttley2005}.  An example of such  extensive  arrangements  is the UK Military Flying Training System (MFTS); here DE\&S use both industry equipment and personnel in a Private Finance Initiative to train armed forces aircrew \parencite[p.~37]{MoD2016}.  Despite the depth of industry links, the MoD has been criticised for failing to manage its contracts proficiently.  The Public Accounts Committee noted that the MFTS contract was late, over-budget and underperforming; a key reason being that data to manage the contract was unavailable as it was ``held in pockets within the Department, and is not routinely analysed'' \parencite[p.~14-16]{Parliament2015}.  

\begin{figure}
\includegraphics[width=\textwidth]{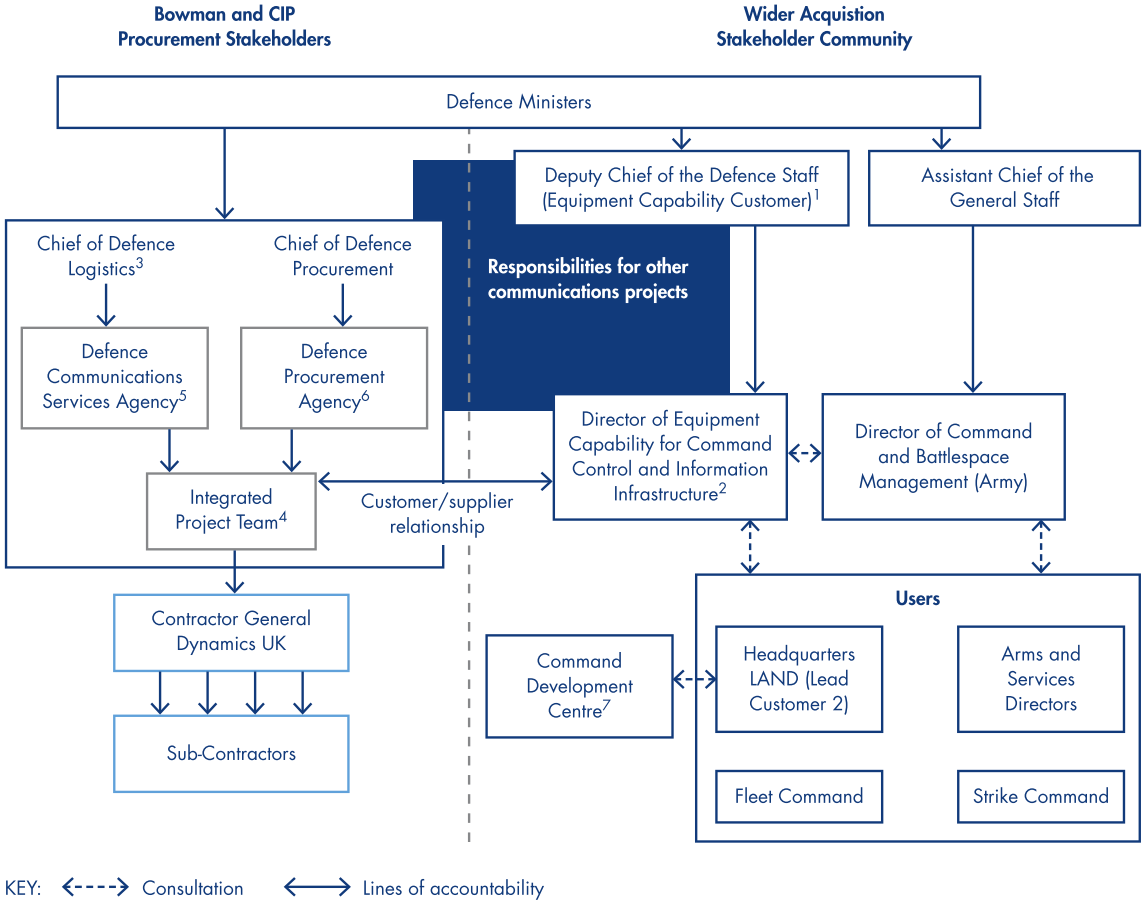}
\caption[Bowman radio procurement stakeholders]{Bowman radio procurement stakeholders \parencite{NationalAuditOffice2006}}
\label{fig:naoBowman}
\centering
\end{figure}

The other element of multiparty is the internal element,  as Oracle comments not all DLT members have to be external.  Again an area that the DSN struggles with; Figure \ref{fig:naoBowman} illustrates what the National Audit Office \parencite*{NationalAuditOffice2006} described as ``complex inter-relationships'' in Bowman procurement which contributed to the failure of the project.  Beck and Muller-Bloch \parencite*{Beck2017}'s case study of a firm that did adopt DLT found that both intra- and inter-organisational boundaries faded due to the decentralised nature of data sharing - this could therefore benefit the DSN whose very definition is as a ``flexible set of [...] connecting points [and] linked nodes'' \parencite[p.~9]{MoDJDP2015}.

The DSN can therefore be seen as a multiparty enterprise -  internally due to its federated nature and externally due to its relationship to suppliers.  SAP in their blockchain adoption advice recommends that one party dominates this multiparty arrangement - possibly because the dominant player might be able to influence others to adopt a new technology.  This would  fit well with the MoD - it is often a major client of suppliers and consequently  is in a position to influence them.  Thus the multiparty factor of DLT adoption could apply to the DSN; so helping MoD overcome its challenges with sharing data over organisational boundaries. 

\paragraph{Auditability}  The next factor that all three vendors agree on is that of auditability.  When one considers the criticality of equipment (e.g. armaments, medical, radiological items) that makes up the DSN, this is a key requirement.  An example can be taken from the Haddon-Cave report into the mid-air explosion of a Nimrod surveillance aircraft over Afghanistan with the loss of 14 lives.  This report tracks the procurement of a rubber seal that was the likely cause of the fuel leak responsible.  This had been sold by the manufacturer for £15, and after having passed through two subsequent sub-contractors, was eventually procured by Defence Equipment \& Support (DE\&S) for £123.50 \parencite[p.~101-102]{HaddonCave2006}. DE\&S assumed at this cost it meant they were procuring the item from an ``accredited aviation supplier'' however neither the manufacturer nor anyone else in the procurement chain were performing any quality control.  This issue might be ameliorated by a DLT, as interviews with technologists involved with the automotive industry revealed.   If DE\&S implemented a system whereby any party who was supplying air-worthiness items had to log dates of manufacture and identifying batch or serial numbers on a blockchain; then any quality checks performed either by the original equipment manufacturer or a third party could also be logged against that item on the same blockchain.  Signatures using public-private key encryption would ensure the quality control was only conducted by authorised parties.  Although this would still leave the digital-physical gap (i.e. how can one be sure the item in the blockchain is the same as the item in hand) it would help in ensuring the provenance of items where many layers of sub-contractors are involved.  Auditability could therefore be a strong driver for DSN adoption of DLT.   

\paragraph{Manual Processes}  All three vendors agree that adopting companies should look to DLT to replace manual processes, specifically where those manual processes introduce errors.  This is a factor that is also identified by Lord Holmes \parencite*{Holmes2017} arguing that both the health sector and real estate markets suffer from excessive manual reconciliation which could be eradicated by DLT.  This problem also exists in the DSN however, as illustrated by the NAO while investigating the MoD's Logistics Information Systems \parencite{NationalAuditOffice2011}.  Process mapping the delivery of a single item to Afghanistan the NAO found data of its movement recorded on four different systems.  To allow full visibility of the consignment the data had to be transferred manually, involving portable hard drives and re-keying, thus is error prone..  Although the situation since 2011 has improved, with the limited introduction of some new Log IS, many examples of manual reconciliation still exist.  However the existence of a manual process in itself is not a strong driver to DLT specifically - the basic premise of almost all IS is that they take an unstructured process and automate it, in the process reaping productivity gains \parencite{Davenport1990}.  This particular use case criteria might therefore be best considered not as a single driver, but one to be taken in consultation with others: when looking for potential DLT use cases a logical starting point are ones that currently have a large proportion of manual effort.

\paragraph{Intermediary}  The next criteria is not universal: SAP and Oracle highlight the role of intermediaries, while IBM does not.  The role of intermediaries is also highlighted in Greenspan's \parencite*{Greenspan2015} analysis; this is not surprising, one of DLT's founding principles was to removes intermediaries, in the case of bitcoin the banks.  This criteria does not necessarily map across to the DSN because the MoD often plays the role of Greenspan's ``trusted intermediary'', with the MoD often playing an outsized role in arrangements between it and its suppliers.  It does not necessarily mean there are no aspects to explore here - one avenue might be a potential `disintermediation' of roles the MoD plays.  For instance the role of the United Kingdom National Codification Bureau (UKNCB) is to assign NATO Stock Numbers (NSN) on behalf of the UK to items procured by the MoD.  But as the majority of the item information is supplied and updated by the manufacturer, it might be more efficient for the MoD and manufacturer to simply share a DL; this use case will be further discussed in Section \ref{sec:codification}.  Thus disintermediation could therefore be an internal effect.

Although the MoD may play an outsized role with its own suppliers, this does not apply in coalition where the DSN works alongside other nations; as key element in DSN doctrine, as illustrated at Figure \ref{fig:DSNRichPicture}.  An example here is the F-35 Joint Strike Fighter (JSF) project where equipment is shared between a multinational alliance of twelve countries.  Each time repairable JSF equipment is transferred between nations it is accompanied by an Electronic Equipment Logbook (EEL) which shows all maintenance performed on that item.  There has been criticism of how this is handled within the JSF's Autonomic Logistics Information System (ALIS) \parencite[p.~53]{Behler2018}, potentially using a distributed system to record and share this information might be more effective.  As ALIS has also been criticised for inappropriately sharing sensitive data \parencite{Seidel2017} a DL could also have the advantage of giving nations more granularity over  data to be   shared (e.g. a nation would share what maintenance was conducted, but might redact where that maintenance was conducted).  Consideration though should be given to whether a standard relational database might be more efficient, which is possible with an agreed central authority to trust; trust being the subject of the next criteria.

\paragraph{Trust}  Trust is not mentioned by all vendors. Oracle mentions it explicitly, while IBM's reference is implicit in referencing contractual relationships.  As  discussed the DSN relies heavily on contractual relationships;  this research discovered trust issues here, particularly involving Contracting for Availability (CfA) - which is where the MoD pays a contractor for  performance of a platform (e.g. days a ship is available for tasking) rather than for the equipment outright \parencite{Caldwell2014}.  One MoD interviewee reported frequently disagreements with industry when reviewing CfA performance indicators because both parties believed their own data more reliable.  Echoes of this can  be found in official publications, for instance NAO's assessment of the MoD's current equipment plan raises ``significant risk around ... the quality of contractor data'' \parencite{Parliament.HouseofCommons2017}.  Trust issues over contractor data therefore have room for improvement, meaning that the DSN meets this criteria for DLT adoption.

\paragraph{Fraud}  IBM and SAP both list fraud minimisation as reasons to adopt DLT; although this is a wide category, with IBM including cyber-attack and human error, and SAP a more general `risk.'  Oracle although not specifically having a fraud criteria refers previously to `corruptible' intermediaries.  Given the sums of money involved in the DSN it is perhaps not surprising that contractor misconduct does occur - forty-four allegations of bribery and corruption involving the equipment budget were made by the MoD between 2011 and 2016 \parencite{PressAssociation2016}.  Similarly the Single Source Regulations Office \parencite*{SSRO2016} stated that £61 million charged by DE\&S contractors may be for ``costs that are not appropriate, attributable and reasonable.''  This is a criteria that again fits the DSN.  Fraud within the DSN is not limited to financial matters such as inflated costs, it can also have a physical impact such as with the supply of counterfeits.  Indeed, 15\% of spare and replacement semiconductors procured by the US military are estimated to be counterfeit \parencite{SemiconductorIndustryAssociation2013}.  Within a defence context counterfeit products raise not only operational issues, due to reliability and quality concerns from unscrupulous sub-contractors using fake items to reduce costs; but also cyber-security risks - counterfeit electrical items can be an attack vector for hostile nation states.  As defence equipment becomes ever more complicated and reliant on contractors, this is a growing problem \parencite{Barnas2016}.  Hsieh and Ravich \parencite*{Hsieh2017} directly address this problem in their analysis of how DLT might be used to protect the Industrial Base from Supply Chain attacks.  They propose that contractors and sub-contractors involved in the manufacture of complex systems are provided with accounts on a blockchain and payment is only rendered when a prime or sub-contractor records value-adding activity on this chain.  Data analytics performed on this chain would then highlight anomalies where bad actors (i.e. front companies) might have opportunity to supply counterfeits.  Although this is a bold vision, there are considerable challenges, for instance influencing all the sub-contractors to join the chain.  Either way fraud is certainly a criteria where DSN and DLT might interact.

\paragraph{Periodicity and Visibility}  Only one vendor, Oracle, provide periodicity and visibility as a criteria within Table \ref{table:DLTAdoptionComparison}.  Both are tangential to DLT implementation - many systems provide real time visibility and are periodically reconciled.  Thus these criteria are excluded from the proposed evaluation framework.

\subsubsection{Academic DLT evaluation frameworks}
\label{sec:Academic}

Having examined a sub-set of commercial advice as to when DLT should be adopted and how this might address DLT challenges; the academic literature will now be reviewed.  This is far less focussed, than the commercial publications, on the solving of particular business problems.  Rather academia focuses on the technical aspects of implementation, this is borne out by previous DLT literature reviews \parencite{Yli-huumo2016, Risius2017}.  Subsequently these frameworks, unlike the vendor supplied ones, will not be compared to the challenges the DSN faces (Research Objective 2, Section \ref{sec:researchQAO}).

\begin{figure}
\includegraphics[width=\textwidth]{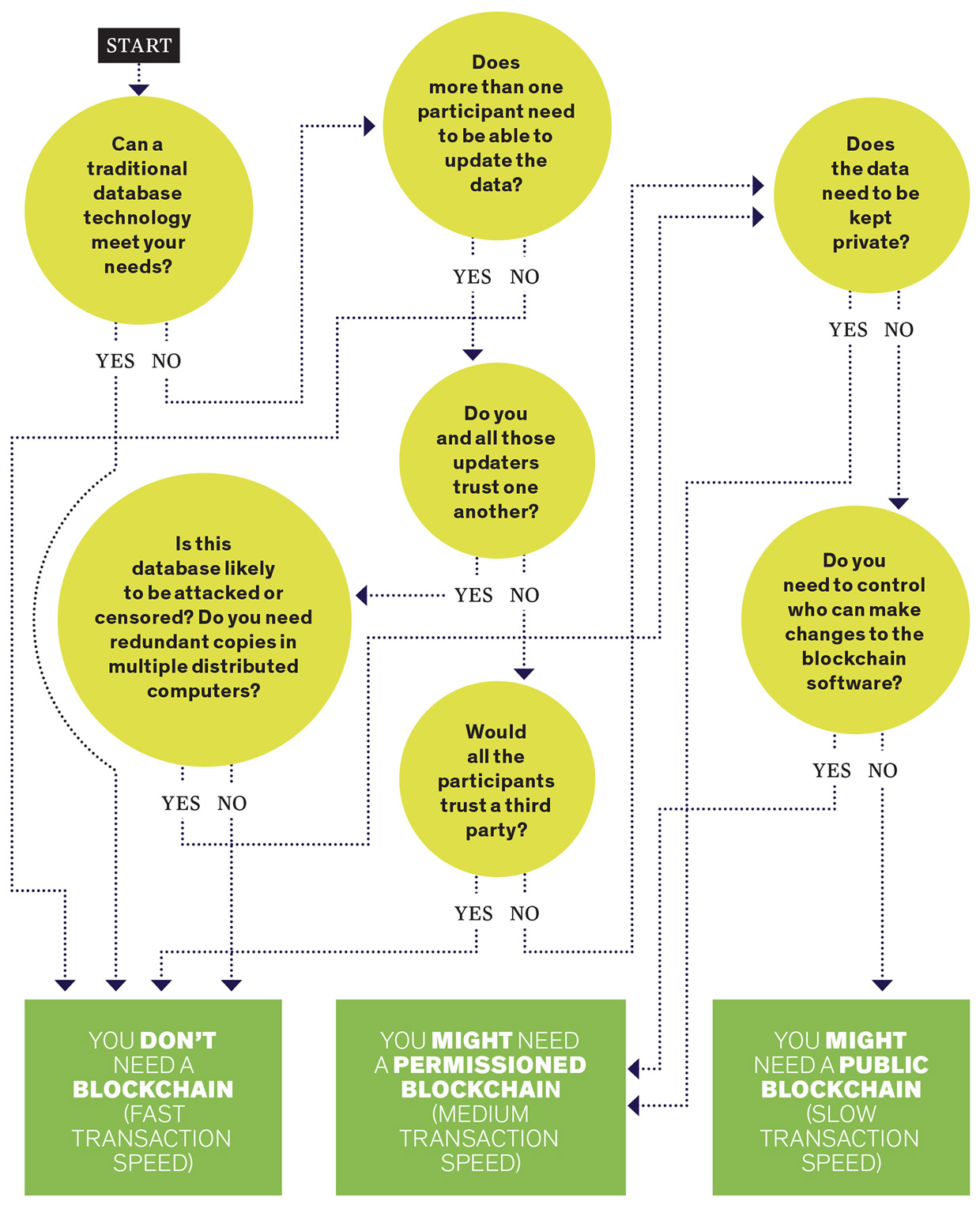}
\caption[Peck's blockchain adoption decision flowchart]{Peck's \parencite*{Peck2017} blockchain adoption decision flowchart}
\label{fig:PeckDoYouNeed}
\centering
\end{figure}

Peck \parencite*{Peck2017} provides a decision tree to assist enterprises to decide if they require a blockchain, and if so, whether public or permissioned, as at Figure \ref{fig:PeckDoYouNeed} (NB - Peck is a journalist, but as she is writing in a technical journal this will be assessed as academic literature).  Of the seven possible paths a user might take through the tree; four result in not needing a blockchain, two in a permissioned blockchain and only one in a public blockchain.  The pathways that lead to DLT adoption are based around censorship resistance and universal access; while if speed, cost, speed, predictability or privacy are important Peck believes you should avoid blockchain.  The logic behind this is unclear, if not actually circular.   If the user answers the starting question, ``Can a traditional database technology meet your needs?'', negatively; the user will still be redirected to the default of ``You don't need a blockchain.''  But in this case what solution is Peck proposing for the enterprise - a non-Computer Based IS (e.g. a filing cabinet)?  Alternatively maybe Peck is assuming that a ``no'' to the traditional database question might be the wrong answer - in which case the user should be redirected to the traditional solution; however if you cannot rely on reliable answers the decision tree concept is flawed.

\begin{figure}
\includegraphics[width=\textwidth]{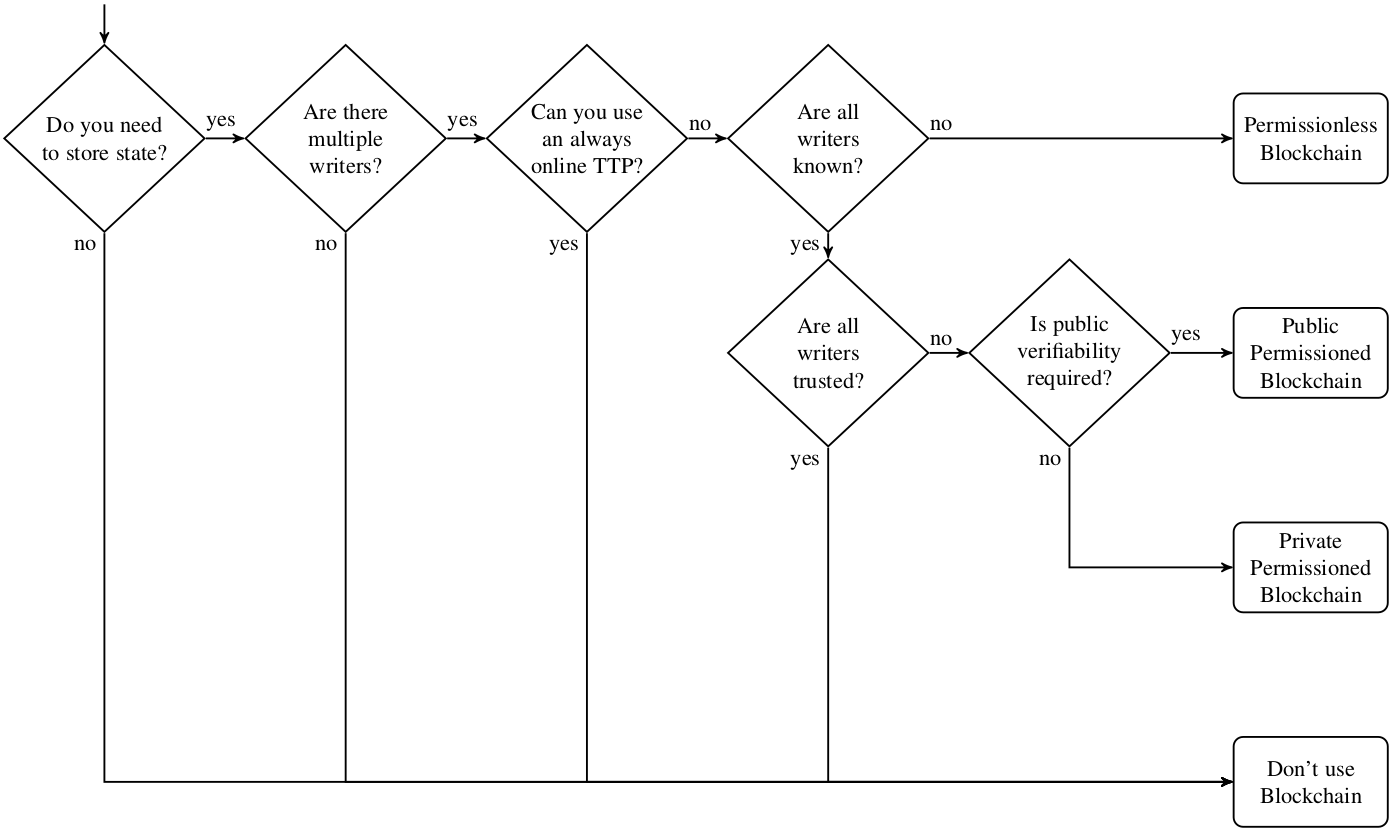}
\caption[Wust \& Gervais' blockchain adoption decision flowchart]{Wust \& Gervais' \parencite*{Wust2017} blockchain adoption decision flowchart}
\label{fig:doyouneedWust}
\centering
\end{figure}

Wust \& Gervais' \parencite*{Wust2017} provide a more precise decision tree at Figure \ref{fig:doyouneedWust} than Peck \parencite*{Peck2017}, with the initial question regards storing state usefully focussing on concepts rather than specific technologies.  With this version four out of seven paths lead to ``don't use blockchain,''  and singular paths lead to private permissioned, public permissioned and permissionless respectively.  One of their key decision points, as highlighted by the industry criteria, revolves around trust.  Here trust is defined as assuming ``no participant is malicious''; this is a far stricter definition than  expressed by industry, Greenspan \parencite*{Greenspan2015} for instance believing mistrust could exist within the same organisation when one unit was unwilling to let another modify its data.  Of particular relevance to this research is the authors' scepticism over supply-chain management DLT because of the digital-physical gap, or the difficulty of confirming the object that exists on-chain matches that in reality; potential solutions to their criticisms are examined in Section \ref{sec:SupplyChainProvenance}.  

\begin{landscape}
\begin{figure}
\includegraphics[width=\paperwidth]{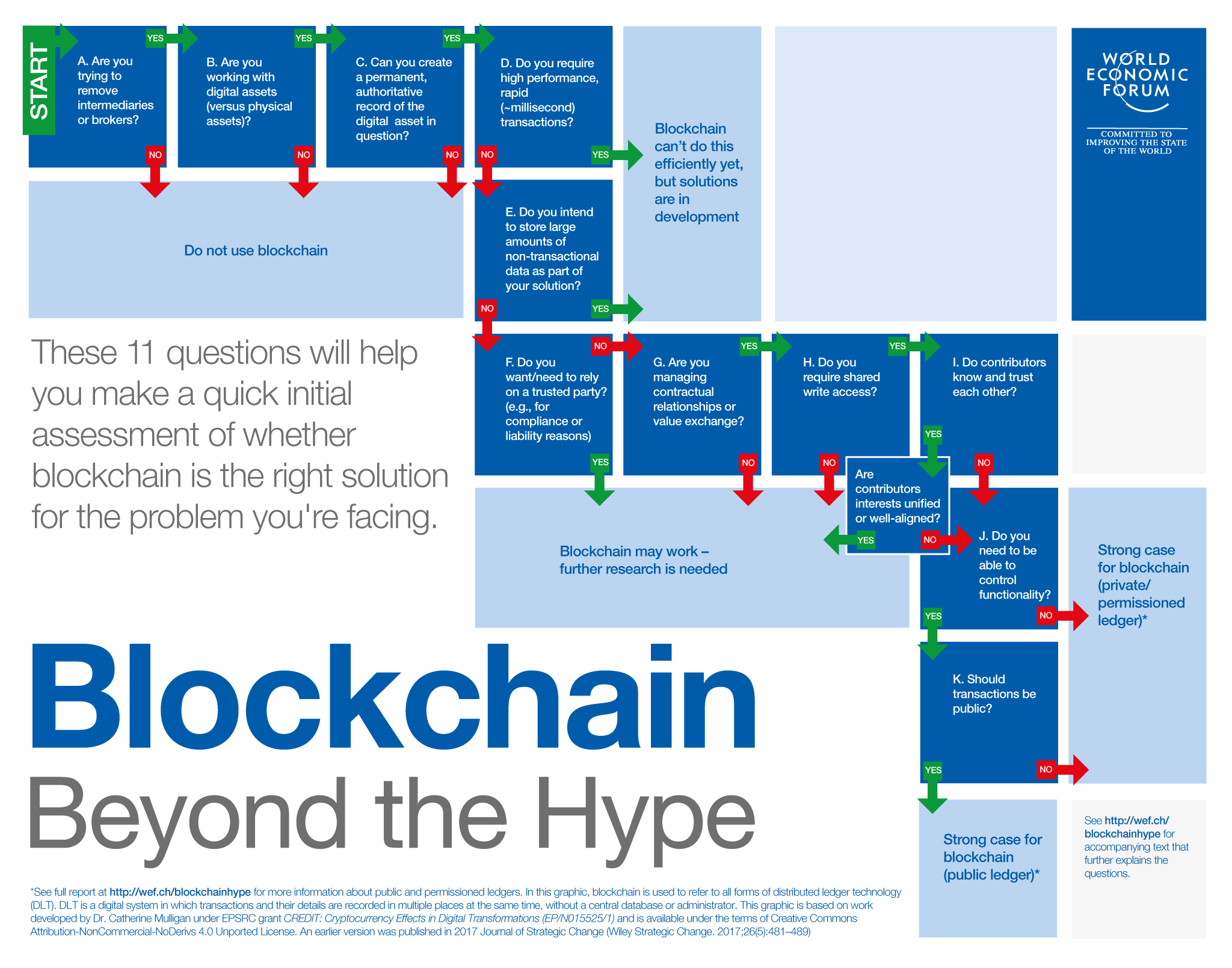}
\caption[World Economic Forum's blockchain adoption decision flowchart]{World Economic Forum's \parencite{Mulligan2018} blockchain adoption decision flowchart}
\label{fig:WEFFlow}
\centering
\end{figure}
\end{landscape}

One of the more authoritative decision trees is published by the World Economic Forum at Figure \ref{fig:WEFFlow}, which is adopted from an earlier academic model \parencite{Maull2017}.  This decision tree is the most sophisticated reviewed with extensive descriptions of decision points and worked examples.  However there are contradictions: the  first step asks ``are you trying to remove intermediaries or brokers?'' yet the use case features an imaging company where ``the intermediaries are actually the boundaries of the firm that currently hold the GPU computational capacity for rendering images.''  It is difficult to imagine a firm  without boundaries - which effectively means that all firms could answer `yes' to the first question.  This does not detract from the use case (which is similar to Angrish et al.'s  \parencite*{Angrish2018} cybermanufacturing example), but does illustrate that an inflexible process to decide where DLT may fit, is unsuited to fluidly-defined DLT.

The academic world is therefore  more reticent to propose DLT as a solution for business problems than the commercial one.  There are a number of possible reasons - the first and most obvious being the lack of commercial incentive.  However, established companies such as IBM, SAP and Oracle are likely to prefer long-standing business relationships over short term gain over; they also potentially have more practical insight into business problems than academics.  A more subtle problem might be the ``not invented here'' syndrome \parencite{Antons2015}.  Narayanan and Clark \parencite*{Narayanan2017} comment that although Nakamoto's white paper was more novel than typical academic research, it was ignored by academia initially because of Nakamoto's rejection of academic norms such as peer-review.  Furthermore they critique that many academics continued to argue that theoretically Bitcoin could not work, even when the reality contradicted this.  Though this can only be taken so far - as the literature review shows, many academics are engaging with DLT.  Lastly the divergence of views might be explained by the difficulty of defining DLT; as discussed there is not yet a standard definition of DLT.  As Peck \parencite*{Peck2017a} admits elsewhere, DLT is like the parable of three blind people describing an elephant: ``one person feels the leg and it’s a tree trunk, another person feels the side of the elephant and it’s a wall, and a third person feels the trunk and it’s a snake.''  Regardless of reason considerable difference exists over when DLT should be applied to processes and when not, and in truth only its practical application will tell.

\section{An evaluation framework for DSN}
\label{sec:DSNevaluationframework}

Synthesising this elephant into one framework for evaluating DSN use cases is therefore difficult. 
However it is worthwhile; given the DSN's scope this research cannot cover all possible use cases, but it can contribute by designing a framework which can be used to shortlist use cases for development (Research Objective 3, Section \ref{sec:researchQAO}). 

A decision tree, following the examples in Section \ref{sec:Academic}, was considered, but rejected.  As Mulligan et al. \parencite*{Mulligan2018} inadvertently demonstrated `yes' or `no' questions can lead to contortions when valid use cases do not conform to a pre-designed question set; not helped by DLT being so widely-defined.  Rather an analogue approach is required which rates use cases on a spectrum of DLT adaptability.

Inspired by the Gartner Magic Quadrant series of data representation \parencite{Whitehorn2007}, the proposed framework takes the form of a plotted chart and is a visual aid.   Utility and ease of implementation form the two axes.  Use cases are plotted against these axes and are represented by a point whose size corresponds with the impact of the use case on the IS landscape of the organisation.  The definition of the terms utility, ease of implementation and impact are:

\begin{description}
\item[Utility.]  The benefit to the enterprise of applying DLT to a use case - this metric recognises that some processes are more suited to DLT-adoption than others.
\item[Ease of implementation.]  Although there might be utility in applying DLT to a process, this recognises that the business change required may be difficult or require substantial resources.
\item[Impact.]  The size of the change this will have on the IS landscape of the enterprise as a whole, for example will it affect everyone within the enterprise, or a small team.
\end{description}

\begin{figure}
\includegraphics[width=\textwidth]{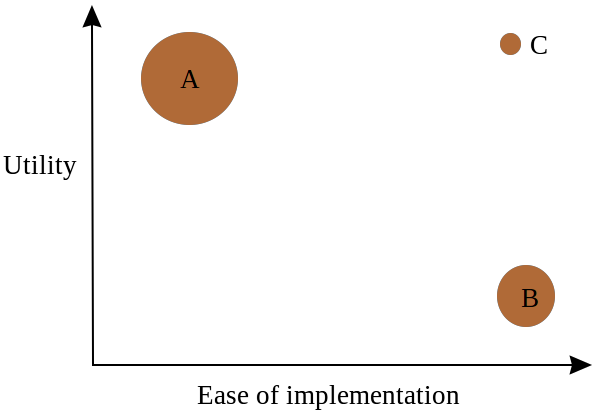}
\caption[Indicative example of evaluation framework]{Indicative example of evaluation framework}
\label{fig:graphdemo}
\centering
\end{figure}

Figure \ref{fig:graphdemo} is an indicative example of how this graph might look when populated.  This graph shows the assessment of use cases A, B and C.  Use Case A is a very good fit for DLT adaptation (utility), and will have a large effect on the enterprise (impact) but will be difficult to implement (ease of implementation).  Use Case B alternatively will have a medium impact and will be very easy to implement - however imposing DLT on the business process will produce few benefits as opposed to another method.  Use Case C is both a good fit for DLT and is easy to implement, however will have a small impact on the enterprise (e.g. minimal users, small revenue stream).    

Using the above exploration of commercial and academic literature, criteria have been selected to assess where use cases fall along each axis and how impact can be assessed.  Wherever possible these criteria utilise some measurable aspect of the enterprise so allowing objectivity.  It is recognised however that when dealing with complex social systems (such as an enterprise), and change to those systems, a completely metric driven approach is difficult to achieve \parencite{VonHayek1989}.  

\subsection{Utility Factors}

This axis measures the enterprise benefit gain by adopting DLT for a given use case.  It draws on the criteria identified above by Goodhue and Thompson's \parencite*{GoodhueThompson1995}, Greenspan \parencite*{Greenspan2015}, IBM \parencite[p.~37]{ManavGupta2017}, SAP \parencite{Roehricht2017} and Oracle \parencite{Goel2017}.  Three factors are selected for measuring DSN utility: multiparty, trust and auditability.  For each a score of between zero and five will be accrued, thus a maximum of 15 (with zero representing least utility).

Before this assessment can be made, there are two obligatory factors.  Firstly Greenspan's \parencite*{Greenspan2015} ``database'' criteria: a structured repository for information must be required.  Use cases without such a repository are awarded a maximum utility score of zero.  The second factor concerns Goodhue and Thompson's \parencite*{GoodhueThompson1995} `Task Equivocality' relating to ``ill-defined'' or ``ad-hoc, non-routine business problems.''  Although the data contained within a DLT might contribute to solving ill-defined business problems, ultimately it is stored in a structured fashion - so there must be an understanding of what problems exist for a structure to be imposed.  Therefore if the use case involves business problems which are poorly defined, DLT is unsuited and a maximum utility score of zero should again be awarded.  Once this is completed the factors can be scored as follows:

\begin{itemize}
\item \textbf{Multiparty}.  The number of organisational boundaries that data is shared across in the course of the business process.  These relate to both internal boundaries (business units) and external (different enterprises). 
\begin{itemize}
\item[0]  If there are no organisational boundaries crossed then a multiparty score of zero is ascribed.
\item[1]  If one internal organisational boundary (i.e. two business units) are crossed then a score of one is ascribed.   
\item[2]  If two internal organisational boundaries (i.e. three business units) are crossed then a score of two is ascribed.   
\item[3]  If three or more internal organisational boundaries or one external boundary (i.e. two enterprises) are crossed then a score of three is ascribed.   
\item[4]  If two external boundaries (i.e. three enterprises) are crossed then a score of four is ascribed.
\item[5]  If three or more external boundaries are crossed then a score of five is ascribed.
\end{itemize}

\item \textbf{Trust}.  Whereas organisational boundaries can be precisely measured, trust is an abstract concept.  The literature on trust in organisations was reviewed \parencite{Dietz2006, McEvily2011} and the subject discussed in interview, which led to the selection of three trust-influencing factors:

\begin{itemize}
\item \textbf{Contractual relationships}.  Commercial relationships will affect trust.  If one organisation has a contractual relationship which includes performance penalties or bonuses it will likely have an impact on trust.  The same will apply when organisations share information with competitors.  Although this typically applies to external relationships, it could also apply to internal DSN business units - especially where written service agreements are used to hold to account.
\item \textbf{Organisational functions}.  If organisations are engaged in  different areas of the enterprise (e.g. accounting, personnel) there may be more reticence to let others alter their data, and therefore less trust.  This builds on the work of Williams \parencite*{Williams2001} examining outgroups in organisations and finding ``different functional areas ... view members of contrasting groups with distrust, suspicion, and animosity.''
\item \textbf{Culture}.  Some organisations are by their very nature less likely to trust others outside of their sphere - within DSN this applies to those business units which routinely deal with classified information (e.g. Special Forces, Submarines, etc) as opposed to more generic areas (e.g. commodity supplies).  Another example of this culture might be expressed by inter-service rivalry \parencite{Barlow1994}.
\end{itemize}

It is appreciated that the majority of these factors are subjective.   However they can be used to form the basis of an assessment of trust for a use case; which can then be followed up by more rigorous investigation.  Once the trust level has been ascertained it should be scored from zero denoting full trust, to five: least trust.

\item \textbf{Auditability}.  This factor is based around the importance of preserving the information stored in a DL.  The constituents here are:

\begin{itemize}
\item \textbf{Controlled Items}.  The DSN manages items that are subject to regulatory control either by domestic or international regulation.  Items that fall into this category are for instance those subject to the Polaris Sales Agreement, International Traffic in Arms Regulations (ITAR), Airworthiness items, Attractive to Criminal or Terrorist Organisations (ACTO), etc.  MoD will wish to assure itself of transactions involving these items, so adding weight to a DLT solution.  Use cases involving these should be awarded two points, depending on the proliferation of these items (e.g. zero for nil items, one for limited amounts of controlled items, two for a use case focussed on controlled items).

\item \textbf{Classification}.  Use cases that concern a classification higher than Official, will also benefit from DLT's immutability, typically (depending on the DLT implemented) by preventing adversaries from reading/writing data without leaving an audit trail.  Use cases featuring classified items should be awarded two points, depending on the extent of classification (e.g. zero for nil items, one for limited amounts of classified items, two for a use case focussed on classified items).

\item \textbf{Theft or Fraud}.  Use cases which are particularly susceptible to theft or fraud (e.g. high value / attractive or items susceptible to corruption) gain an additional one point.  This is relatively judgement based due to criminality's pervasive nature, however as there is policy ascribing certain items valuable and attractive, there is some basis for decisions to be made on. 
\end{itemize}

\end{itemize}

\subsection{Ease of implementation}

While the y-axis of Figure \ref{fig:graphdemo} represents how  suited a particular use case is to DLT adoption, the x-axis represents ease of DLT implementation.  This measure will allow the enterprise to consider factors such as time and cost, prior to deciding on what project to implement.  Studies were therefore analysed which defined taxonomies of constraints to successful implementation of IS \parencite{Yeo2002, Al-ahmad2009} - although what constitutes success \parencite{Agarwal2006, Shaul2013} was considered out of scope.  One common theme was the importance of soft factors e.g. project manager \parencite{Mohd2011}, team communication \parencite{Mohan2011}.  Despite this soft factors are not measured in the axis as they are typically reflective of management within the whole enterprise rather than specific to DLT use cases.  It is recognised that further studies may wish to consider this - for instance by measuring resistance to change or project manager skills.

Instead to determine ease of implementation constraints, two approaches were used.  Firstly adaptive change cases proposed to the Defence Logistics Directorate's Information Systems Working Group were analysed for common themes which had expedited or stalled implementation.  Secondly implementation factors were elicited from interviewees.  The above methods, in combination with an analysis of the above literature, resulted in four factors, again with a maximum score of 15 points (with zero points representing most difficulty):

\begin{itemize}

\item \textbf{Contractorisation}.  There are typically four hierarchies of software development within the DSN: end user computing (where a non-technical user creates scripts such as Visual Basic in Excel), software developers within the employ of MoD, independent developers contracted by a business unit on a project and lastly underpinning contracts with a third party for a suite of IS.  Each of these methods typically involves more bureaucracy than the previous and act as a drag on implementation.  Although it is recognised there is currently a dearth of DLT development expertise \parencite{Stein2018}, this is likely to change over time as the technology becomes more familiar; meaning all hierarchies, with the possible exception of end-users, will be able to develop DLT solutions.  Where a business unit has in house developers three points should be scored; independent developers: two; and underpinning contract: one or zero points depending on the contractual relationship.

\item \textbf{Manual process}.  The existence of a manual process was an adoption factor highlighted by all three blockchain vendors in Table \ref{table:DLTAdoptionComparison}.  However within this DSN evaluation framework it is used as an implementation rather than utility measure.  It does not feature on the utility axis as computer IS typically aims to improve a paper based process by digitising it \parencite[p.~12]{stair2012information}; this is therefore far from limited to DLT.  It does feature on the implementation axis as within a manual system it is easier to understand the flow of information and processes, than when that process is embedded in code which requires specialised skill sets to interpret.  To grade against this metric the researcher should note the number of manual transactions used (e.g. paper forms, faxes, phone calls), as opposed to digital and use that ratio to score from zero to four - i.e. all manual: four; half manual: two; all digital: zero.

\item \textbf{System age}.  The National Audit Office criticise the DSN for the antiquated nature of its logistic systems, noting that two of its main inventory IS began service in the 1980s \parencite{NationalAuditOffice2011}.  The issue of legacy IS interfacing with new projects is considered a critical success factor by Fui-Hoon Nah, Lee-Shang Lau \& Kuang \parencite*{FuiHoonNah2001} and was revealed as a point of contention with interviewees.  Therefore if the oldest system in the use case is less than five years-old four points are added, and with every five years one implementation point should be removed - so a 12 year-old system would have two points, and a 20 plus year-old system zero points.

\item \textbf{Complexity}. Beese et al \parencite*{Beese2016} propose that IS complexity makes adaptive change more difficult; although as Xia and Lee \parencite*{XIA2005} discuss there is not necessarily one agreed way to measure this.  Complexity has many definitions \parencite{Geraldi2011}, but for the purposes of this dissertation Beese et al's  structural complexity (as opposed to dynamic) is most useful; and can be sub-divided into number (size), variety and interdependence of systems \parencite*{Beese2016}.  This measurement will be a count of the numbers of systems required to integrate with the DLT implemented - four points for a simple web-front end (as comes standard with blockchains such as IBM Hyperledger's Fabric), three points for one to five systems, two points for six to ten systems, one point for eleven to twenty systems and zero points for over twenty.  Both DSN service catalogues (see Section \ref{sec:Impact}) can be used to ascertain this figure.

\end{itemize}

Although the factors for both utility and ease of implementation are rudimentary, it is nevertheless adequate to provide a scale for comparison of use cases.  In practice the scale is intended to be adapted as circumstances dictate.

\subsection{Impact}
\label{sec:Impact}

This last metric recognises that high utility and ease of implementation does not necessarily entail impact on the overall enterprise.  An example of this is a system which has low business criticality - for instance a trivial application for recording office chores; or one which has only a few users.

Naturally business impact will be specific to the enterprise in question.  If organisations are following IT service management best practice, as specified by the  Information Technology Infrastructure Library (ITIL), they will be maintaining a service catalogue, which should capture application business value \parencite[p.~104]{Axelos2012}.  The DSN has two applicable service catalogues: the Defence Application Register (DAR) listing all applications used within MoD, and the Support Chain Information Services Architectural Repository (SCISRA) which covers applications provided by SCIS.  Between them these contain the following pertinent information:

\begin{itemize}

\item \textbf{Users}.  An approximate number of users for an application and associated data such as number of licences and installations.

\item \textbf{Business Priority}.  A range of metrics for deciding business priority including ``criticality,'' ``impact,'' ``business importance,'' ''importance rating'' and ``deployment priority''.  Although there is considerable overlap these attempt to rate whether an application supports a business unit's primary function, and the inconvenience of a workaround were it not available; the last two also factor in the number of users an application has.

\item \textbf{Contractual Support}.  The level of support an application receives when it has been contracted out to third-party support, particularly Boeing Defence UK \parencite*{Boeing2018}.

\end{itemize}

It is unlikely that either the DAR nor SCISRA are entirely current, however they would be a good first source of information for measuring business impact.

It is proposed that ``importance rating'' above (which is scored 0 - 130) should be used to ascertain impact; by measuring across business priority and users it captures in a granular way the whole enterprise footprint of an application.  For the purpose of visually representing this data, as at Figure \ref{fig:graphdemo}, the impact score should be used as the circle's area, which can be used to ascertain the radius and circumference.  For instance a DAR Importance Rating of 55 would equate to a radius of 4.18 (i.e. 4.18 = \begin{math} \sqrt{\frac{55}{\pi}} \end{math}) .  The circle could then be drawn with a radius of 4.18mm or 4.18cm, or any other scale.  Where a use case covers more than one system then the highest rated Importance Rating should be used.  In cases where there is no application currently in use (e.g. a manual solution) then the DAR scoring matrix should be used to estimate an Importance Rating.

This mechanism should be adjusted as required - especially should service catalogues be found out of date.  However striving for an objective criteria when measuring business impact should improve the rigour of a DLT use case selection process.  

\section{Results of questionnaire}
\label{sec:Resultsofquestionnaire}

\begin{figure}
\includegraphics[width=\textwidth]{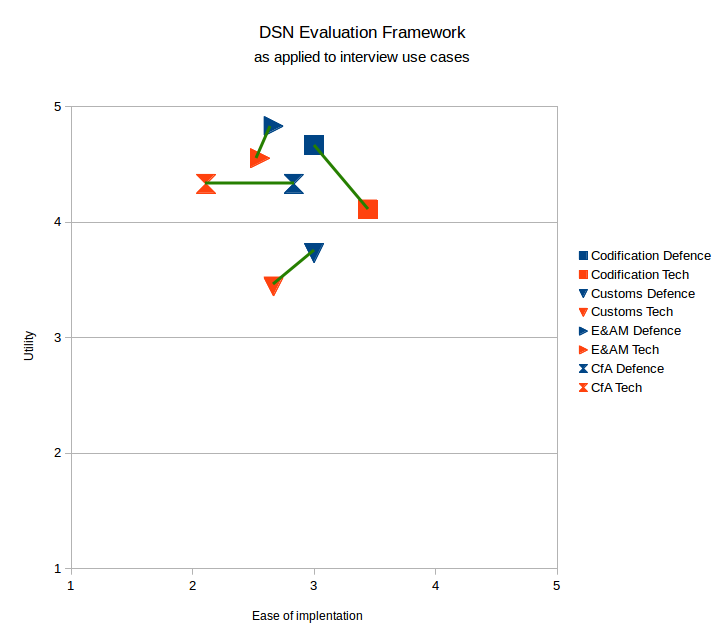}
\caption{DSN evaluation framework applied to interview use cases}
\label{fig:useCaseGraph}
\centering
\end{figure}

The proposed evaluation framework is designed to identify where DLT might be best adopted in the DSN.  However as generic use cases were used to prompt discussion within interviews, a lightweight version of the framework has been used to assess these.  Instead industry and MoD interviewees rated the utility and ease of implementation of the use cases (see Appendix \ref{ch:questionnaire}) based on their experience.  As the use cases are hypothetical; no measure of impact (Section \ref{sec:Impact}), which relies on an importance rating provided by the business, is given.  Data collected from these interviews was averaged amongst the two sectors and is presented at Figure \ref{fig:useCaseGraph}.

The ease of implementation axis ranges from one (very difficult) to five (very easy), while utility ranges from one (not at all useful) to five (very useful).  All use cases were rated above moderately useful (three) by all interviewees.  Three use cases (Customs, E\&AM, CfA) shared a similar pattern - compared to technology-sector interviewees, defence considered them easier to implement and greater (or equal) utility.  One possibility is that defence interviewees, being more aware of the problems the DSN faces, are more open to solutions.  Alternatively this may suggest a weakness in the research: the pro-DLT nature of the introduction videos might have led to a bias in defence interviewees, while technology interviewees may have a more balanced understanding of DLT.   

Codification (Use Case 1) was the exception, with defence-sector considering it more difficult to implement than the technology-sector (although continuing to believe it had greater utility).  Reasons for this will be considered below in Section \ref{sec:codification}.

The easiest use case to implement as rated by the technology-sector was Codification (Use Case 1), for defence-sector it was joint first with Customs (Use Case 2). Contracting for Availability (Use Case 4) was considered the hardest to implement by technology-sector, E\&AM by defence-sector.  Both defence and technology considered Engineering and Asset Management (Use Case 3) the most useful and Customs (Use Case 2) the least.

\section{Generic use cases exploration}
\label{sec:genericusecaseexploration}

This  section will discuss the qualitative data collected - the interviewees'  feedback on use cases.

\subsection{Codification - Use Case 1}
\label{sec:codification}

Codification was considered by technology-sector interviewees the easiest use case to apply to DLT.  A major factor in this was the lack of systems complexity - the Codification Support Information System (CSIS) is one relational database exclusively owned by MoD.  This was coupled with existing business processes - MoD already contracts in DEFCON 117 \parencite{MinistryofDefence2013} for this information to be provided - as one interviewee stated: ``we are masters of our own destiny.''  This would act as solid foundations for a DLT pilot - it could be accomplished internally and then expanded.  Another factor in favour of implementation was that the database contains NATO Stock Numbers (NSNs) which represent classes of real world objects, but avoids the complications of trying to track specific, unique items.  

The defence-sector gave higher ease of implementation ratings than technology-sector interviewees in all but one use case: the Codification use case (Figure \ref{fig:useCaseGraph}).  Partly this was because the UKNCB respondent, being intricately aware of the complexities of codification (e.g. other processes relying on the CSIS database), gave this a lower ease of implementation score.  However other defence-respondents were also sceptical of data sharing between MoD, prime-contractors and sub-contractors when competition existed in the network over this data (especially pricing); and therefore believed this would make implementation difficult.  All however rated the use case highly on usefulness - this was due to the multitude of third parties (e.g. Original Equipment Manufacturers (OEM)) involved in the network.

Interviews revealed there could be surprisingly little trust between DE\&S Delivery Teams (DT) as regards item data; for instance one commodity-based team might have no requirement for an item being certified (e.g. airworthiness), but this requirement might not be shared by a DT responsible for supporting complex equipment.  This led to a situation where although CSIS might have one value, a DT could amend this data on a Base Inventory System otherwise.  Participants believed DLT could achieve data integrity across systems, by ensuring all changes are immutably recorded and that consensus ensures only those with relevant permissions are allowed to update - although it might be argued a well designed relational database would also achieve this.  These findings agree with academic research \parencite{Banerjee2018} suggesting DLT would be a good fit for master data management - although authored by a vendor of DLT.

One factor that resulted in easy implementation - that MoD controlled the system and can demand that suppliers provide details, also attracted criticism regarding the absence of disintermediation.  Greenspan \parencite*{Greenspan2015} argues DLT is redundant if a``trusted intermediary'' exists - yet this in theory is the role of UKNCB.  There was inconsistency in responses to this, some interviewees stated industry does not always trust the MoD; for instance items are not codified because industry is unwilling to provide data containing intellectual property (e.g. technical drawings) or commercially sensitive information (e.g. price).  Others believed that although the MoD might be a trusted intermediary there was an issue when contractors were collecting codification data on behalf of other contractors, as in industry coalitions.   Potentially DLT could partly alleviate both problems by giving OEMs verifiable control over who sees what data - for instance price could be seen by the MoD, but not by other commercial companies.  It is unrealistic however to expect a technological panacea - industry is likely to wish to keep secrets regardless.  

The NAO has reported that codification has not occurred because of the desire to reduce costs \parencite[p.~31]{Parliament.HouseofCommons2017a}.  Interviewees considered that a UKNCB-run DLT might allow seamless integration of OEM's databases of items manufactured or procured, so reducing cost.  The emergent nature of DLT means that it is hard to predict whether DLT would be more cost-effective than a relational database, though it is possible. 

Where the disintermediation argument works in DLT's favour  is codification at a supranational level.  All NSN issued by the UK have the NATO Country Code `99' applied indicating the country of original codification, this is then shared with other participating countries via the NATO Master Catalogue of References for Logistics (NMCRL).  This is a loosely coupled system as participating countries need not be NATO members; simply approved `partner' countries - for instance non-NATO Singapore has the country code of `32' \parencite{NATO2018a}.  Using DLT the intermediary role of NATO could be considerably reduced, with associated cost savings.  In this scenario manufacturers in each country would submit transactions (e.g. the creation of a new item of supply) onto a DL; these transactions would then be checked against the business logic imposed by the relevant national codification bureau (e.g. all fields contain valid data, the item is unique, etc).  Passing the business logic would result in the transaction being cryptographically signed with that country's identifying code and submitted to a ledger visible to all NATO partners.  Information could be encrypted for release to specified parties - for instance detailed technical drawings might be submitted to the chain, but only viewable by the originating country and the OEM.  This use case deserves further investigation - either at the national or supranational level. 

\subsection{Revenue \& Customs - Use Case 2}

This use case was rated as the least useful by both sectors.  It was considered the second easiest to implement by the technology-sector, and joint second by defence.  Much of this lack of utility was due to this use case being the `odd one out' of the four - Revenue \& Customs is something that happens to the MoD, as a transactional cost of doing business, rather than something MoD sets out to achieve.  Correspondingly it was felt there was little point in MoD forging ahead with something outside its control.  A world-class DL could be established by DE\&S to capture this data; but if no one else could be persuaded to use it (e.g. HMRC, shipping lines, etc), it would be of no use.

Despite this lack of utility, it rated highly for implementation - coming second in line (although far behind codification).  The reasoning behind this was that interviewees strongly felt that much of the current manual process (e.g. bills of lading) was ripe for digital conversion.  Resultantly there was a strong feeling amongst interviewees that if another organisation was to lead on this, the MoD should be keen to participate.  Several interviewees were already aware of organisations pursuing similar endeavours (apart from the IBM-Maersk trial mentioned in the questionnaire) and that MoD would gain advantage by joining these schemes when they had reached scale, rather than piloting its own.

\subsection{Engineering \& Asset Management - Use Case 3}

This use case rated as the most useful for both sectors.  Part of this might have been cultural - the MoD has been criticised for being more focused on equipment and engineering rather than the logistics tail supporting that equipment \parencite[p.~8]{Parliament.HouseofCommons2017a}, and this may have been reflected amongst interviewees.  Although this seems unlikely as a reason alone, given that interviewee selection aimed to maximise those with broad experience of each use case. 

Several interviewees referred to current issues affecting Engineering \& Asset Management (E\&AM) systems across DE\&S.  JAMES (Joint Asset Management and Engineering Solutions), which records maintenance on land managed equipment, is used by the Armed Forces but is not implemented by contractors who manage third line warehousing on behalf of the MoD - which means MoD does not have visibility of its assets throughout the reverse supply chain.  One defence-sector interviewee highlighted that  errors occurred where MoD had lost sight of DE\&S assets managed by industry, and pointed to recent changes in logistic policy to prevent this.  Interviewees felt DLT, by integrating contractor systems with those used by the MoD, might allow this visibility.

IBM made this point most strongly during discussions.  Their view was that DLT is unlikely to replace current E\&AM systems, but rather it would act as back-end glue allowing assets to transfer seamlessly from the MoD to partners and back; so ensuring the integrity of records.  This solution may be more palatable than trying to assert one system across a plethora of suppliers, and allows parties to employ systems suited to their own needs.  There is evidence of this approach being investigated by industry, such as Rolls Royce with aero-engines \parencite{Bryan2018}.

This factor also explained why this use case was rated second most difficult by the tech-sector, and most difficult by defence.  Attempting to connect a DL back-end with established E\&AM systems would be challenging, likely requiring cross industry co-operation.  This could be difficult in a marketplace where vendors use competitive differentiation - rather than making their products interchangeable with rivals.  Were this to be achieved however all felt that the gains across the DSN would be significant.

\subsection{Contracting for Availability - Use Case 4}

The Contracting for Availability (CfA) use case was considered the most complicated of all - and therefore scored lowest in ease of implementation by the technology-sector, and second most difficult by defence.  One defence-interviewee argued that even routine commercial activities were difficult within MoD, let alone more ambitious plans. 

Examples of CfA were elicited during interview to confirm this use case.  One current contract with BAE pays a daily rate when a ship is available for tasking.  However when Operational Defects occur then the MoD gains `credits,' with more serious defects gaining more credits, which are subtracted from the daily rate.  Credits normally commence on the date a ship signals a defect and end when the ship signals the defect is rectified - however credits in some circumstances also cease when the contractor can prove dispatch of the part rectifying the defect.  Interviewees confirmed that a source of frustration was having to audit large amounts of hard-copy paper to compile credits from a variety of sources at the end of the reporting period.  The advantage of a DL which all parties, including the contractor, could append to, was instantly seen and considered a strong case for the technology.

This use case is not without flaws. A system which results in winners and losers will create incentives for people to try and game the system - for instance by claiming that stores were dispatched earlier than they were.  This might especially be the case where a penalty is imposed by a smart contract; if one can fabricate the data with no audit process, other than smart contract, then that incentivises malfeasance.  Defence and Security Technological Laboratories \parencite{Colley2016} raise similar concerns over incentives encouraging data to be generated to meet targets, rather than record the truth.

A further complication if this use case is to be taken to its ultimate conclusion and have payment conducted on the blockchain  using a cryptocurrency is what might be referred to as the `locked in problem.'  Take a scenario where the MoD were to agree a £12 million availability contract and was going to disburse that money using smart contract alone.  That would require the MoD providing \pounds12 million in the DL monitoring contract up-front; if they were not prepared to do this then there would be little assurance that they would abide by the terms of the smart contract - the SC might ask them to pay, but they could decide not to.  However putting these funds in at the start of the contract is problematic because it means they cannot be spent elsewhere.

CfA was therefore seen as very useful, but highly complex to implement in a reliable way.  

\section{Wider use case exploration}
\label{sec:Widerusecaseexploration}

A number of interviewees commented that the generic use cases presented were similar, with one technologist going as far as to state they were ``obvious.''  This is not necessarily negative, it is logical for discussion to begin with areas that are better understood.  The generic use cases acted as a jumping-off point for qualitative evidence gathering on other areas for consideration - the following section expands on this.

\subsection{Supply chain provenance}
\label{sec:SupplyChainProvenance}

An oft cited use case for DLT is supply chain provenance - i.e. verifying the authenticity of items and equipment.  The use of DLT within defence for provenance has been proposed by a number of parties including Barnas \parencite*{Barnas2016}, Colley \parencite*{Colley2016} and Hsieh \& Ravich \parencite*{Hsieh2017}.  It was not however selected as one of the four generic use cases within this research because of the perceived `digital-physical gap' problem.  This problem, discussed in Section \ref{sec:twosidestothecoin} as part of the share versus prove debate and raised by academia in Section \ref{sec:Academic}, is that although a digital object (e.g. a circuit board's identifying serial number) might exist in immutable form in a DL, the circuit board in hand is not necessarily the genuine article.  Although checking  the serial number of a circuit board against a serial number in a DL, as proposed by Barnas \parencite*{Barnas2016}, is better than no form of verification; any adversary (e.g. hostile nation states, criminals) capable of creating either a fake or malicious circuit board, will likely be capable of creating fake serial numbers.  

Several methods  could be used to mitigate the digital-physical gap:

\begin{itemize}

\item \textbf{Bubble-tag}\texttrademark ~ \parencite{Prooftag2017}.  An adhesive sticker (Figure \ref{fig:bubbletag}) incorporating a polymer layer that when attached to an item chaotically creates a pattern of bubbles which can be read electronically \parencite{Patraucean2010}.

\begin{figure}
\centering
\includegraphics[width=2.5cm]{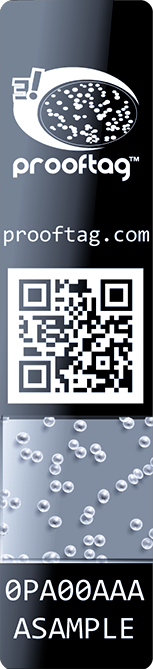}
\caption[Sample bubble tag]{Sample bubble tag\texttrademark ~ \parencite{Prooftag2017}}
\label{fig:bubbletag}
\centering
\end{figure}

\item \textbf{Q-ID}\texttrademark ~ \parencite{QuantumBase2017}.  A graphene layer (Figure \ref{fig:qid}) with atomic scale imperfections that reflect light in such a way that an identifying reference can be created \parencite{Roberts2015}. 

\begin{figure}
\centering
\includegraphics[width=18cm]{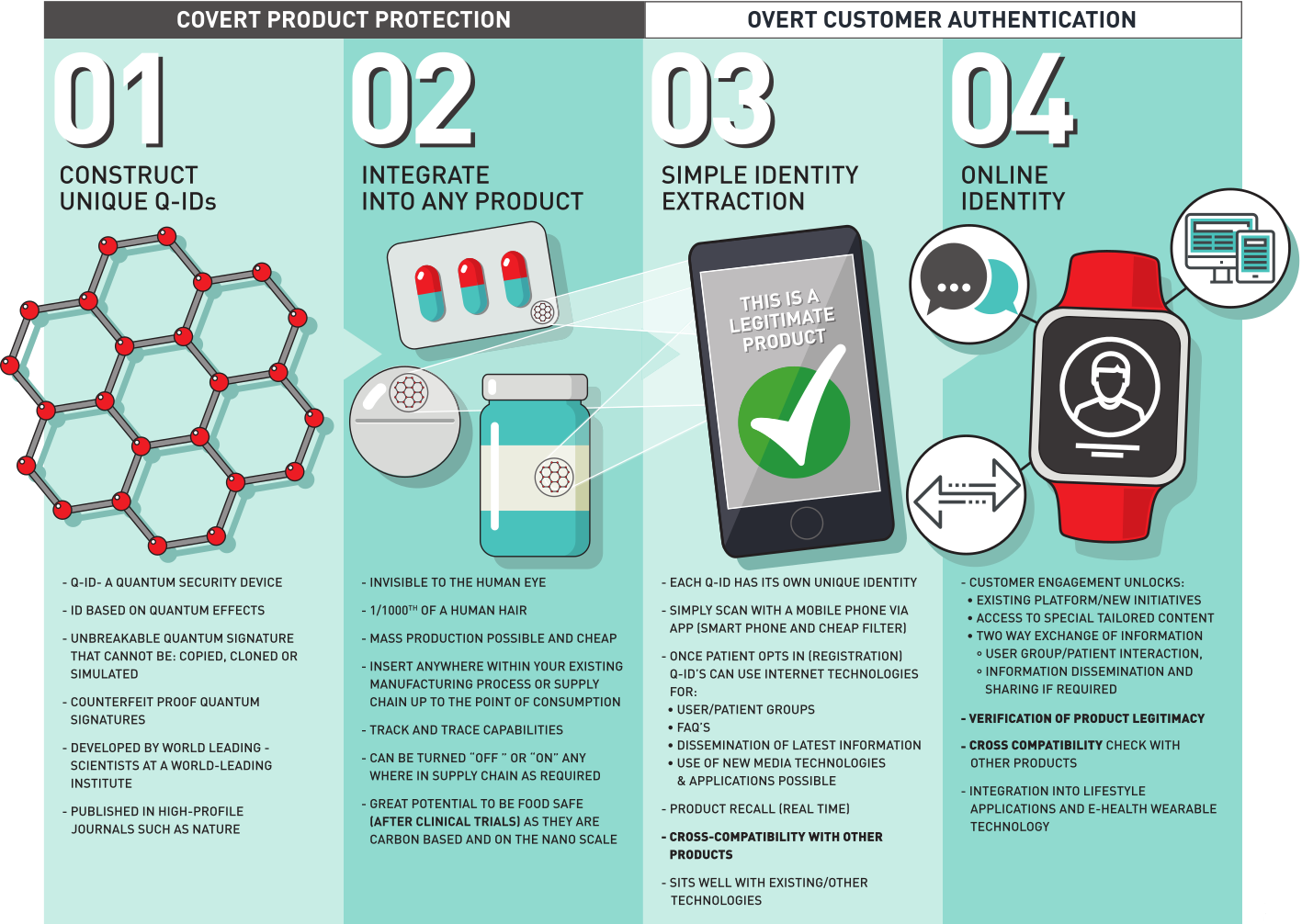}
\caption[Q-ID Infographic]{Q-ID Infographic \parencite{QuantumBase2017}}
\label{fig:qid}
\centering
\end{figure}

\item \textbf{CryptoSeal} \parencite{Chronicled2018}. A tamper-proof Radio Frequency Identification tag (Figure \ref{fig:cryptoseal}) read by Near Field-Communication enabled devices; as used by Thales in a DLT proof of concept to track sensor equipment.

\begin{figure}
\centering
\includegraphics[height=2.5cm]{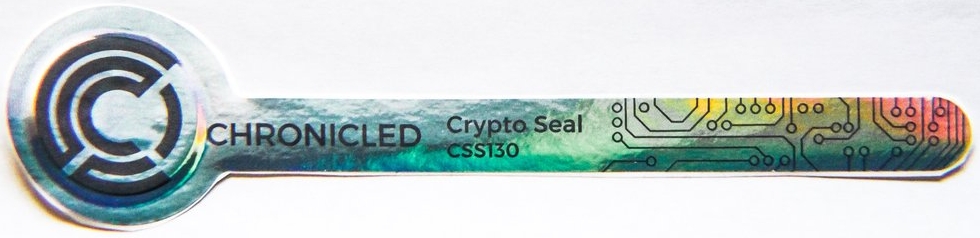}
\caption[CryptoSeal]{CryptoSeal \parencite{Chronicled2018}}
\label{fig:cryptoseal}
\centering
\end{figure}

\end{itemize}

Even more promising solutions exist for assemblies containing integrated circuits, as incorporated into the mentioned Thales DLT pilot .  Here software can be used to prove the existence of a physical unclonable function (PUF).  Due to variations in manufacturing all integrated circuits will have different physical manifestations - for instance nanosecond differences in logic gate operations.  As these characteristics are embedded in the circuit they form a signature that is  difficult to replicate or change as they are caused by a process outwith the manufacturer's control.  Although one attack vector might be to simply copy the signature once it has left the circuit, this is made more complicated by the fact that more than one test can be run: using different combinations of gates, different temperatures etc.  Again this facility is available commercially \parencite{Intrinsic2018} and has been academically assessed \parencite{Gu2017}.

The above examples are far from a comprehensive overview of the market; for instance it only links physical items to DLT, rather than processes.  These exist too however - the Fishface case study captured video-feed and GPS onto a DLT to record fishermen's catches \parencite{ZYenGroup2018} for fisheries protection purposes.   This summary however does indicate that the digital-physical gap is bridgeable.  

Ultimately it could be argued at a philosophical level the digital-physical gap can never be truly overcome; for instance how can one be certain that the original equipment manufacturer is producing what they claim to be producing.  However realistically few security solutions are flawless, and the techniques explored above considerably mitigate risk.  In view of this supply chain provenance is a use case that should  merit further exploration.

\subsection{Certification}
\label{sec:certification}

Closely tied to provenance, DLT could also play a role in certification; as noted by Defence Research and Development Canada \parencite{Willink2018}.  The defence enterprise is involved in many regulated activities: e.g. nuclear engineering, medicine, aviation and of course the application of lethality.  To ensure these activities are conducted in accordance with external laws and internal regulations, there are a plethora of certificates that apply to items of supply: from air worthiness conformity to hazard data sheets for cleaning products.  These are typically issued in paper form by an authority and accompany the item through its life-cycle.  This process is error prone: paper certificates can easily separate from the relevant item, especially as they cross organisational boundaries (e.g. store, repair, calibrate, etc).  This causes a range of issues: in the best case items are quarantined until a certificate can be sourced, in the worst case they are used in contravention of their design intent.

Recording certificates for items on a DL can assist with this, due to DLT's strengths in public-private key cryptography and the sharing of information across boundaries.  Imagine a scenario where a submarine depth gauge is due for periodic calibration: first a transaction is made in the ledger as the item is passed from depot to third party calibrator; the third party calibrator then records on the ledger that the item has been calibrated, the name of the individual who calibrated it and when that calibration expires, signing it with a private key so making it irrefutable.  When the item passes back to the unit, again recorded in the ledger, the calibration certificate is visible to all.  Should a problem be revealed later it is obvious where blame lies.  Furthermore as the certificate is digitised, information can easily be queried from the chain - it is a simplistic matter for the Commanding Officer to be presented with a list of all items requiring calibration prior to patrol.

The UKNCB interviewee raised a related problem.  When a new item of supply is codified, for instance a washer supplied by BAE, if certification is required a flag is raised on the relevant record on the codification system.  However when this record is passed to a base inventory system a project manager might append to that record an item from a different manufacturer performing the same function (for instance a washer bought off the shelf from B\&Q), but does not have the same certification or quality assurance guarantees.  This can result in units demanding what they believe is a certified part from one manufacturer and receiving a non-certified part from another manufacturer.  This problem can be prevented by encoding the business logic in the DL of parties authorised to add new part numbers to a NSN record.  Although this problem can be addressed in a relational database too, DLT's difference is that the certification flag could be established during the OEM creating the item record and promulgated seamlessly to all DL connected systems, rather than relying on post-event capture by the MoD.

Although not included in the generic use cases, following the course of research it is believed a strong use case for DLT is the management of certification in the DSN.

\subsection{Additive Manufacturing}

Additive manufacturing (AM), or 3D printing, is where computer technology is used to solidify material to create an object in three dimensions.  It has the potential to revolutionise the Defence Support Network as spare parts could be manufactured at the front line to satisfy immediate demands, rather than being stored and then shipped from a rear echelon area \parencite{Campbell2011a}. AM has inherent risks however - including maliciously altering the design files or copying them without authorisation (so stealing intellectual property).  Both the US Department of Defence \parencite{Dobesh2017} and Marine Corps \parencite{Daugherty2017} believe these risks might be mitigated by DLT, which they are piloting.  Dobesh \parencite*{Dobesh2017} proposes a DL as a ``ubiquitous data bus,'' meaning design files can be encrypted and only viewable by those with the right privileges, with the immutable nature of DL preventing alteration prior to manufacture.

This references back to Figure \ref{fig:ShareProveVenn}'s share vs prove Venn diagram.  An AM design file is a digital artefact, in much the same way as cryptocurrency, and therefore sits at the intersection where a DL can perform both the functions of sharing the artefact whilst at the very same time validating it.  AM  within the DSN is at an embryonic stage, therefore this use case is not an immediate one; although  has considerable future potential.

\subsection{MoD Coin}

Possibly the most unconventional use case is that inspired by discussion with Agility Sciences: MoD Coin.  Here when the MoD contracts a good or service it issues to the contractor a token, MoD Coin, stored on a DL.  That token is redeemable when the contractor produces the good or service to an agreed specification, at which point the token holder can present it to Defence Business Services for exchange with fiat currency.  This then allows the contractor who has been issued the token to raise funds to procure sub-assembly parts or services - they might for instance use them as collateral on a loan from a bank.  Alternatively the token could be paid directly to the manufacturer of the sub-assembly, who would then be able to redeem it when the final product was delivered.  This latter might also be used to gain business intelligence into the MoD network of sub-contractors, below the prime-contractor level; although it would be questionable as to how deep into the chain this reached - it seems unlikely every sheet of bubble-wrap is going to be accounted for via this method.

More sophistication could be added to this by other services utilising the MoD coin's DL.  For instance suppliers could be reviewed: positive for early delivery, negative if late or not to specification.  This would help DE\&S Delivery Teams evaluate suppliers used previously by others.  The objects procured using MoD Coin (assuming the object itself was recorded on chain) could, as previously discussed, have certifications (e.g. tolerances, etc) recorded against that unique item or batch; this could even be applied by a third party who cryptographically signs to prove quality control - a condition of the MoD Coin being redeemed might be that this certification is present; both verification and payment could then be accomplished via smart contract.  Indeed the ultimate conclusion of this is that contractors might even use MoD Coin as a guarantee to issue their own coin (e.g. BAE or Babcock Coin), on the same basis as fractional reserve banking, knowing that it is unlikely all their suppliers would redeem those coins at the same time.

This vision is an exciting possibility, but also the most ambitious use case laid out in this research.  There would be many barriers to adoption - not least government accounting rules, but also whether they would be accepted by suppliers.  Despite these barriers this could be worthy of further research when this technology is considerably more mature.

\subsection{Experimentation}

If the last use case was unconventional, this final one is counter-intuitive: it does not start with a use case.  Instead this approach proposes individual business units experiment by adopting DLT to capture information (alongside more traditional methods), learn from their application and observe where use cases emerge.  This is contrary to traditional MoD processes which begin with capability analysis and requirements; not surprisingly this strand was inspired by interviews outside the Defence sphere (namely Z/Yen Group who principally serve the financial industry).  Although this use case has risks, specifically that tax payers money will be expended without deliverables; it is feasible - MoD has the organic capability to experiment in this way.  After considering DLT's grand visions, this concept is refreshing in proposing emergent change arising out of needs identified from the user-base.  This also tallies with Holmes \parencite*{Holmes2017} who suggests that digital disruption is only learnt in organisations by doing and that small scale pilots are essential to de-risk, optimise and develop.  Whether this theme is taken forward will likely be the result of individual decisions in various business units, although buy-in by management will still be required; this research hopes that will be forthcoming.  

\chapter{Conclusion}

DLT emerged from anarchic beginnings, but has the potential to impact all sectors of society, the economy and government.  Its value comes from allowing different organisations to reach consensus on shared data, so removing much of the transactional friction common in business processes.  This dissertation thematically reviewed the literature; reported results from interviews with technical and defence-sector employees on the use of DLT in DSN and proposed a framework for evaluating the utility, ease of implementation and impact of DL use cases. 

DLT's utility is not yet proven however, with few existing real-life applications in industry. Indeed even its definition remains open (Section \ref{sec:searchstrategy}), Figure \ref{fig:ShareProveVenn} illustrated this showing DLs can look and act radically different from each other and solve divergent problems.  The fountainhead of this activity - Bitcoin - although potentially revolutionary, is unlikely to have direct applications with the DSN.  The evolutionary branch of Bitcoin's domesticated off-spring, the permissioned DL, is however worthy of closer examination as regards the DSN.

This therefore is a conditional endorsement.  Applications of permissioned DLT outside of test environment are rare, with the exception of purely `prove' deployments such as Guardtime KSI Blockchain.  If efficiency gains were more obvious there would be larger scale deployment; much current interest is doubtless driven by hype and greed.  However given that this is an emerging technology this is expected.  

Even if large scale DLT adoption occurs there is no guarantee that the DSN will prove a fertile ground for it.  One of the key tenets of DLT is disintermediation.  While this attribute is strongly suited to scenarios involving transactions without central authority and  trust - perfect for digital cash for example; it is less easy to apply to government departments.  After all in this instance the trusted central authority to disintermediate, MoD, is the very entity seeking to use DLT - a serpent eating its own tail.

On the other hand there are pros for adoption in the DSN too.  DLT is likely to be particularly useful when it is applied to situations where assets or services extend outside the boundaries of the MoD into industry or allied nations; this is one of the hallmarks of the DSN.  Likewise the immutable nature of DLT is particularly useful considering the regulatory environment much of the DSN takes place in.

Therefore the MoD should approach DLT cautiously, running pilot projects before making large scale investment.  Due to the pros and cons of adopting DLT in the DSN, an evaluation framework is required to objectively assess pilot use cases.  This research has proposed such a framework, which measures for utility, ease of implementation and impact; it is hoped this will help identify DSN processes for DLT adoption.  It may also form the basis of assessment in other areas of MoD, or related industrial sectors.

This research did not aim to identify specific DLT use cases, rather generic use cases were selected for data gathering.  Of these use cases studied, codification (Section \ref{sec:codification}) stood out, both on utility and ease of implementation, as a worthy contender for further study.  Although not included in the questionnaire, it became apparent in the course of research that a use case involving certification (Section \ref{sec:certification}), possibly coupled with supply chain provenance (Section \ref{sec:SupplyChainProvenance}), also deserves further investigation - as it could greatly aid the challenges the enterprise faces in this direction.  It should be emphasised however that the use cases covered in this research are the genesis for further work, such as feasibility analysis, and not a definitive statement.

\section{Strengths and limitations}

Exploring the potential of DLT is timely given high-levels of government, business and academic interest.  As a result there is a plethora of research being published, and interviewees have been willing to contribute.

Conducting this research from within the MoD has also been a benefit - previous experience meant that all use cases proposed were identified by interviewees as being useful (Figure \ref{fig:useCaseGraph}).  Selecting interviewees from both defence and technology-sectors allowed a variety of insights to be gained which was useful for considering utility and implementation factors and meant that enough data was collected to be confident in providing concrete recommendations.  The unique contribution of this research is proposing a framework to consider how DLT might be adopted, variations of this framework might have wider application than the current vogue towards binary-choice flowcharts.

As an emergent technology, DLT research is being generated at a rapid rate: Bano et al \parencite*{Bano2017} calculates one paper is produced every day and a half, thus one inevitable limitation is that this research will have a short shelf life.

Another limitation concerns the sample. Participants might have a positive bias towards DLT: technology interviewees were naturally bullish regarding its long term success; defence interviewees might have been biased due to the pro-DLT introduction videos. It must be noted however that feedback from interviewees indicate they were unlikely to be blindly positive, as critical discussion was held on the relative merits and drawbacks of DLT in different areas. Future research could involve a bigger sample and creating a more `neutral' introduction to DLT to reduce bias. Partly this bias is a function of its novelty - as time passes and there are more deployments of DLs a stronger evidence base will be produced as to where this fails or succeeds. 

Another lesson for future research is that much evidence may not be in English - considering the global activity in this - particularly in Asia (see Section \ref{sec:searchstrategy}); any following analysis of current work in this field would have to account for this.

\section{Recommendations}

\begin{enumerate}

\item The MoD should pilot use cases of DLT within the DSN to establish whether efficiency gains can be made.  A cautious approach is recommended because of the emergent nature of DLT.  Wide scale adoption or investment would be unwise at this stage.

\item Pilot projects should be selected carefully because DLT will not be suitable for all use cases.  An evaluation framework would assist in objectively assessing use cases for DLT adoption.  This research has proposed such a framework (Section \ref{sec:DSNevaluationframework}); which measures for utility, ease of implementation and impact; either this framework or a similarly adapted one could assist in selecting pilot use cases.

\item Although the aim of this research was not the selection of pilot projects, it is noted that use cases involving codification (Section \ref{sec:codification}), certification (Section \ref{sec:certification}) and supply chain provenance (Section \ref{sec:SupplyChainProvenance}) appear particularly worth of further investigation.  

\item Once a use case has been selected for piloting then a private and permissioned DL is more likely to prove successful for enterprise use than a public and permissionless one.  This is due to both classification issues and the requirement within permissionless DLs for there to be an incentive to prevent malicious actors (Section \ref{sec:conclusionsAndImplications}).  As DLT covers a wide variety of mechanisms for storing data, further consideration will be required to choose the appropriate DL for the selected use case.

\item Further academic research into organisations that have trialled DLT would benefit the literature greatly.  Much real world enterprise use of this technology is currently disseminated only via press releases or corporate communications, so lacking objective assessments of benefits and challenges.  Gaining access to conduct this research is likely to prove difficult given that participants may prove reticent to discuss failure, which could occur frequently in this emergent field.

\end{enumerate}

\appendix 

\chapter{Interview questionnaire use cases} 
\label{ch:questionnaire}

\section{Use case 1 - Codification}

\paragraph{Problem}
All items demanded within the Joint Supply Chain (JSC) should be codified with a NATO Stock Number.  Unfortunately this is often not the case causing ‘grit’ within the JSC.  Non-codified items cannot be demanded in an automated way and resultantly cannot be accounted for, performance managed or have stock holding levels optimised.  The amount of non-codified items being demanded from deployed platforms can approach 50\% of total demands (HMS DARING’s Current Operations\textbackslash Sidereal\textbackslash J4\textbackslash SC PM\textbackslash dated 20 Oct 13).  A contributing factor to not codifying items is financial cost to codify each item and a workflow penalty (e.g. completing manual forms).

\paragraph{Use case}
UK National Codification Bureau (UKNCB) maintain a blockchain for the codification of items.  Commercial companies wishing to codify items provide information on these items which are accepted by UKNCB as valid onto the blockchain.  Contents of blockchain are visible to all participants – except where deliberately restricted e.g. price for commercial confidentiality.  Companies are able to update blockchain as items become obsolete or superseded, with minimal oversight required. Ideally companies integrate their blockchain node into internal databases of suppliers so allowing frictionless update.  A more developed version of this use case would include sharing this data with NATO partners, so allowing visibility of NATO Stock Numbers outside of 99 (UK) only.

\paragraph{Advantages}
As process moves from manual system of updating database to a more automated one, cost of transactions reduce; more items are put forward for codification.

\paragraph{Disadvantages}
UKNCB already have a mature technology for recording codification (eISIS), will require some form of integration.

\section{Use case 2 - Revenue \& Customs}

\paragraph{Problem}
Dispatching items overseas to meet deployed units, especially by sea freight, can generate large amounts of paperwork for import / export purposes.  The MoD also has a requirement to provide information to UK authorities (HMRC) on items it is exporting – the WATERGUARD project has been set up specifically for this.  Consignments are frequently delayed in customs, so resulting in late delivery.

\paragraph{Use case}
A blockchain is used to provide information to all authorised parties on the dispatch and handling of sea freight containers, whilst also providing data to HMRC.  This use case has already been established by a pilot project run by IBM and Maersk, Defence might implement these mechanisms within their own processes and reporting.

\paragraph{Advantages}
Financial savings as less paperwork required to be raised by MoD, operational effect as units receive demands in quicker time.

\paragraph{Disadvantages}
The proportion of sea-freight dispatched by MoD is relatively small compared to commercial parties (e.g. Maersk), it is uncertain if other participants will be prepared to change ways of working for small proportion of total.

\section{Use case 3 - Engineering \& Asset Management}

\paragraph{Problem}
Repairable items are frequently required to leave the custody of the MoD for third-line repair and calibration.  MoD would wish to understand the location of these items at all times (especially items Attractive to Criminal and Terrorist Organisations (ACTO)) and what modifications have been performed.  Likewise Original Equipment Manufacturers (OEM) would wish to understand what maintenance has been performed on items at front-line.

\paragraph{Use case}
As transactions are performed, such as movement of items from MoD to OEM and maintenance / calibration they can be recorded on the blockchain, likewise information such as end of life dates, maintenance frequency etc could also be registered for the asset.  This could be integrated with existing systems within MoD and OEM (e.g. JAMES/UMMS).   A more developed version of this use case would involve smart contracts whereby any item out of date for service is automatically flagged, or any breach of contract (for instance MoD not returning within certain time) could be notified to legal/commercial teams.

\paragraph{Advantages}
Both MoD and OEM have more confidence in equipment state, which possibly increases reliability as maintenance regime better understood.  ACTO incidents (and other losses) could decrease as whereabouts of items more thoroughly understood.

\paragraph{Disadvantages}
Given that E\&AM systems are already established, integration could be required – if not risk ‘swivel-chair’ entry to enable this use case.  One might argue current losses and maintenance recording are cultural as much as technological, therefore a purely technological fix may not result in significant improvement.

\section{Use case 4 - Contracting for Availability}

\paragraph{Problem}
Equipment is often contracted by MoD on the basis of availability or reliability.  Currently manual processes are used to measure this availability / reliability and can result in disagreements with suppliers over the true picture (financial penalties typically leading to a low-trust environment).

\paragraph{Use case}
To record in DLT events which signify contract Performance Indicators e.g. dispatch of repairable items or advice by contractor, time equipment spent offline, capability losses, etc.  Smart contracts can be used to notify commercial/legal teams of contract compliance or otherwise – very advanced model could even include automated payment (although likely to require considerable running in period).

\paragraph{Advantages}
Financial savings as less manual intervention and human error is involved in overseeing contracts.  As both parties better understand terms of contract they are entering into, future contracts may be optimised as risks or costs are more transparent.

\paragraph{Disadvantages}
Commercial confidentiality may mean unwillingness to clearly define terms of contract.  One could argue that a level of contract obfuscation may be deliberate for some parties to ensure Performance Indicators consistently met, which may make some reluctant to enter this form of monitoring.

\end{document}